%% file: main.tex
\documentclass[superscriptaddress,twocolumn]{revtex4-2}
\usepackage[caption=false]{subfig}
\usepackage{float}
\usepackage{amsmath, amssymb, amsthm, wasysym}
\usepackage{pgfplots}
\pgfplotsset{compat=1.14}
\usepackage{graphicx}
\graphicspath{{figures/}}
\usepackage{bm}
\usepackage{xcolor}
\usepackage{comment}
\usepackage[utf8]{inputenc}
\input{preamble}
\input{gp_macros}

\input{tikz_macros}

\usepackage{tikz-feynman}
\usepackage{caption}
\usepackage{hyperref}
\hypersetup{
  colorlinks = true,
  linkcolor =blue,
  anchorcolor = red,
  citecolor = blue,
  urlcolor = blue
}

\tikzstyle{circ}=[shape=circle, inner sep=2pt, outer sep =0.5pt, line width=0.4mm, draw, node contents=]

\tikzstyle{dashhorizontal}=[shape=rectangle, inner sep=0pt, minimum height=1pt, minimum width=5pt, draw,fill,node contents=]
\tikzstyle{dashhorizontallarge}=[shape=rectangle, inner sep=0pt, minimum height=1pt, minimum width=10pt, draw,fill,node contents=]
\tikzstyle{dashrotate}=[shape=rectangle, inner sep=0pt, minimum height=0.2pt, minimum width=10pt,rotate=75, draw,fill,node contents=]
\tikzstyle{dashvertical}=[shape=rectangle, inner sep=0pt, minimum height=5pt, minimum width=1pt, draw,fill,node contents=]
\tikzstyle{dashverticallarge}=[shape=rectangle, inner sep=0pt, minimum height=10pt, minimum width=1pt, draw,fill,node contents=]
\tikzstyle{dot}=[fill,circle,inner sep=0pt,outer sep=0pt,minimum size=8pt,label={[label distance=0cm]#1}]
\tikzset{baseline={(current bounding box.center)}}

\makeatletter
\newcommand{\polarity}[1][]{\@ifempty{#1}{\mu}{\mu_{#1}}}
\newcommand{\density}[1][]{\@ifempty{#1}{\rho}{\rho_{#1}}}
\newcommand{\densityRight}[1][]{\@ifempty{#1}{\density_{+}}{\density_{#1\,+}}}
\newcommand{\densityLeft}[1][]{\@ifempty{#1}{\density_{-}}{\density_{#1\,-}}}
\makeatother
\newcommand{\densityLR}{\density_{\pm}}
\newcommand{\ABindex}[1]{[#1]}
\newcommand{\current}{J}
\newcommand{\currentOptim}{\Optim{\current}}
\newcommand{\fullPot}{\Phi}
\newcommand{\extPot}{U}
\newcommand{\extPotOptim}{\Optim{\extPot}}
\newcommand{\indicator}{\mathcal{I}}
\newcommand{\action}{\mathcal{A}}
\newcommand{\matrixHarm}{\underline{\underline{A}}}
\newcommand{\actionHarm}{\action_0}
\newcommand{\actionPert}{\action_1}
\newcommand{\aveHarm}[1]{\left\langle #1 \right\rangle_0}
\newcommand{\invBareProp}[1]{\Gamma_{#1}}
\newcommand{\invBarePropO}[1]{\Gamma^0_{#1}}
\newcommand{\Appref}[1]{App.~\ref{sec:#1}}
\newcommand{\SMref}{Suppl.~\ref}
\newcommand\textmicrometer{~{\textmu}m}
\newcommand\micrometer{%
\ifmmode\textnormal{\textmicrometer}%
\else\textmicrometer%
\fi}
\newcommand{\nuRoc}{\nu_{\text{\tiny{roc}}}}
\newcommand{\nuRocSmall}{\nu^<_{\text{\tiny{roc}}}}
\newcommand{\maxPowNu}{\NC_\nu}
\newcommand{\nuRocOptim}{\Optim{\nuRoc}}
\newcommand{\nuRocSmallOptim}{\Optim{\nuRocSmall}}
\newcommand{\nuOptim}{\Optim{\nu}}
\newcommand{\peclet}{P\kern-0.1em e}
\newcommand{\qeclet}{Qe}
\newcommand{\pqRatio}{\kappa}
\newcommand{\efficiency}{\eta}
\newcommand{\Optim}[1]{\widehat{#1}}
\newcommand{\matrixM}[1]{\underline{\underline{M_{#1}}}\,}
\newcommand{\SSpropline}{black, line width=0.2mm, -}
\newcommand{\SSpropdashed}{black, line width=0.2mm, -,boson}
\newcommand{\SSpropgluon}{black, line width=0.2mm,gluon, -}

\begin{document}

\title{Optimal Ratchet Potentials for Run-and-Tumble particles}

\author{Zigan Zhen}
    \affiliation{Department of Mathematics, 180 Queen's Gate, Imperial College London, SW7  2AZ  London}
    \affiliation{Centre for Complexity Science, Imperial College London}
      \email[Correspondence email address: ]{g.pruessner@imperial.ac.uk}
\author{Gunnar Pruessner}
    \affiliation{Department of Mathematics, 180 Queen's Gate, Imperial College London, SW7  2AZ  London}
    \affiliation{Centre for Complexity Science, Imperial College London}
    \email{z.zhen19@imperial.ac.uk}
\date{\today} 

\begin{abstract}
Run-and-Tumble particles, mimicking the behaviour of microorganisms like \latin{E.~coli}, are a paradigmatic model of active matter.
Due to self-propulsion, their random and undirected motion can be rectified in a ratchet potential. Using perturbative field theory, we determine the shape of the potential that produces the maximum particle current as a function of the particles' parameters.
\end{abstract}

\maketitle
Active matter systems, operating away from equilibrium, are composed of 
agents that consume energy from the environment to exert mechanical work \cite{Marchetti2013Jul}.
Even when their motion is isotropic in free space,
they can break detailed-balance by producing a spontaneous and directional motion in asymmetric  environments \cite{Doering1992Oct, Magnasco1993Sep, Astumian1994Mar, Pavliotis2005Sep, Galajda2007Dec, Martin2021Mar, Angelani2011Dec, Baek2019, Angelani2009Jan,Koumakis2014Jul}
such as moving a ratchet wheel
\cite{DiLeonardo2010May}.
Such directed motion is known to control activated events in glassy systems \cite{Berthier2013May}, transport biological molecules \cite{Hagan1989Oct, Lindner2001Mar, Ghosh2013Jun, Bressloff2013Jan, Brenner1990Dec, Woillez2019Jun, Bijnens2021Mar, Walter2021Jan} and  revert the Ostwald process in active fluids \cite{Tjhung2018Sep}.
Compared to the many observational studies, relatively little theoretical progress has been made to quantify and optimise non-equilibrium transport from first principles
\cite{Astumian1994Mar,Angelani2009Jan, Berger2009Mar}.

In the present work, we find the periodic ratchet potential that optimises the steady-state current of one-dimensional Run-and-Tumble (RnT) particles \cite{Tailleur2008May}. These are diffusive particles that move ballistically by self-propulsion until they change direction instantaneously and spontaneously under a Poisson process \cite{Zhang2021Jun}, taking place symmetrically in both directions.
Feynman famously used a ratchet to illustrate that useful work cannot be extracted from diffusive particles \cite{Feynman1963}.
That the random motion of RnT particles can be rectified at all is somewhat counter-intuitive and possible only because the spatial scale of the external potential is commensurate with the particles' mean free path \cite{Cates2012Mar,Doering1992Oct, Magnasco1993Sep, Astumian1994Mar}.

Rectified unidirectional motion of RnT particles
in an asymmetric sawtooth potential has been investigated both theoretically and experimentally in \cite{DiLeonardo2010May, Astumian1994Mar, Angelani2011Dec, Koumakis2014Jul}.
The locomotion of \latin{E. coli} can be rectified by means of microfluidic funnels,
which were further studied through theoretical models and numerical simulations \cite{Ghosh2013Jun}. Such studies have illustrated the far-reaching technological implications of active particles for example for drug delivery or for separating particle species by activity.
Despite the efforts made in experimental and numerical studies \cite{Bijnens2021Mar, Coppola2021Jul}, not a lot of light has been shed on the design principles of optimal transportation of active particles in ratchets.
Extending well beyond
existing work on RnT particles in specific potentials \cite{Astumian1994Mar, Angelani2011Dec}, 
in the following
we consider completely general potentials in a perturbative field theory, without making any approximations. While the perturbation theory can equally be implemented using classical methods, the field theory provides a systematic, diagrammatic framework to keep track of the contribution at every order.
We subsequently use a numerical scheme to find the potential, such as \fref{p=1_q=1}, that maximises the particle current.
\begin{figure}
\resizebox{.9\linewidth}{!}{
\centering
\newcommand{\figOlab}[1]{\textcolor{gray}{\small{#1}}}
\input{Fig_p=1_q=1}
}
\caption{
Optimal potential $\extPotOptim(x)/D$ for $D=1$, $w=1$, $L=1$ and $\gamma=1$, so that $\peclet=\qeclet=\pqRatio=1$,
resulting in the current $\currentOptim L^2/D=0.03789\ldots$ to the right and $\nuRocSmallOptim=1.167\ldots$.
The details of the numerical scheme employed to produce the plot are discussed in \SMref{plotting_procedure}.  
}
\flabel{p=1_q=1}
\end{figure}
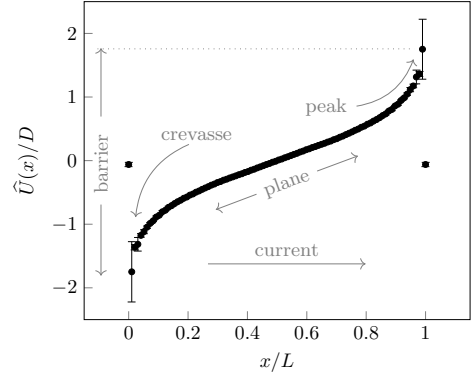

\paragraph*{Model and method.}
In this work, we study a single RnT particle with position $x(t)\in[0,L)$ in a potential $\fullPot(x)$ on a ring with circumference $L$. The motion is governed by the Langevin equation
\begin{equation}
    \dot x(t)=- \fullPot'(x)+wu(t)+\sqrt{2D}\xi(t)\elabel{Langevin},
\end{equation}
where
$\fullPot'(x)$ is the derivative of $\fullPot(x)$,
$w$ is the particle's self-propulsion speed,
$u(t)$ is a telegraphic noise switching between $1$ and $-1$ with Poissonian rate $\gamma$
and $\xi(t)$ is a unit Gaussian white noise with correlator $\ave{\xi(t)\xi(t')}=2D \delta(t-t')$
and diffusion constant $D$. The motion \Eref{Langevin} is subject to periodic boundary conditions with period $L$ and $x\in[0,L)$, so that the process has a steady state and effectively $\fullPot(x)=\fullPot(x+L)$.
Due to the self-propulsion, the system is driven out of equilibrium and its invariant measure is generically not of Boltzmann form $\rho(x)\propto e^{-\fullPot(x)/D}$.
\paragraph*{Field theory.}
In what follows, we calculate the steady-state density $\density(x)$ of an RnT particle \Eref{Langevin}, as well as its steady-state current $J$ in a perturbation theory in the coupling $\nu$ to the arbitrary potential $\extPot(x)$, such that $\fullPot(x)=\nu \extPot(x)$. To this end, we cast the RnT dynamics in the language of a Doi-Peliti field theory \cite{Doi1976Sep,Peliti1985Sep,Cardy2008Dec,Tauber2014Mar,Zhang2021Jun,Garcia-MillanPruessner:2022} as detailed in \SMref{FT}. Observables are more conveniently expressed in frequencies $\omega$ instead of direct time $t$ and discretised modes $k_a=2\pi a/L$ instead of real space, \Eref{def_Fourier}.

The propagators can then be written in a perturbation theory about $\nu=0$. Using the potential vertices \Eref{bauble_vertices},
each propagator picks up $2^n$ contributions to order $n$ in the perturbation theory, for example
\begin{align}\elabel{example_phi_phitilde}
\nonumber &\ave{\phi_a(\omega)\phitilde_b(\omega')}\\
&\corresponds
\begin{tikzpicture}[scale=0.6]
    \begin{feynman}
    \vertex (m1) ;
    \vertex [left=1 em of m1] (w) ;
    \vertex [below=0.5 em of m1] (rc) ;
    \vertex [right=1 em of m1] (w0) ;
    \diagram* {
            (m1) -- [white, line width=0.4mm, -, dashed] (rc),
             (w) -- [black, line width=0.2mm, -] (m1) -- [black, line width=0.2mm, -] (w0),
    };
    \end{feynman}
    \end{tikzpicture}
    +
    \begin{tikzpicture}[scale=0.6]
    \begin{feynman}
    \vertex (m1) ;
    \vertex [left=2 em of m1] (w) ;
    \vertex [below=1.2em of m1] (rc) ;
    \vertex [right=2 em of m1] (w0) ;-
    \diagram* {
        (w) -- [black, line width=0.2mm, -] (m1) -- [black, line width=0.2mm, -] (w0),
        (m1) -- [black, line width=0.4mm, -, dashed] (rc),
    };
     \vertex [dashhorizontal, below=0.75em of m1];
     \vertex [dashvertical, left=0.5em of m1];
    \vertex [circ, below=1.2em of m1];
    \vertex [xshift=0.3cm, yshift=-0.3cm] ;
    \end{feynman}
    \end{tikzpicture}
    +
    \begin{tikzpicture}[scale=0.6]
    \begin{feynman}
    \vertex (m1) ;
    \vertex [left=1 em of m1] (b) ;
    \vertex [left=1 em of b] (a);
    \vertex [below=1.2em of m1] (V) ;
    \vertex [right=1 em of m1] (c) ;
    \vertex [right=1 em of c] (d) ;
    \diagram* {
        (a) -- [black, line width=0.2mm, -] (b) -- [black, line width=0.2mm, boson] (m1)--[black, line width=0.2mm, boson] (c)--[black, line width=0.2mm, -](d),
        (m1) -- [black, line width=0.4mm, -, dashed] (V),
    };
    \vertex [circ, below=1.2em of m1];
    \vertex [dashhorizontal, below=0.75em of m1];
     \vertex [dashvertical, left=0.5em of m1];
    \vertex [xshift=0.3cm, yshift=-0.3cm] ;
    \end{feynman}
    \end{tikzpicture}
\\
&\nonumber\quad
+\begin{tikzpicture}[scale=0.6]
    \begin{feynman}
    \vertex (m1) ;
    \vertex [left=1 em of m1] (b) ;
    \vertex [left=1 em of b] (a);
    \vertex [below=1.2em of m1] (V) ;
    \vertex [right=1 em of m1] (c) ;
    \vertex [right=1 em of c] (d) ;
    \vertex [below=1.2em of d] (V2) ;
    \vertex [below=1.2em of d] (V2) ;
    \vertex [right=1 em of d] (e) ;
    \vertex [right=1 em of e] (f) ;
    \diagram* {
        (a) -- [black, line width=0.2mm, -] (b) -- [black, line width=0.2mm] (m1)--[black, line width=0.2mm] (c)--[black, line width=0.2mm, -](d),
        (m1) -- [black, line width=0.4mm, -, dashed] (V),
        (d) -- [black, line width=0.4mm, -, dashed] (V2),
        (d) -- [black, line width=0.2mm, -] (e) -- [black, line width=0.2mm] (f),
    };
    \vertex [circ, below=1.2em of m1];
    \vertex [circ, below=1.2em of d];
    \vertex [dashhorizontal, below=0.75em of m1];
    \vertex [dashhorizontal, below=0.75em of d];
     \vertex [dashvertical, left=0.5em of m1];
     \vertex [dashvertical, left=0.5em of d];
    \vertex [xshift=0.3cm, yshift=-0.3cm] ;
    \end{feynman}
    \end{tikzpicture}+\begin{tikzpicture}
    \begin{feynman}
    \vertex (m1) ;
    \vertex [left=1 em of m1] (b) ;
    \vertex [left=1 em of b] (a);
    \vertex [below=1.2em of m1] (V) ;
    \vertex [right=1 em of m1] (c) ;
    \vertex [right=1 em of c] (d) ;
    \vertex [below=1.2em of d] (V2) ;
    \vertex [below=1.2em of d] (V2) ;
    \vertex [right=1 em of d] (e) ;
    \vertex [right=1 em of e] (f) ;
    \diagram* {
        (a) -- [black, line width=0.2mm, -] (b) -- [black, line width=0.2mm] (m1)--[black, line width=0.2mm] (c)--[black, line width=0.2mm, -,boson](d),
        (m1) -- [black, line width=0.4mm, -, dashed] (V),
        (d) -- [black, line width=0.4mm, -, dashed] (V2),
        (d) -- [black, line width=0.2mm, -,boson] (e) -- [black, line width=0.2mm] (f),
    };
    \vertex [circ, below=1.2em of m1];
    \vertex [circ, below=1.2em of d];
    \vertex [dashhorizontal, below=0.75em of m1];
    \vertex [dashhorizontal, below=0.75em of d];
     \vertex [dashvertical, left=0.5em of m1];
     \vertex [dashvertical, left=0.5em of d];
    \vertex [xshift=0.3cm, yshift=-0.3cm] ;
    \end{feynman}
    \end{tikzpicture}\\
    &\nonumber\quad
    +\begin{tikzpicture}
    \begin{feynman}
    \vertex (m1) ;
    \vertex [left=1 em of m1] (b) ;
    \vertex [left=1 em of b] (a);
    \vertex [below=1.2em of m1] (V) ;
    \vertex [right=1 em of m1] (c) ;
    \vertex [right=1 em of c] (d) ;
    \vertex [below=1.2em of d] (V2) ;
    \vertex [below=1.2em of d] (V2) ;
    \vertex [right=1 em of d] (e) ;
    \vertex [right=1 em of e] (f) ;
    \diagram* {
        (a) -- [black, line width=0.2mm, -] (b) -- [black, line width=0.2mm,boson] (m1)--[black, line width=0.2mm,boson] (c)--[black, line width=0.2mm, -](d),
        (m1) -- [black, line width=0.4mm, -, dashed] (V),
        (d) -- [black, line width=0.4mm, -, dashed] (V2),
        (d) -- [black, line width=0.2mm, -] (e) -- [black, line width=0.2mm] (f),
    };
    \vertex [circ, below=1.2em of m1];
    \vertex [circ, below=1.2em of d];
    \vertex [dashhorizontal, below=0.75em of m1];
    \vertex [dashhorizontal, below=0.75em of d];
     \vertex [dashvertical, left=0.5em of m1];
     \vertex [dashvertical, left=0.5em of d];
    \vertex [xshift=0.3cm, yshift=-0.3cm] ;
    \end{feynman}
    \end{tikzpicture}+\begin{tikzpicture}
    \begin{feynman}
    \vertex (m1) ;
    \vertex [left=1 em of m1] (b) ;
    \vertex [left=1 em of b] (a);
    \vertex [below=1.2em of m1] (V) ;
    \vertex [right=1 em of m1] (c) ;
    \vertex [right=1 em of c] (d) ;
    \vertex [below=1.2em of d] (V2) ;
    \vertex [below=1.2em of d] (V2) ;
    \vertex [right=1 em of d] (e) ;
    \vertex [right=1 em of e] (f) ;
    \diagram* {
        (a) -- [black, line width=0.2mm, -] (b) -- [black, line width=0.2mm,boson] (m1)--[black, line width=0.2mm,boson] (c)--[black, line width=0.2mm, -,boson](d),
        (m1) -- [black, line width=0.4mm, -, dashed] (V),
        (d) -- [black, line width=0.4mm, -, dashed] (V2),
        (d) -- [black, line width=0.2mm, -,boson] (e) -- [black, line width=0.2mm] (f),
    };
    \vertex [circ, below=1.2em of m1];
    \vertex [circ, below=1.2em of d];
    \vertex [dashhorizontal, below=0.75em of m1];
    \vertex [dashhorizontal, below=0.75em of d];
     \vertex [dashvertical, left=0.5em of m1];
     \vertex [dashvertical, left=0.5em of d];
    \vertex [xshift=0.3cm, yshift=-0.3cm] ;
    \end{feynman}
    \end{tikzpicture}+\order{\nu^3}.
\end{align}

\paragraph*{Steady-state density and current.}
In the present work we focus on the steady state, \SMref{SteadyStateDist}. We write the steady-state density $\densityRight$ of right-moving particles as a power series in $\nu^n$ with coefficients $\densityRight^{(n)}$,
\begin{multline}
    \densityRight(x)=\sum^\infty_{n=0}\nu^n\densityRight^{(n)}(x)\\
    =\lim_{t_0\to -\infty}
    \ave{\phi(x,t) \phitilde(x_0,t_0)}
\end{multline}
and similarly for the density of left moving particles
$\densityLeft$, so that $\density(x)=\densityRight(x)+\densityLeft(x)$.
For finite tumble rate $\gamma$, the steady-state densities $\densityLR$ are independent of the initial state.
As discussed in \SMref{SteadyStateDist},
the limit $t_0\to-\infty$ effectively amputates the right, incoming leg in the diagrams,  \Eref{example_phi_phitilde}, so that they readily provide us with a diagrammatic expansion of the steady state density, order by order in $\nu$.

Calculating the $a$th Fourier coefficient $\density_a^{(n)}$ of the steady-state density $\density$ to order $n$ in $\nu$ is now a matter of some well-organised algebra
(\SMref{SteadyStateDist}). To this end,
we introduce the polarity $\polarity(x)=\densityRight(x)-\densityLeft(x)$
and express the steady-state current as
\begin{equation}\elabel{J_from_den_and_pol}
    \current=
      w \polarity(x)
    - (D \partial_x + \nu \extPot'(x))
    \density(x) \ .
\end{equation}
Defining the matrix
(\SMref{recurrenceDensPol})
\begin{equation}\elabel{def_M}
    \matrixM{a} =
    \frac{k_a \indicator_a}{\invBarePropO{a}\invBarePropO{-a}-\gamma^2}
    \begin{pmatrix}
    -D k_a^2 + 2\gamma &  \imag w k_a\\
    \imag w k_a      & -Dk_a^2
    \end{pmatrix}
\end{equation}
with $\invBarePropO{a}=\invBareProp{a}(0;0)=Dk_a^2-\imag wk_a+\gamma$, \Eref{def_invBareProp},
and $\indicator_a=1-\delta_{a,0}$, \Eref{def_indicator},
the density and the polarity at order $n$ in $\nu$
are the
elements
of the vector
\begin{widetext}
\begin{equation}\elabel{def_denPol}
    \begin{pmatrix}
      \density_{a_1}^{(n)} \\
      \polarity_{a_1}^{(n)}
    \end{pmatrix} =
    \sum_{a_2,\ldots,a_n}
    W_{a_1-a_2}W_{a_2-a_3}\ldots W_{a_{n-1}-a_n} W_{a_n}
    \matrixM{a_1}
    \matrixM{a_2}
    \ldots
    \matrixM{a_n}
    \begin{pmatrix}
      1 \\
      0
    \end{pmatrix}
\end{equation}
with $W_a=\extPot_a k_a/L$. Using \Eref{J_from_den_and_pol}, the steady-state current at order $n$ in $\nu$ is correspondingly
\begin{equation}\elabel{J_from_M}
J^{(n)}=-\frac{\imag}{L}
\sum_{a_1,\ldots,a_{n-1}}
W_{-a_1}W_{a_1-a_2}W_{a_2-a_3}\ldots W_{a_{n-2}-a_{n-1}}W_{a_{n-1}}
\begin{pmatrix}
1\\0
\end{pmatrix}^\transpose
\matrixM{a_1}
\matrixM{a_2}
\ldots \matrixM{a_{n-1}}
\begin{pmatrix}
1\\0
\end{pmatrix}.
\end{equation}
\end{widetext}
This concludes our derivation.
Calculating the current is now a matter of performing the matrix multiplications and summation in \Eref{J_from_M} for a potential given in terms of its modes $\extPot_a$ and summing these contributions order by order in $\nu$, so that
\begin{equation}\elabel{J_from_sum}
    \current=\sum_{n=0}^\infty \nu^n J^{(n)} \ .
\end{equation}

From the symmetries of $\matrixM{a}$, \Eref{def_M}, and the form of \Eref{J_from_M}, it follows that there is no current to first order in $\nu$ and generally no contributions to the steady-state current in even powers of $\nu$,
\SMref{curr_odd_in_nu}.
As a result, the current reverts if the potential is inverted, $\extPot\to-\extPot$. This is not a trivial insight, because the steady-state \emph{density} of the inverted potential generally bears no resemblance to that of the original potential, \fref{examplePotForInversion}, which may have been overlooked \cite{Reimann2001May}.
Several other properties of the steady-state current $J$ arising from \Eref{J_from_sum} are discussed in detail in \SMref{SteadyStateDist}.  Firstly, in a potential
even about
$x^*$, \ie $\extPot(x^*+x)=\extPot(x^*-x)$, the steady-state current, of course, vanishes, because the potential fails to provide even just a preferred direction \cite{Razin2020Sep}, \SMref{even_pot}.
The steady-state current also vanishes if the potential $\extPot(x)$ has only odd modes, \ie $\extPot_a=0$ for all even $a$, which renders it ``supersymmetric"
\cite{Reimann2001May},
$\extPot(x)=-\extPot(x+L/2)$, \SMref{superSym_pot}.

The steady-state current in any potential parameterised by $\extPot_a$ is given by \Erefs{J_from_M} and \eref{J_from_sum}. Any numerical scheme
can cope only with a finite number of modes,
as the sums in \Eref{J_from_M} need to terminate, and similarly for the maximum order of $\nu$ entering in \Eref{J_from_sum}.
Nevertheless, including hundreds of modes to calculate the current to hundreds of orders in $\nu$, in principle poses little numerical difficulty.
To confirm the correctness of our scheme, we calculate the current for a plain ratchet and compare to \cite{Astumian1994Mar}, which
can be done
to large extent in closed form, \SMref{AstumianBier}, owing to the piece-wise linearity of the potential.
\fref{comparisonAB}
demonstrates perfect agreement of the steady-state current calculated in both schemes for all $\nu<\nuRocSmall$. Beyond that point the power series \Eref{J_from_sum} eventually diverges.
In principle, the
radius of convergence $\nuRoc$ is determined, say, through the root test on the odd terms of the series \Eref{J_from_sum},
$\nuRoc=1/\limsup_{m\to\infty}|\current^{(2m+1)}|^{1/(2m+1)}$,
but in the present numerical procedures based on gradient descent, we used the more conservative estimate of the minimal radius of convergence of the derivative of \Eref{J_from_sum},
\begin{equation}\elabel{def_nuRoc}
    \nuRocSmall=\min\left(\big|m \current^{(m)}\big|^{-\frac{1}{m-1}}
    : m=3,5,\ldots,n\right)\ ,
\end{equation}
with $n$ the highest order calculated.

\paragraph*{Optimising the potential.}
The current \Eref{J_from_sum}
picks up as $\nu^{3}$, but eventually falters as $\nu$ becomes so large that $\nu\extPot$ is too steep for the
self-propulsion speed to overcome the potential. 
Somewhere in-between lies the \emph{optimal potential that maximises the current}, passable for the RnT particles in one direction, but (almost) impassable in the other direction.

To find the best such \emph{shape}, we fix $\nu=1$, rendering $\fullPot(x)$ equal to $\extPot(x)$ for the remainder of this section, and, leaving $\nu$ untouched, instead we find the modes $\extPot_a$ that maximise the steady-state current $\current$, \Erefs{J_from_M} and \eref{J_from_sum}, using \texttt{frprmn} \cite{Press2007Sep},
\SMref{Numerical_procedure}. Once such an ``optimal potential" $\extPotOptim(D,w,L,\gamma;x)$
as a function of the parameters
is found in terms of its modes
$\extPotOptim_a$, we determine
its current
$\currentOptim$
for this potential, as well as, \latin{a posteriori},
its radius of convergence $\nuRocOptim$ according to \Eref{def_nuRoc}, to confirm that it exceeds unity.
Since
$\extPot(x)$ is real,
$\extPot_a^*=\extPot_{-a}$, each pair $(\extPot_a,\extPot_{-a})$ may be written as an amplitude and a phase. To avoid degeneracy, we further fix the phase of the lowest mode so that $\extPot_1$ is purely imaginary.
Further, we
de-dimensionalise the problem, leaving only two parameters, the P{\'e}clet number
$\peclet=w L/D$
and
$\qeclet=\gamma L^2/D$.
In this parameterisation, length is measured in units of $L$, diffusion and the potential in units of $D$ and time therefore in units of $L^2/D$. We think in the following of $L$ and $D$ as being fixed, so that $\peclet$ parameterises the self-propulsion velocity, $\qeclet$ the tumbling rate and
$\extPotOptim(D,w,L,\gamma;x)=D\extPotOptim(1,\peclet,1,\qeclet;x/L)$.
While any particular choice of the underlying parameterisation should make no difference to the optimisation scheme,
making different choices for $D,w,L$ and $\gamma$ for the same
$\peclet$ and $\qeclet$ gives us a way to overcome some practical, numerical limitations.

\paragraph*{Results.}
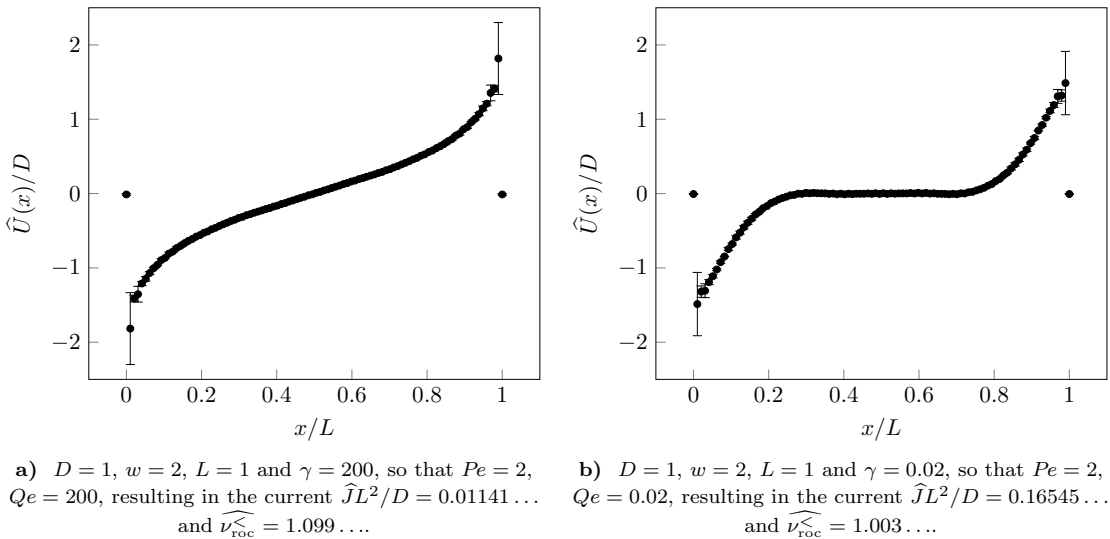
\begin{figure*}[t]
\resizebox{\linewidth}{!}{
\subfloat[\flabel{peclet=2_qeclet=200}
$D=1$, $w=2$, $L=1$ and $\gamma=200$, so that $\peclet=2$, $\qeclet=200$,
resulting in the current $\currentOptim L^2/D=0.01141\ldots$ and
$\nuRocSmallOptim=1.099\ldots$.
]{
\input{p=2_q=100_with_error_bar}
}
\subfloat[\flabel{peclet=2_qeclet=0.02}
$D=1$, $w=2$, $L=1$ and $\gamma=0.02$, so that $\peclet=2$, $\qeclet=0.02$,
resulting in the current $\currentOptim L^2/D=0.16545\ldots$ and $\nuRocSmallOptim=1.003\ldots$.
]{
\input{p=2_q=point_zero_1_with_error_bar}
}
}
\caption{Shapes of potential maximising the steady-state current at the parameter values indicated.
Technical details of the plots in \SMref{plotting_procedure}.
\flabel{more_shapes}
}
\end{figure*}
The resulting shape of the optimal potential is surprisingly stable across the range of $\peclet$ and $\qeclet$ explored. \fref{p=1_q=1} shows a typical shape, \fref{more_shapes} shows two more for somewhat more extreme parameter values.
The shape resembles that of the letter N, but it is not a simple, piece-wise linear ratchet. The resulting current in the figures above takes place from left to right. The deep crevasse towards $x=0$ in \fref{p=1_q=1}, towards the left followed by a steep rise of the potential towards the left and over to the right by periodicity, traps particles and prevents them from moving left. There is always a shallower path towards the right,
that terminates in a sharp peak towards $x=L$,
rendering the whole shape inversion symmetric about $(L/2,0)$. An apparent discontinuity at $x=0$ results in the Gibbs phenomenon, \SMref{plotting_procedure}.

The P{\'e}clet-number $\peclet$ indicates whether
\emph{free} particle movement is dominated by diffusion, $\peclet<1$, or ballistic motion, $\peclet>1$. The additional complications of tumbling and potential modify its r{\^o}le. On the large scale, diffusion is enhanced by $L\gamma^2$. On the small scale, diffusion $D$ sets the scale of the maximum barrier height the particle can overcome. Without potential, ballistic transport dominates on the scale of the system, if $w/\gamma>L$, \ie $\pqRatio=\peclet/\qeclet>1$.

To rectify the flow optimally, the basic design principles we derive from \frefs{p=1_q=1} and \ref{fig:more_shapes} are,
firstly, the need for a deep crevasse, \fref{p=1_q=1},
to set a high hurdle to a RnT particle attempting to pass through the potential towards the left.
Similarly, a sharp peak at around $x=L$ increases the barrier height.
The rise of the potential required as a barrier grows linearly in $D$.
The second design principle is the shallower incline of the potential
towards the right, where
the potential difference of the steep crevasse is spread out across the whole length $L$ of the potential, making it much less of an obstacle for particle movement towards the right.

Apart from the deep crevasse and the sharp peak on either end of the shallow incline, the intermediate ``plane'', \fref{p=1_q=1},
may be expected to have a slope of around $w$, so that it is just about overcome by a right-moving RnT particle.
However, for this long section to be passed \emph{ballistically} within time $1/\gamma$, its slope would have to be less than $w-L\gamma$. This can be realised only when $w$ is sufficiently large, as the slope is necessarily positive.
If ballistic transport dominates over diffusion, $\peclet>1$, but ballistic transport itself is too slow to overcome any significant potential slope within the time available between tumbles, $\pqRatio=\peclet/\qeclet=w/(L\gamma)\ll1$, we expect a flat plane maximising the steady-state current. Together with the need of a high barrier around $x=0$, $\peclet>1$ and $\qeclet\gg1$ result in sharp crevasses and peaks, \fref{peclet=2_qeclet=200}.

If, on the other hand, ballistic transport is strong, $\pqRatio>1$ and $\peclet>1$, the slope of the potential in the plane is not constrained by the ballistic transport but rather by the need to connect (only) between the bottom and the top of the barrier face, \ie the slope will scale like $D/L$. In this case, $\peclet>1$ and $\qeclet\ll1$, the intermediate slope is steeper and the crevasses widened,
\fref{peclet=2_qeclet=0.02}, producing a relatively large current.

If transport is primarily diffusive, $\peclet<1$, large $\qeclet=\gamma L^2/D$ effectively enhances the diffusion on spatial scales greater than $w/\gamma$. Transport in this parameter region is typically poor.

In general, a high barrier helps rectify the steady-state current. The need for a high barrier requires potential differences so large, that the slope
of the potential $\propto D/L$
in the plane exceeds the self-propulsion velocity $w$. This phenomenon can be observed already in
a piece-wise linear ratchet with the steady-state current $\current$ plotted in
\fref{comparisonAB}, where the maximum current is attained when the slope of the long ``plane'' section is around $3.9/0.9$, far exceeding the self-propulsion velocity of $w=1$.

The periodic nature of the setup allows for
any $\extPotOptim(D,w,L,\gamma;x)$ to be
periodically repeated $\ell$ times and investigated as a candidate for the optimal potential
$\extPotOptim(D,w,\ell L,\gamma;x)$. Such a potential would have
only modes $\extPot_\ell$, $\extPot_{2\ell}$ populated with all other
modes
$\extPotOptim_{1},\ldots,\extPotOptim_{\ell-1},\extPotOptim_{\ell+1},\ldots$ vanishing, as the first mode
of $\extPotOptim(\ldots,L,\ldots)$
gets mapped to the $\ell$th mode
of $\extPotOptim(\ldots,\ell L,\ldots)$, the second to the $2\ell$-th mode and so on.
This consideration provides the lower bound
    $\currentOptim(D,w,\ell L,\gamma) \ge \currentOptim(D,w,L,\gamma)/\ell$
as the steady-state current of the $\ell$-fold repeat of the potential is an $\ell$th of the original, and thus
\begin{equation}
    \currentOptim(1,\ell\peclet,1,\ell^2 \qeclet) \ge
    \ell \currentOptim(1,\peclet,1, \qeclet) \ .
\end{equation}

\paragraph*{Discussion.}
The optimal rectified current $\currentOptim$ is generally small compared to the unidirectional current $w/L$. Only for $\peclet\to\infty$ and $\qeclet\downarrow0$ the particles whizz through the potential in one direction, getting stuck on any tiny snag of a conventional ratchet in the other, so that $\currentOptim=(1/2)w/L$. 
A key question to address in future research is whether an external potential provides the best rectification or whether there are other
passive devices
that do a better job. What are the limits of rectification?

An experimental verification of our work requires a fine-tuned potential landscape.
Microorganisms have been subjected to a potential well by optical trapping
\cite{Xin2014Oct}, bidirectional molecular motors by magnetic trapping \cite{Fallesen2017Nov}, but simpler effective potential, due to boundary interaction \cite{Vizsnyiczai2020May,Sipos2015Jun} and a transversal microfluidic flow may be equally feasible.

Typical values for \latin{E.~coli} are $D=0.2\micrometer^2/s$ for the \emph{thermal} diffusion constant, $w=40\micrometer/s$ for the self-propulsion speed and $\gamma=1s^{-1}$ for the tumble rate \cite{Singh2017Oct, Dyer2021May}, producing values between
$\peclet=1000$ and $\qeclet=125$ for $L=5\micrometer$
and
$\peclet=2\cdot10^4$ and $\qeclet=5\cdot10^4$ for $L=100\micrometer$.
For these parameters we expect the optimal potential to be akin to \fref{peclet=2_qeclet=200}. Because the radius of convergence drops significantly with increasing Peclet number, we are not able to determine the shape directly.

The field theoretic formalism above can be extended: Firstly, it provides a route to the steady-state entropy production via Gaspard's \cite{Gaspard2004Nov, Cocconi2020Nov} approach, which, however, lies beyond the scope of the present work.
While it is fairly straight-forward to calculate it for the full Markov process, determining it while ignoring the particle species is notoriously difficult \cite{Garcia-Millan2021Jun}.
One might expect, however, that maximised current coincides with maximised entropy production.
Secondly, the lack of convergence as shown in \fref{comparisonAB} for $\nu>\nuRocSmall$
is a challenge that needs to be overcome, possibly with the help of
renormalisation. Thirdly,
although many experimental settings are well captured in one longitudinal and one transversal spatial dimension, extending the present framework to two dimension plus one transversal direction is an important generalisation.

This work determines the properties of the potential that optimally rectifies the steady-state current of RnT particles. We have shown that generally, the profile deviates significantly from an ordinary ratchet.
Our quantitative and qualitative findings provide
the design principles for rectification devices for microorganisms and for the cogs and wheels of an active engine \cite{Pietzonka2019Nov}.

\paragraph*{Acknowledgements}
We thank Martin Bier for helpful discussions
and acknowledge communications with Peter Reimann.
ZZ thanks Huiyao Zheng for technical support.
\bibliographystyle{apsrev4-2}
\bibliography{main}
\clearpage

\appendix

\renewcommand{\thefigure}{\Alph{section}\arabic{figure}}
\counterwithin*{figure}{section}
\onecolumngrid

\section{Numerical optimisation and plotting procedure}
In this section, we discuss the
numerical procedures employed to
determine the optimal shape $\extPotOptim(x)$ of the potential and
to subsequently
generate the plots of the potential $\extPot(x)$ in \frefs{p=1_q=1} and \ref{fig:more_shapes}. The optimisation scheme outlined in the main text results in the potential being characterised by its modes $\extPot_a$ and translating a finite number of them to a potential $\extPot(x)$ in real space is marred by ambiguity. We will outline the process of generating the plots, the origin of the error bars, the Gibbs phenomenon and the resolution of troughs and peaks.

\subsection{Optimisation procedure}
\label{Numerical_procedure}
In the following we outline the
numerical procedure to determine the optimal potential
$\extPotOptim(x)$, \ie the potential that maximises the steady-state current $\current$. The numerical scheme draws on \Erefs{J_from_M} and \eref{J_from_sum}, which effectively produce the steady-state current $\current$ for any given, finite set of $2A+1$ modes of the potential $\extPot_a$ with $a=-A,-A+1,\ldots,A$. The numerical scheme is greatly improved by expressing the partial derivative of $\current$ with respect to any of the modes $\extPot_a$. While tedious to determine, given the structure of \Eref{J_from_M}, a closed-form expression is readily available.

While the finite number of modes, $2A+1$, limits the summation to be performed in \Eref{J_from_M}, the order of $\nu$ needs to be limited to $\maxPowNu$ in the sum \Eref{J_from_sum}.

As $\extPot(x)$ is real, the modes $\extPot_a$ are complex conjugate pairs, $\extPot^*_a=\extPot_{-a}$. The $0$-mode $\extPot_0$ does not enter at all, so that for any given $A$ the maximisation scheme determines $A$ complex numbers, $\extPot_a$ with $a=1,2,\ldots,A$.

If $\extPotOptim(x)$ extremises the current, so does $\extPotOptim(L-x)$ and in fact any translation of the two, $\extPotOptim(x^*\pm x)$. To break symmetry, we demand $\extPot_1$ be purely imaginary, leaving the maximisation scheme with $2A-1$ degrees of freedom. We found that generally the maximisation scheme is not sensitive to the details of the initialisation.
We have also verified that the maximum current is not obtained with periodically repeated potentials, ``higher harmonics", where only modes are populated that are multiples of some $a^*$, \ie $\extPot_{a^*}$, $\extPot_{2a^*}$, $\extPot_{3a^*}$ are non-zero and all other modes vanish.

We found numerically that the steady-state current is maximised if all $\extPot_a$ are purely imaginary up to a small numerical error. As a result the optimal potential obeys $\extPotOptim(L/2+x)=-\extPotOptim(L/2-x)$. Demanding this throughout the maximisation scheme reduces the degrees of freedom to $A$, without changing the optimal shape $\extPotOptim(x)$ or the optimal current $\currentOptim$.

We used \texttt{frprmn} \cite{Press2007Sep} to obtain the results reported in the present work. The following heuristic proved efficient:
\begin{enumerate}
    \item[0.] Initialisation:  $A=50$, $\maxPowNu=75$ and
    $\extPot_a$ for $a=-50,-49,\ldots,50$ are initialised with $U_a=\imag/(2\pi a)$, which are the Fourier coefficients of a piece-wise periodic linear ratchet $U(x)=x$ for $x \in [0,L)$.
    \item[1.] Given the other parameters, $D,w,L,\gamma$, as well as \Erefs{J_from_M} and \eref{J_from_sum} with the sum running up to $\maxPowNu$, the optimal $\extPotOptim_a$ are determined using \texttt{frprmn} \cite{Press2007Sep}.
    \item[2.] The results are written into a file and $A$ is increased by $50$. These additional degrees of freedom are initialised with $0$. Those determined so far serve as initial values for the optimisation with increased $A$. The procedure returns to step~1.
\end{enumerate}

Throughout the optimisation procedure, we determine the apparent radius of convergence of the power series \Eref{J_from_sum} as the value of $\left(\current^{(n)}\right)^{-1/n}$ for the largest (odd) $n$ that produces a finite numerical value.
If the optimisation procedure strays into a parameter region where this radius of convergence
drops below unity, the supposed current becomes numerically unreliable. In this case, the optimisation is considered unattainable within the present perturbative framework and the optimisation is abandoned, \ie we do not pursue the optimsation for these parameter values further and no results are shown in the present work.
If, on the other hand, an optimal potential $\extPotOptim(x)$ is eventually found, we also determine the radius of convergence $\nuRocSmallOptim$, \Eref{def_nuRoc}, \latin{a posteriori} to ensure that it is greater than unity.

\subsection{Plotting procedure}
\label{plotting_procedure}
The plots \frefs{p=1_q=1} and \ref{fig:more_shapes} are generated as follows. With $\maxPowNu=75$ the modes $\extPotOptim_a$ are determined for $A=100$, $A=150$ and $A=200$. Increasing $A$ gives the optimisation procedure more degrees of freedom, which affects \emph{all} modes, \eg in general, $\extPotOptim_a$ determined with $A=100$
differs from $\extPotOptim_a$ determined with $A=150$
for all $a$, in particular at $a\le100$. \fref{Fig_modes_convergence} 
shows how $\extPotOptim_a$ changes with $A$, with the largest (relative) change visible for $a$ close to $A$, whereas $\extPotOptim_a$ for smaller $a$ suggest convergence.

\begin{figure}
\input{Fig_modes_convergence}
\caption{Imaginary part of the modes $\extPotOptim_a$ 
allowing for $A=50,100,150,200$ modes in the optimisation procedures (filled symbols) and, also, as extracted from the linear interpolation of the potential shown in \fref{p=1_q=1} (open squares). Parameters are $D=1$, $w=1$, $L=1$ and $\gamma=1$, so that $\peclet=\qeclet=\pqRatio=1$.
}
\flabel{Fig_modes_convergence}
\end{figure}
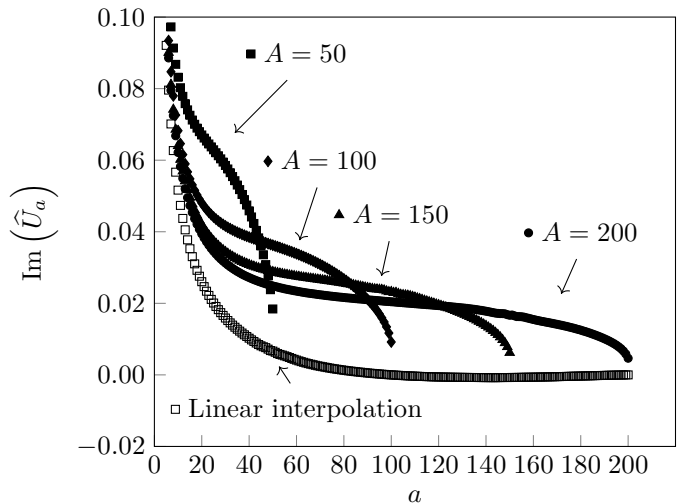

\begin{figure}
\resizebox{.9\linewidth}{!}{
\subfloat[$A=50$.]{
\input{Fig_p=1_q=1_dense_A=50}
}
\subfloat[$A=200$.]{
\input{Fig_p=1_q=1_dense_A=200}
}}
\caption{
The optimal potential $\extPotOptim(x)$ of \fref{p=1_q=1} shown with higher resolution of $2000$ points and $A=50,200$ as indicated. The peaks around $x=0$ and $x=L$ are likely to be artefacts, as discussed in the text.
}
\flabel{p=1_q=1_dense}
\end{figure}
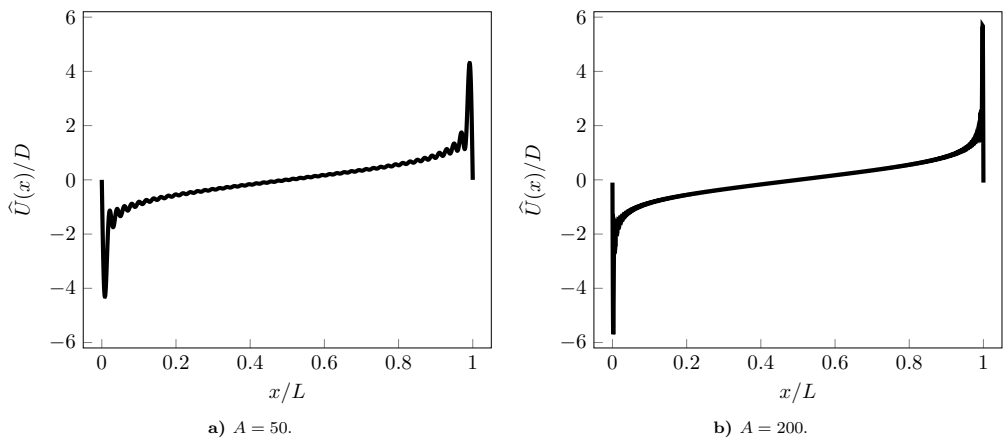

Evaluating \Eref{extPot_from_modes} in the form
\begin{equation}\elabel{extPot_from_modes_modified}
    \extPotOptim(x) = \frac{1}{L} \sum_{a=-A}^A \exp{\imag k_a x} \extPotOptim_a
\end{equation}
at numerically densely chosen positions $x$ reveals an apparent undulation throughout and turning into some distinct peaks close to $x=0$ and $x=L$ reminiscent of the Gibbs phenomenon \cite{Gibbs1898Dec}, \fref{p=1_q=1_dense}. The problem persists even at large $A$, although the undulations away from $x=0,L$ become less noticeable.

\begin{figure}
\input{Fig_modified_sum}
\caption{Detail of the potential $\extPot_B(x)/D$ of \Eref{extPot_from_modes_with_B}
around the peak near $x/L=1$ based on $A=100$ modes using
$B=20,40,60,80$ and $100$ modes respectively in the sum. Parameters: $D=1$, $w=1$, $L=1$ and $\gamma=1$, so that $\peclet=\qeclet=\pqRatio=1$.
}
\flabel{modified_sum}
\end{figure}
This is confirmed in a modified sum,
\begin{equation}\elabel{extPot_from_modes_with_B}
    \extPot_B(x) = \frac{1}{L} \sum_{a=-B}^B \exp{\imag k_a x} \extPotOptim_a
\end{equation}
where the summation runs only over $B\le A$ modes, \ie it takes into account fewer modes in the plot than are available in the given numerical optimisation of the first $A$ modes. As shown in \fref{modified_sum}, the resulting sharp peaks around $x=0$ and $x=L$ are present even for moderate values of $B$, although they, of course, change shape. This suggests that their presence in \fref{p=1_q=1_dense} is not due to the relatively big error of $\extPotOptim_a$ at $a$ around $A$, but due to the Gibbs phenomenon, which arises
when a Fourier-sum of a discontinuous function is terminated after finitely many terms, even when the infinite sum is exact. In other words, some features around $x=0$ and $x=L$  are not due to poor numerical estimates of $\extPotOptim_a$ as $A$ is finite, but instead caused by the Gibbs phenomenon, \ie using a finite number of modes to describe a discontinuous function.

The plots shown in \frefs{p=1_q=1} and \ref{fig:more_shapes} are designed to show what we know about
the potential $\extPotOptim(x)$ and quantify its convergence. To achieve this, we determine the modes of the best potential for $A=100$, $A=150$ and $A=200$ and evaluate \Eref{extPot_from_modes_modified}
at values of $x$ incommensurate with all $L/A$ in an attempt to avoid ``resonances" and with spacing greater than the largest $L/A$. In \frefs{p=1_q=1} and \ref{fig:more_shapes} we choose $x$ at multiples of $L/97$. The error bars in these plots indicate the range of $\extPotOptim(x)$ for the different $A$ at the corresponding $x$. This procedure avoids artefacts due to the Gibbs phenomenon at $x=0$ and $x=L$ as well as the undulations throughout finite $A$.
However, given the comparative sparseness of the points, sharp features are not well resolved, in particular those close to $x=0$ and $x=L$.

One might be tempted to linearly interpolate the points in the resulting plots, say \fref{p=1_q=1}. However, as shown in \fref{Fig_modes_convergence}, determining the resulting modes $\extPot_a$ by Fourier-transform of such a linear interpolation, shows that they are generally a very poor match with the modes that gave rise to the plot in the first place, even when their inverse Fourier transform reproduces \fref{p=1_q=1} perfectly. Only the very low modes up to $|a|=5$ agree within $10\%$.
As a result,
the current of $200$ such modes produces only
$\current L^2/D=0.01872...$,
and is thus
reduced by a factor $0.5$ compared to the optimal current obtained originally.

The plotting procedure used in the main text thus provides a useful representation of the potential in real space with suitable indication of how reliably we have determined certain features. The ambiguity arises because the optimisation procedure produces modes, as shown in \fref{Fig_modes_convergence} rather than an estimate of the shape in real space.

\section{Field theory}\label{FT}
In this section, we give a derivation of the path-integral formulation of RnT motion in a periodic potential using Doi-Peliti field theory. The coupled Fokker-Planck equations describing an RnT particle in a periodic potential $\nu\extPot(x)$ are, similar to \Erefs{AB_FP},
\begin{subequations}
\elabel{FP_left_right}
\begin{align}
\partial_t\densityRight(x,t)=&-\partial_x[(w-\nu\extPot'(x))\densityRight(x,t)]-\gamma(\densityRight(x,t)-\densityLeft(x,t))+D\partial_x^2 \densityRight(x,t)\nonumber\\
\partial_t\densityLeft(x,t)=&-\partial_x[(-w-\nu\extPot'(x))\densityLeft(x,t)]-\gamma(\densityLeft(x,t)-\densityRight(x,t))+D\partial_x^2 \densityLeft(x,t)
\end{align}
\end{subequations}
where $\densityRight(x,t)$ and $\densityLeft(x,t)$ are the densities of right-moving and left-moving particles respectively, as a function of the position $x$ and time $t$. The dashed potential $\extPot'(x)$ denotes its derivative.
We further introduce the
total particle density
$\density(x,t)=\densityRight(x,t)+\densityLeft(x,t)$ and the polarity $\polarity(x,t)=\densityRight(x,t)-\densityLeft(x,t)$, so that the particle current
\begin{equation}\elabel{def_current_app}
    \current(x,t)= \big(w -\nu\extPot'(x) - D \partial_x \big)\densityRight(x,t)
                  +\big(-w -\nu\extPot'(x)- D \partial_x \big)\densityLeft(x,t)
\end{equation}
can be re-written as
\begin{equation}
    \current(x,t)=w\polarity(x,t)-\density(x,t) \nu\extPot'(x)-D\partial_x\density(x,t)
\end{equation}
and the Fokker-Planck equation as
\begin{align}
\partial_t\rho(x,t)=-\partial_x[
w\polarity(x,t)-\density(x,t) \nu\extPot'(x)-D\partial_x
]=-\partial_x \current(x,t)\nonumber\\
\partial_t \polarity(x,t)=-\partial_x[w\density(x,t)-\polarity(x,t) \nu\extPot'(x)]-2\gamma \polarity(x,t)+D\partial_x^2 \polarity(x,t)
\ .
\end{align}
The Fokker-Planck equation describes the evolution of a particle density. \latin{A priori}, it does not enforce the particle entity of the constituent degrees of freedom, nor is it concerned with it --- the \emph{densities} $\densityLR$ above are completely unconstrained by being due to particles and could equally describe, say, a temperature profile. To cast it into a Doi-Peliti field theory, the Fokker-Planck equation is normally re-written as a master equation by discretising space and interpreting it as the evolution of the \emph{probability (density)} of finding a \emph{single particle} at a particular position, before generalising it to the evolution of \emph{multiple, indistinguishable particles} and the probability of obtaining a particular \emph{occupation number configuration}.
In the canonical procedure \cite{Cardy2008Dec,Tauber2005Apr}
the field-theoretic action is then obtained
by
expressing the evolution in terms of ladder operators, turning them into conjugate fields and taking the continuum limit. This procedure invariably reproduces the Fokker-Planck operator in the action $\action$ \cite{Garcia-MillanPruessner:2022}, which after rearranging can be expressed as the sum $\action=\actionHarm+\actionPert$ of a harmonic part
\begin{align}\elabel{def_actionHarm}
\actionHarm&=\int_0^L \dint{x} \int\dint{t}
\phitilde(x,t)
(\partial_t + w\partial_x-D\partial_x^2 +r)
\phi(x,t)
+
\psitilde(x,t)
(\partial_t - w\partial_x-D\partial_x^2 + r)
\psi(x,t)
\\
\nonumber&\qquad
+\gamma\big(\phitilde(x,t) -\psitilde(x,t)\big)\big(\phi(x,t)-\psi(x,t)\big)
\end{align}
and a perturbative part,
\begin{align}
\elabel{def_actionPert}
\actionPert&=\int_0^L \dint{x} \int\dint{t}
\nu\big(\partial_x\phitilde(x,t)\big) \extPot'(x)\phi(x,t)
+
\nu\big(\partial_x\psitilde(x,t)\big) \extPot'(x)\psi(x,t)
\end{align}
where $\phi(x,t)$ and $\psi(x,t)$ are the annihilation fields of a right-moving and a left-moving particle respectively and $\phitilde(x,t)$ and $\psitilde(x,t)$ are, correspondingly, the Doi-shifted \cite{Cardy2008Dec} creation fields of a right-moving and a left-moving particle so that $\phi^\dagger =\tilde\phi +1$ and $\psi^\dagger =\tilde\psi+1$. We have included in the harmonic part a positive \emph{mass} $r$ as to regularise the infrared and restore causality. The mass is a mere technicality and will be taken to $0^+$ whenever suitable.
The perturbative part of the action $\actionPert$ in \Eref{def_actionPert} incorporates the perturbative parameter $\nu$ in front of the potential $\extPot(x)$ to guide the perturbation theory that is to follow.

The expectation of an observable in terms of fields is taken via the path-integral \cite{Garcia-Millan2021Jun}
\begin{align}
\ave{\bullet}&=\int \mathcal D[\phi,\tilde\phi,\psi,\tilde\psi]\bullet e^{- \action[\phi,\tilde\phi,\psi,\tilde\psi]}.
\end{align}
Only at $\nu=0$ this path integral can generally be taken in closed form,
\begin{align}
\aveHarm{\bullet}&=\int \mathcal D[\phi,\tilde\phi,\psi,\tilde\psi]\bullet e^{- \actionHarm[\phi,\tilde\phi,\psi,\tilde\psi]}
\end{align}
about which the action in the perturbative part is expanded, so that
\begin{equation}\elabel{expansion_pert}
    \ave{\bullet}=\sum_{n=0}^\infty \frac{1}{n!} \aveHarm{\bullet (-\actionPert)^n} \ .
\end{equation}
To make the action local in the fields, we introduce a Fourier representation in the form
\begin{align}\elabel{def_Fourier}
    \phi(x,t)=
    \frac{1}{L}\sum_{a=-\infty}^\infty \exp{\imag k_a x}
    \int_{-\infty}^\infty \dintbar{\omega} \exp{-\imag \omega t}
    \phi_a(\omega)
\end{align}
and correspondingly for all other fields and observables, with
$\dbar{\omega}=\plaind{\omega}/(2\pi)$ and
$k_a=2\pi a/L$,
which readily accommodates the periodic nature of $x$. For example the potential in real space is expressed as
\begin{equation}\elabel{extPot_from_modes}
    \extPot(x) = \frac{1}{L} \sum_{a=-\infty}^\infty \exp{\imag k_a x} \extPot_a
    \quad\text{ with }\quad
    \extPot_a=\int_0^L \dint{x} \exp{-\imag k_a x} \extPot(x)
\end{equation}
in terms of its \emph{modes} or coefficients $\extPot_a$.

The bare propagators are easily determined from \Eref{def_actionHarm} in the form
\begin{equation}
\actionHarm=
\frac{1}{L} \sum_a \int\dintbar{\omega}
\begin{pmatrix}
\phitilde_{-a}(-\omega)\\
\psitilde_{-a}(-\omega)
\end{pmatrix}^\transpose
\matrixHarm_a(\omega)
\begin{pmatrix}
\phitilde_{a}(\omega)\\
\psitilde_{a}(\omega)
\end{pmatrix}
\end{equation}
with
\begin{equation}
    \matrixHarm_a(\omega) =
    \begin{pmatrix}
    -\imag\omega + Dk_a^2 + \imag w k_a + r + \gamma & -\gamma \\
    -\gamma&-\imag\omega + Dk_a^2 - \imag w k_a + r + \gamma
    \end{pmatrix}
\end{equation}
by calculating the inverse
\begin{equation}\elabel{inverse_matrixHarm_a}
    \Big(\matrixHarm_a(\omega)\Big)^{-1} \deltabar(\omega+\omega') L \delta_{a+b,0}=
    \begin{pmatrix}
    \ave{\phi_a(\omega)\phi_b(\omega')}
      & \ave{\phi_a(\omega) \psi_b(\omega')} \\
    \ave{\psi_a(\omega)\psi_b(\omega')}
      & \ave{\psi_a(\omega) \phi_b(\omega')}
    \end{pmatrix}
\end{equation}
with the Kronecker $\delta$-function denoted by
$\delta_{a+b,0}$
and
the Dirac
$\delta$-function by
$\deltabar(\omega)=2\pi\delta(\omega)$.
The propagators are then found to be
\begin{subequations}
\elabel{all_bare_propagators}
\begin{align}
\elabel{bare_propagator_phiphi}
    \aveHarm{\phi_a(\omega)\phitilde_b(\omega')}&=
    \frac{\invBareProp{a}(\omega;r)\deltabar(\omega+\omega')L\delta_{a+b,0}}{\invBareProp{a}(\omega;r)\invBareProp{-a}(\omega;r)-\gamma^2}
    &\corresponds
   \begin{tikzpicture}[circ/.style={shape=circle, inner sep=2pt, outer sep =0.5pt, line width=0.4mm, draw, node contents=}]
    \begin{feynman}
    \vertex (m1) ;
    \vertex [left=2.5 em of m1] (w) ;
    \vertex [below=1.5em of m1] (rc) ;
    \vertex [right=2.5 em of m1] (w0) ;
    \diagram* {
        (w) -- [black, line width=0.2mm, -] (m1) -- [black, line width=0.2mm, -] (w0),
    };
    \vertex [above=0.5em of w] {$\phi_a(\omega)$};
    \vertex [above=0.5em of w0] {$\tilde\phi_b(\omega')$};
    \end{feynman}
    \end{tikzpicture}\\
\elabel{bare_propagator_psipsi}
    \aveHarm{\psi_a(\omega)\psitilde_b(\omega')}&=
    \frac{\invBareProp{-a}(\omega;r)\deltabar(\omega+\omega')L\delta_{a+b,0}}{\invBareProp{a}(\omega;r)\invBareProp{-a}(\omega;r)-\gamma^2}
    &\corresponds \begin{tikzpicture}[circ/.style={shape=circle, inner sep=2pt, outer sep =0.5pt, line width=0.4mm, draw, node contents=}]
    \begin{feynman}
    \vertex (m1) ;
    \vertex [left=2.5 em of m1] (w) ;
    \vertex [below=1.5em of m1] (rc) ;
    \vertex [right=2.5 em of m1] (w0) ;
    \diagram* {
        (w) -- [black, line width=0.2mm, boson] (m1) -- [black, line width=0.2mm, boson] (w0),
    };
    \vertex [above=0.5em of w] {$\psi_a(\omega)$};
    \vertex [above=0.5em of w0] {$\tilde\psi_b(\omega')$};
    \end{feynman}
    \end{tikzpicture}
\\
\elabel{bare_propagator_phipsi}
\aveHarm{\phi_a(\omega)\psitilde_b(\omega')}&=
    \frac{\gamma\deltabar(\omega+\omega')L\delta_{a+b,0}}{\invBareProp{a}(\omega;r)\invBareProp{-a}(\omega;r)-\gamma^2}
    &\corresponds  \begin{tikzpicture}[circ/.style={shape=circle, inner sep=2pt, outer sep =0.5pt, line width=0.4mm, draw, node contents=}]
    \begin{feynman}
    \vertex (m1) ;
    \vertex [left=2.5 em of m1] (w) ;
    \vertex [below=1.5em of m1] (rc) ;
    \vertex [right=2.5 em of m1] (w0) ;
    \diagram* {
        (w) -- [black, line width=0.2mm, -] (m1) -- [black, line width=0.2mm, boson] (w0),
    };
    \vertex [above=0.2em of w] {$\phi_a(\omega)$};
    \vertex [above=0.5em of w0] {$\tilde\psi_b(\omega')$};
    \end{feynman}
    \end{tikzpicture}
    \\
\elabel{bare_propagator_psiphi}
\aveHarm{\psi_a(\omega)\phitilde_b(\omega')}&=
    \frac{\gamma\deltabar(\omega+\omega')L\delta_{a+b,0}}{\invBareProp{a}(\omega;r)\invBareProp{-a}(\omega;r)-\gamma^2}
    &\corresponds  \begin{tikzpicture}[circ/.style={shape=circle, inner sep=2pt, outer sep =0.5pt, line width=0.4mm, draw, node contents=}]
    \begin{feynman}
    \vertex (m1) ;
    \vertex [left=2.5 em of m1] (w) ;
    \vertex [below=1.5em of m1] (rc) ;
    \vertex [right=2.5 em of m1] (w0) ;
    \diagram* {
        (w) -- [black, line width=0.2mm, boson] (m1) -- [black, line width=0.2mm, -] (w0),
    };
    \vertex [above=0.5em of w] {$\psi_a(\omega)$};
    \vertex [above=0.5em of w0] {$\tilde\phi_b(\omega')$};
    \end{feynman}
    \end{tikzpicture}
\end{align}
\end{subequations}
where we have introduced
\begin{equation}\elabel{def_invBareProp}
    \invBareProp{a}(\omega;r)=-\imag\omega+Dk_a^2-\imag wk_a+r+\gamma
\end{equation}
to ease notation.

After Fourier-transforming,
the perturbative part of the action, \Eref{def_actionPert} reads
\begin{equation}\elabel{def_actionPert_Fourier}
\actionPert = - \frac{1}{L^3}\sum_{a,b,c}
L \delta_{a+b+c,0}
\int\dintbar{\omega}
\nu\big(k_a \phitilde_a(\omega)\big)
k_b \extPot_b \phi_c(-\omega)
+
\nu\big(k_a \psitilde_a(\omega)\big)
k_b \extPot_b \psi_c(-\omega)
\end{equation}
which effectively destroys momentum conservation, as any mismatch $a+c\ne0$ can be made up by the Fourier modes $b$ of the potential, $\extPot_b$. Each order in the perturbative expansion thus requires a summation over the momenta, similar to integrating a loop.

The ensuing diagrammatics significantly simplifies the bookkeeping.
The
external potential vertex may be written as
\begin{equation}\elabel{bauble_vertices}
\begin{tikzpicture}
    \begin{feynman}
    \vertex (m1) ;
    \vertex [left=1 em of m1] (w) ;
    \vertex [below=1.5em of m1] (rc) ;
    \vertex [right=1 em of m1] (w0) ;-
    \diagram* {
        (w) -- [black, line width=0.2mm, -] (m1) -- [black, line width=0.2mm, -] (w0),
        (m1) -- [black, line width=0.4mm, -, dashed] (rc),
    };
    \vertex [circ, below=1.5em of m1];
    \vertex [dashhorizontal, below=0.75em of m1];
    \vertex [dashvertical, left=0.5em of m1];
    \vertex [xshift=0.3cm, yshift=-0.3cm] ;
    \vertex [above=0.5em of w] {$\tilde\phi_a$};
     \vertex [xshift=0cm, yshift=-1cm] {$\nu \extPot_b$};
    \vertex [above=0.5em of w0] {$\phi_c$};
    \end{feynman}
    \end{tikzpicture}
    \quad
    \begin{tikzpicture}
    \begin{feynman}
    \vertex (m1) ;
    \vertex [left=1 em of m1] (w) ;
    \vertex [below=1.5em of m1] (rc) ;
    \vertex [right=1 em of m1] (w0) ;
    \diagram* {
        (w) -- [black, line width=0.2mm, boson] (m1) -- [black, line width=0.2mm, boson] (w0),
        (m1) -- [black, line width=0.4mm, -, dashed] (rc),
    };
    \vertex [circ, below=1.5em of m1];
    \vertex [xshift=0.3cm, yshift=-0.3cm] ;
     \vertex [dashhorizontal, below=0.75em of m1];
     \vertex [dashvertical, left=0.5em of m1];
    \vertex [above=0.5em of w] {$\tilde\psi_a$};
    \vertex [above=0.5em of w0] {$\psi_b$};
    \vertex [xshift=0cm, yshift=-1cm] {$\nu \extPot_c$};
    \end{feynman}
    \end{tikzpicture}
\end{equation}
with the bauble representing the external potential, which supplies the missing momentum. The short, thick dashes across (amputated) propagators indicate diagrammatically the factors $k_a$ and $k_b$, \Eref{def_actionPert_Fourier}.
\Eref{bauble_vertices} corrects the propagators  as shown in \Eref{example_phi_phitilde}. Unfortunately, the (Dyson) summation can generally not be performed in closed form, but rather needs to be done order by order in the perturbative parameter $\nu$. How this is done to determine steady-state expectations is described in the next section.

\section{Steady-state density and current}
\label{SteadyStateDist}
In the following, we present the details of the derivation of the steady-state particle densities
\begin{subequations}
\begin{align}
    \densityRight(x) & =
    \lim_{t_0\to-\infty} \ave{\phi(x,t)\phitilde(x_0,t_0)} =
    \lim_{t_0\to-\infty} \ave{\phi(x,t)\psitilde(x_0,t_0)}\\
    \densityLeft(x) & =
    \lim_{t_0\to-\infty} \ave{\psi(x,t)\phitilde(x_0,t_0)} =
    \lim_{t_0\to-\infty} \ave{\psi(x,t)\psitilde(x_0,t_0)} \ ,
\end{align}
\end{subequations}
independent of $t$ and $x_0$ given the limit to the steady state and ergodicity due to the perturbation theory about vanishing potential at positive tumbling rate $\gamma$. We will firstly give a rather general argument about the effect of the limit $t_0\to-\infty$ on diagrams, before focusing on the present field theory.

\subsection{General arguments}
The diagrammatic expansion of the propagators, \Eref{example_phi_phitilde} or

\begin{align}
\ave{\phi_a(\omega)\phitilde_b(\omega')}&
=
\begin{tikzpicture}[dot/.style={fill,circle,inner sep=0pt,outer sep=0pt,minimum size=8pt,label={[label distance=0cm]#1}},baseline={(current bounding box.center)}]
    \begin{feynman}
    \vertex (m1) ;
    \vertex [left=2 em of m1] (w) ;
    \vertex [below=0.5 em of m1] (rc) ;
    \vertex [right=2 em of m1] (w0) ;
    \diagram* {
            (m1) -- [white, line width=0.4mm, -, dashed] (rc),
             (w) -- [black, line width=0.2mm, -] (m1) -- [black, line width=0.2mm, -] (w0),
    };
    \end{feynman}
    \end{tikzpicture}
    +
    \ \begin{tikzpicture}
    \begin{feynman}
    \vertex (m1) ;
    \vertex [left=2 em of m1] (w) ;
    \vertex [below=1.5em of m1] (rc) ;
    \vertex [right=2 em of m1] (w0) ;-
    \diagram* {
        (w) -- [black, line width=0.2mm, -] (m1) -- [black, line width=0.2mm, -] (w0),
        (m1) -- [black, line width=0.4mm, -, dashed] (rc),
    };
    \vertex [circ, below=1.5em of m1];
    \vertex [dashhorizontal, below=0.75em of m1];
     \vertex [dashvertical, left=0.5em of m1];
    \vertex [xshift=0.3cm, yshift=-0.3cm] ;
    \end{feynman}
    \end{tikzpicture}
    +
    \begin{tikzpicture}[circ/.style={shape=circle, inner sep=2pt, outer sep =0.5pt, line width=0.4mm, draw, node contents=}, baseline={(current bounding box.center)}]
    \begin{feynman}
    \vertex (m1) ;
    \vertex [left=1 em of m1] (b) ;
    \vertex [left=1 em of b] (a);
    \vertex [below=1.5em of m1] (V) ;
    \vertex [right=1 em of m1] (c) ;
    \vertex [right=1 em of c] (d) ;
    \diagram* {
        (a) -- [black, line width=0.2mm, -] (b) -- [black, line width=0.2mm, boson] (m1)--[black, line width=0.2mm, boson] (c)--[black, line width=0.2mm, -](d),
        (m1) -- [black, line width=0.4mm, -, dashed] (V),
    };
    \vertex [dashhorizontal, below=0.75em of m1];
     \vertex [dashvertical, left=0.5em of m1];
    \vertex [circ, below=1.5em of m1];
    \vertex [xshift=0.3cm, yshift=-0.3cm] ;
    \end{feynman}
    \end{tikzpicture}+\ldots\\
\ave{\psi_a(\omega)\psitilde_b(\omega')}&
=
\begin{tikzpicture}[dot/.style={fill,circle,inner sep=0pt,outer sep=0pt,minimum size=8pt,label={[label distance=0cm]#1}},baseline={(current bounding box.center)}]
    \begin{feynman}
    \vertex (m1) ;
    \vertex [left=2 em of m1] (w) ;
    \vertex [below=0.5 em of m1] (rc) ;
    \vertex [right=2 em of m1] (w0) ;
    \diagram* {
            (m1) -- [white, line width=0.4mm, -, dashed] (rc),
             (w) -- [black, line width=0.2mm, boson] (m1) -- [black, line width=0.2mm, boson] (w0),
    };
    \end{feynman}
    \end{tikzpicture}
    +
    \begin{tikzpicture}[circ/.style={shape=circle, inner sep=2pt, outer sep =0.5pt, line width=0.4mm, draw, node contents=}, baseline={(current bounding box.center)}]
    \begin{feynman}
    \vertex (m1) ;
    \vertex [left=2 em of m1] (w) ;
    \vertex [below=1.5em of m1] (rc) ;
    \vertex [right=2 em of m1] (w0) ;-
    \diagram* {
        (w) -- [black, line width=0.2mm, boson] (m1) -- [black, line width=0.2mm, boson] (w0),
        (m1) -- [black, line width=0.4mm, -, dashed] (rc),
    };
    \vertex [circ, below=1.5em of m1];
    \vertex [xshift=0.3cm, yshift=-0.3cm] ;
    \vertex [dashhorizontal, below=0.75em of m1];
     \vertex [dashvertical, left=0.5em of m1];
    \end{feynman}
    \end{tikzpicture}
    +
    \begin{tikzpicture}[circ/.style={shape=circle, inner sep=2pt, outer sep =0.5pt, line width=0.4mm, draw, node contents=}, baseline={(current bounding box.center)}]
    \begin{feynman}
    \vertex (m1) ;
    \vertex [left=1 em of m1] (b) ;
    \vertex [left=1 em of b] (a);
    \vertex [below=1.5em of m1] (V) ;
    \vertex [right=1 em of m1] (c) ;
    \vertex [right=1 em of c] (d) ;
    \diagram* {
        (a) -- [black, line width=0.2mm, boson] (b) -- [black, line width=0.2mm, --] (m1)--[black, line width=0.2mm, --] (c)--[black, line width=0.2mm, boson](d),
        (m1) -- [black, line width=0.4mm, -, dashed] (V),
    };
    \vertex [circ, below=1.5em of m1];
    \vertex [xshift=0.3cm, yshift=-0.3cm] ;
    \vertex [dashhorizontal, below=0.75em of m1];
     \vertex [dashvertical, left=0.5em of m1];
    \end{feynman}
    \end{tikzpicture}+\ldots
\end{align}
is most easily written in Fourier space, where the steady-state limit $t_0\to -\infty$ has the effect of restricting any incoming frequency and momentum to $\omega'=0$ and $b=0$. We discuss this mechanism, which is very widely applicable, in the following.

Firstly, any tree-like contribution to, say, $\ave{\phi_a(\omega)\phitilde_b(\omega')}$ has an incoming right bare propagator as well as a number of internal bare propagators, which each ``carry" all of $\omega'$.
Given the pre-factor of $\deltabar(\omega+\omega')$ of each diagram, taking the Fourier transform in $t$ and $t_0$ of such a diagram amounts to identifying the poles $p_i$ in $\omega'$ throughout, so that
in direct time, the $n$th order contribution to the propagator generally has the structure
\begin{equation}
\aveHarm{\phi_a(t)\phitilde_b(t_0) \actionPert^n}
= \sum_i \exp{\imag (t-t_0) p_i} \FC_{ab}(p_i)
\end{equation}
with some (complicated) $\FC_{ab}(p_i)$ determined by the residues. By causality, all poles in $\omega'$ have positive imaginary part, provided the mass $r$ is positive. Taking the limit $t_0\to -\infty$ selects those poles that vanish in the limit $r\downarrow 0$. In tree diagrams, these poles are entirely given by the bare propagators, featuring as external or internal legs.

The poles of the bare propagators \Eref{all_bare_propagators} are determined by
\begin{equation}\elabel{propagator_poles}
0=
\invBareProp{a}(\omega;r)
\invBareProp{-a}(\omega;r)
-\gamma^2
=
\Big(\imag\omega'+Dk_b^2+r+2\gamma\Big)
\Big(\imag\omega'+Dk_b^2+r\Big)
+(wk_b)^2
\end{equation}
using $b=-a$ from $\delta_{a+b,0}$ and
$\omega'=-\omega$ from $\deltabar(\omega'+\omega)$ in the numerators of \Erefs{all_bare_propagators}.
It is immediately clear that any $\imag\omega'+r$ that solves this equation cannot be arbitrarily small for any $\Zset \ni b\ne0$. This leaves us with requiring $b=0$, which has one pole that vanishes at $r\downarrow 0$, namely $\omega'=\imag r$.
Taking $t_0\to -\infty$ after Fourier transforming and $r\downarrow0$ therefore leaves only those propagators that have $a=-b=0$.

Because of the structure of the vertex, \Eref{def_actionPert_Fourier},
the perturbative contributions
to the full propagators due to tree-level diagrams such as \Eref{example_phi_phitilde} all have dashed internal and outgoing propagators, \ie they all, except the incoming leg, carry a pre-factor $k_a$, so that they each vanish for vanishing momentum. Taking therefore the limit $t_0\to-\infty$, leaves (of the many residues there might be), only those that do not carry such an extra factor $k_a=0$ of a momentum. In the present field theory, the only such bare propagator is the incoming leg. We will next determine its value in the limit $t_0\to-\infty$ and the effect that this limit has on the rest of the diagram. We can safely assume that this pole is simple, because if it was repeated by being equally the pole of another propagator, its residue  must vanish, because any such other propagator is guaranteed to carry a factor of $k_a$.

At $a=0$ the poles \Eref{propagator_poles} are conveniently written as $(-\imag\omega+r+2\gamma)(-\imag\omega+r)=0$,
leaving a factor of $2\gamma$ in the denominator as a residue. As far as the numerators of the propagators are
concerned, \Erefs{bare_propagator_phiphi} and \eref{bare_propagator_psipsi} produce a factor of
$\invBareProp{0}(-\imag r;r)\deltabar(\omega+\omega')L\delta_{a+b,0}$ and \Erefs{bare_propagator_phipsi} and \eref{bare_propagator_psiphi} a factor of $\gamma\deltabar(\omega+\omega')L\delta_{a+b,0}$, which are in fact the same as $\invBareProp{0}(-\imag r;r)=\gamma$, \Eref{def_invBareProp}.

This mechanism of taking the Fourier transform in $\omega'$ and $\omega$ and taking the limit $t_0\to-\infty$ after $r\downarrow0$ thus has the effect of
replacing any incoming bare propagator such as $\aveHarm{\phi_a(t)\phitilde_b(t_0)}$ by $\gamma L\delta_{a+b,0}\delta_{b,0}/(2\gamma)$,
where $\delta_{b,0}$ enforces that only $b=0=-a$ ever contributes.
Taking further the inverse Fourier-sum over $b$, \Eref{def_Fourier}, and transforming $t$ back to $\omega$ then gives
\begin{equation}\elabel{stationarity_mechanism_app}
    \lim_{t_0\to -\infty}
    \lim_{r\downarrow 0}
    \aveHarm{\phi_a(\omega)\phitilde(x_0,t_0)}
    =
\frac{1}{L} \sum_b \exp{\imag k_b x_0}
\int \dint{t} \exp{\imag\omega t}
    \lim_{t_0\to -\infty}
    \lim_{r\downarrow 0}
    \aveHarm{\phi_a(t)\phitilde_b(t_0)}
    = \frac{1}{2}\deltabar(\omega)\delta_{a,0}
\end{equation}
which is identically the same for all bare propagators \Erefs{all_bare_propagators}.
Within a diagram the effect of the limit $t_0\to-\infty$ is thus a pre-factor determined by the diagram and an amputation, as both $\omega$ and $a$ are forced to vanish. Because the internal field $\phi_a(\omega)$ will feature in an integral $\int\dintbar{\omega}$, the $\deltabar(\omega)$ will simply be integrated out, whereas the sum over the spatial modes, $L^{-1} \sum_a$ will leave behind a factor $1/L$. In summary, the limit $t_0\to -\infty$ in a bare propagator is given by \Eref{stationarity_mechanism_app} and in all the diagrams considered here, amounts to an amputation of the incoming leg and a multiplication by of the diagram by $1/(2L)$.

After the removal of the incoming leg and thus setting effectively $-\imag\omega+r=0$ throughout a diagram, some residues of the form
\begin{equation}
    k_c^2 \left(
    \invBareProp{c}(-\imag r;r)\invBareProp{-c}(-\imag r;r)-\gamma^2
    \right)^{-1}
    =\frac{k_c^2}{(Dk_c^2)^2+(w^2+2D\gamma)k_c^2}
\end{equation}
feature with index $c$ still to be summed over.
At $k_c=0$ such residues are in fact repeated poles of $\omega'$, that are, however, bound to vanish as argued above.
Cancelling $k_c^2$ in numerator and denominator for $c=0$ nevertheless ignores the fact that the term that gives rise to it,
\begin{equation}
k_c^2 \left(
    \invBareProp{c}(\omega;r)\invBareProp{-c}(\omega;r)-\gamma^2
    \right)^{-1}
    =\frac{k_c^2}
    {
    \Big(-\imag\omega+Dk_c^2+r+2\gamma\Big)
\Big(-\imag\omega+Dk_c^2+r\Big)
+(wk_c)^2
}
\end{equation}
strictly
vanishes at $c=0$.
This is the reason for the introduction of the indicator function
\begin{equation}\elabel{def_indicator}
    \indicator_a=1-\delta_{a,0}=
    \begin{cases}
    0 & \text{for}\ a=0\\
    1 & \text{otherwise}
    \end{cases} \ .
\end{equation}
As $c\in\Zset$, this is merely a matter of bookkeeping and algebra, not a matter of complex analysis or exchange of limits, as is illustrated by the example
\begin{subnumcases}{\elabel{illustration_indicator}
    \lim_{t_0\to-\infty}
    \lim_{r\downarrow0}
    \int \dintbar{\omega'}
    \exp{-\imag\omega't_0}
    \frac{Dk_c^2}{\imag\omega'+Dk_c^2+r}
    \frac{1}{\imag\omega'+r}
    =}
    0 & for $c=0$\\
    1 & otherwise
\end{subnumcases}
for any $c\in\Zset$.

\subsection{Steady-state density of Run-and-Tumble particles in a periodic potential}
On the basis of the mechanism outlined above, we can write down a perturbative expansion \Eref{expansion_pert} of the steady-state density, for example
\begin{multline}
\lim_{t_0\to-\infty}
    \ave{\phi_a(t)\phitilde(x_0,t_0)} =
\lim_{t_0\to-\infty}
    \ave{\phi_a(t)\psitilde(x_0,t_0)}
    \\ =
\lim_{t_0\to-\infty}
    \aveHarm{\phi_a(t)\phitilde(x_0,t_0)}
+
\lim_{t_0\to-\infty}
    \aveHarm{\phi_a(t)\phitilde(x_0,t_0)(-\actionPert)}
+ \ldots
\end{multline}
where we have trivially taken the inverse Fourier transform from $\omega$ to $t$. The limit
$\lim_{t_0\to-\infty}
    \aveHarm{\phi_a(t)\phitilde(x_0,t_0)}=\delta_{a,0}/2$
is given by \Eref{stationarity_mechanism_app}, whereas the higher order terms are most easily calculated using the diagrammatics. The lowest order corrections are of the form 
\begin{subequations}
\elabel{first_order_correction_diagrams_all}
\begin{align}
\elabel{first_order_correction_diagrams}
&    \lim_{t_0\to-\infty} \aveHarm{\phi_a(t)\phitilde(x_0,t_0)(-\actionPert)}
\corresponds
\begin{tikzpicture}[circ/.style={shape=circle, inner sep=2pt, outer sep =0.5pt, line width=0.4mm, draw, node contents=}, baseline={(current bounding box.center)}]
    \begin{feynman}
    \vertex (m1) ;
    \vertex [left=2 em of m1] (w) ;
    \vertex [below=1.5em of m1] (rc) ;
    \vertex [right=2 em of m1] (w0) ;-
    \diagram* {
        (w) -- [\SSpropline] (m1),
        (m1) -- [black, line width=0.4mm, -, dashed] (rc),
    };
    \vertex [circ, below=1.5em of m1];
    \vertex [xshift=0.3cm, yshift=-0.3cm] ;
    \vertex [dashhorizontal, below=0.75em of m1];
     \vertex [dashvertical, left=0.5em of m1];
    \end{feynman}
    \end{tikzpicture}
    +
    \begin{tikzpicture}[circ/.style={shape=circle, inner sep=2pt, outer sep =0.5pt, line width=0.4mm, draw, node contents=}, baseline={(current bounding box.center)}]
    \begin{feynman}
    \vertex (m1) ;
    \vertex [left=1 em of m1] (b) ;
    \vertex [left=1 em of b] (a);
    \vertex [below=1.5em of m1] (V) ;
    \vertex [right=1 em of m1] (c) ;
    \vertex [right=1 em of c] (d) ;
    \diagram* {
        (a) -- [\SSpropline] (b) -- [ \SSpropdashed] (m1),
         (m1) -- [black, line width=0.4mm, -, dashed] (V),
    };
    \vertex [circ, below=1.5em of m1];
    \vertex [xshift=0.3cm, yshift=-0.3cm] ;
    \vertex [dashhorizontal, below=0.75em of m1];
     \vertex [dashvertical, left=0.5em of m1];
    \end{feynman}
    \end{tikzpicture}\\
\elabel{first_order_correction_plugged_in}
& \corresponds (-k_a \indicator_a)
\frac{\invBarePropO{a} }{\invBarePropO{a}\invBarePropO{-a}-\gamma^2}
\Big( \nu k_a \extPot_a \Big) \frac{1}{2L}
+
(-k_a \indicator_a)
\frac{\gamma \indicator_a}{\invBarePropO{a}\invBarePropO{-a}-\gamma^2}
\Big( \nu k_a \extPot_a \Big) \frac{1}{2L}\\
\elabel{first_order_correction_simplified}
& =
-
\frac{(Dk_a^2-\imag wk_a + 2\gamma) \indicator_a}{D^2k_a^2+2D\gamma+w^2}
\frac{\nu \extPot_a}{2L}
\end{align}
\end{subequations}
where, again, the Fourier transform in $\omega$ is easily taken as no $\omega$ flows through the diagrams \Eref{first_order_correction_diagrams}, reducing any $\invBareProp{a}(\omega;r)$ in \Eref{all_bare_propagators} to $\invBareProp{a}(0;r)$ and further to $\invBarePropO{a}=\invBareProp{a}(0;0)=Dk_a^2-\imag wk_a+\gamma$ as $r\downarrow0$.
To arrive at \Eref{first_order_correction_plugged_in}, we have firstly inserted the dash on the outgoing propagator due to the perturbation, $(-k_a \indicator_a)$, with the indicator function due to the $t_0\to-\infty$ mechanism, secondly the propagators \Erefs{bare_propagator_phiphi} and \eref{bare_propagator_phipsi} respectively, both evaluated at $\omega=0$ and $r\downarrow0$, thirdly the derivative of the potential due to the perturbation and finally a factor $1/(2L)$ due to the $t_0\to-\infty$ mechanism. The final line \Eref{first_order_correction_simplified} is then a matter of algebra, using $\invBarePropO{a}\invBarePropO{-a}-\gamma^2=k_a^2(D^2k_a^2+2D\gamma+w^2)$.

As all bare propagators will be evaluated at $\omega'=0$ and $r\downarrow0$ in the following,
we introduce $X_a=Y_a^*$ and $Z_a$ to ease notation,
\begin{subequations}
\elabel{def_XYZ}
\begin{align}
\elabel{def_X}
    \aveHarm{\phi_a(\omega)\phitilde_b(\omega'=0)} k_{-a} \indicator_a &=
    \frac{D k_a^2 - \imag w k_a + \gamma}{k_a^2(D^2k_a^2+2D\gamma+w^2)} \deltabar(\omega) L \delta_{a+b,0} k_{-a} \indicator_a
    = X_a \deltabar(\omega) L \delta_{a+b,0} \\
\elabel{def_Y}
    \aveHarm{\psi_a(\omega)\psitilde_b(\omega'=0)} k_{-a} \indicator_a &=
    \frac{D k_a^2 + \imag w k_a + \gamma}{k_a^2(D^2k_a^2+2D\gamma+w^2)} \deltabar(\omega) L \delta_{a+b,0} k_{-a} \indicator_a
    = Y_a \deltabar(\omega) L \delta_{a+b,0} \\
\elabel{def_Z}
    \aveHarm{\phi_a(\omega)\psitilde_b(\omega'=0)} k_{-a} \indicator_a &=
    \frac{\gamma}{k_a^2(D^2k_a^2+2D\gamma+w^2)} \deltabar(\omega) L \delta_{a+b,0} k_{-a} \indicator_a
    = Z_a \deltabar(\omega) L \delta_{a+b,0} \\
\elabel{def_Zdash}
    \aveHarm{\psi_a(\omega)\phitilde_b(\omega'=0)} k_{-a} \indicator_a &=
    \frac{\gamma}{k_a^2(D^2k_a^2+2D\gamma+w^2)} \deltabar(\omega) L \delta_{a+b,0} k_{-a} \indicator_a
    = Z_a \deltabar(\omega) L \delta_{a+b,0}
\end{align}
\end{subequations}
where we have anticipated the factor $k_{-a}\indicator_a$ that each propagator will pick up.

\subsubsection{Recurrence relation for 
density \texorpdfstring{$\density(x)$}{r}
 and 
polarity \texorpdfstring{$\polarity(x)$}{m}
}
\label{recurrenceDensPol}
In the following, we obtain a recurrence relation of the orders in $\nu$ of  $\rho(x)$ and $\polarity(x)$ using their diagrammatic representation. To this end, we introduce a diagrammatic notation for the $b$th Fourier coefficient of the $(n-1)$th order steady-state density of right-moving particles
\begin{multline}\elabel{densityRight_order_nM1}
    \nu^{n-1}\densityRight[b]^{(n-1)}
    =\lim_{t_0\to-\infty} \aveHarm{\phi_b(t)\phitilde(x_0,t_0)\frac{(-\actionPert)^{n-1}}{(n-1)!}}
    =\lim_{t_0\to-\infty} \aveHarm{\phi_b(t)\psitilde(x_0,t_0)\frac{(-\actionPert)^{n-1}}{(n-1)!}}
    \\ \corresponds
    \begin{tikzpicture}
    \begin{feynman}
    \vertex (m1) ;
    \vertex [left=2 em of m1] (w) ;
    \vertex [left=2 em of w] (w1) ;
    \vertex [below=3em of m1] (rc) ;
     \vertex [right=2 em of m1] (w5) ;
    \vertex [right=2 em of w5] (w0) ;
    \vertex [below=3 em of w0] (a0);
    \vertex [right=2 em of m1] (z);
    \vertex [right=1 em of w] (z1);
    \vertex [left=1 em of m1] (z2);
    \vertex [right=1.5 em of m1] (z3);
    \vertex [left=1.5 em of w0] (z4);
    \vertex [above=-0.4em of w5]{...};
    \vertex [dashrotate, above=0 em of z3];
    \vertex [dashrotate, above=0 em of z4];
    \diagram* {
        (w1) -- [\SSpropline] (w)--[\SSpropgluon] (z1),
        (z2)-- [\SSpropgluon] (z3),
        (z4)-- [\SSpropgluon] (w0),
        (m1) -- [black, line width=0.4mm, -, dashed] (rc),
        (w0) -- [black, line width=0.4mm, -, dashed] (a0),
    };
    \vertex [circ, below=3 em of m1];
    \vertex [circ, below=3 em of w0] ;
    \vertex [below=4 em of w0] (c);
    \vertex [dashverticallarge, left=1em of m1];
    \vertex [dashhorizontallarge, below=1.5em of m1];
    \vertex [dashhorizontallarge, below=1.5em of m1];
    \vertex [above=0.2em of w1] {$b$};
     \vertex [dashhorizontallarge, below=1.5em of w0];
     \vertex [dashhorizontallarge, below=1.5em of w0];
     \vertex [below=1.5em of w5]{$...$};
    \vertex [xshift=0.3cm, yshift=-0.3cm] ;
    \end{feynman}
    \end{tikzpicture}
\end{multline}
and the $(n-1)$th order steady-state density of left-moving particles
\begin{multline}\elabel{densityLeft_order_nM1}
    \nu^{n-1}\densityLeft[b]^{(n-1)}
    =\lim_{t_0\to-\infty} \aveHarm{\phi_b(t)\phitilde(x_0,t_0)\frac{(-\actionPert)^{n-1}}{(n-1)!}}
    =\lim_{t_0\to-\infty} \aveHarm{\phi_b(t)\psitilde(x_0,t_0)\frac{(-\actionPert)^{n-1}}{(n-1)!}}
    \\ \corresponds
    \begin{tikzpicture}
    \begin{feynman}
    \vertex (m1) ;
    \vertex [left=2 em of m1] (w) ;
    \vertex [left=2 em of w] (w1) ;
    \vertex [below=3em of m1] (rc) ;
     \vertex [right=2 em of m1] (w5) ;
    \vertex [right=2 em of w5] (w0) ;
    \vertex [below=3 em of w0] (a0);
    \vertex [right=2 em of m1] (z);
    \vertex [right=1 em of w] (z1);
    \vertex [left=1 em of m1] (z2);
    \vertex [right=1.5 em of m1] (z3);
    \vertex [left=1.5 em of w0] (z4);
    \vertex [above=-0.4em of w5]{...};
    \vertex [dashrotate, above=0 em of z3];
    \vertex [dashrotate, above=0 em of z4];
    \diagram* {
        (w1) -- [\SSpropdashed] (w)--[\SSpropgluon] (z1),
        (z2)-- [\SSpropgluon] (z3),
        (z4)-- [\SSpropgluon] (w0),
        (m1) -- [black, line width=0.4mm, -, dashed] (rc),
        (w0) -- [black, line width=0.4mm, -, dashed] (a0),
    };
    \vertex [circ, below=3 em of m1];
    \vertex [circ, below=3 em of w0] ;
    \vertex [below=4 em of w0] (c);
    \vertex [dashverticallarge, left=1em of m1];
    \vertex [dashhorizontallarge, below=1.5em of m1];
    \vertex [dashhorizontallarge, below=1.5em of m1];
    \vertex [above=0.2em of w1] {$b$};
     \vertex [dashhorizontallarge, below=1.5em of w0];
     \vertex [dashhorizontallarge, below=1.5em of w0];
     \vertex [below=1.5em of w5]{$...$};
    \vertex [xshift=0.3cm, yshift=-0.3cm] ;
    \end{feynman}
    \end{tikzpicture} \ .
\end{multline}
As indicated by the dots,
both diagrams, \Erefs{densityRight_order_nM1} and \eref{densityLeft_order_nM1}, contain $n-1$ potential ``baubles" and both are written in terms of ``gluonic" propagators $\begin{tikzpicture}[baseline=-2.5pt]
\begin{feynman}
\vertex (m1) ;
\vertex [right=2 em of m1] (w) ;
\diagram* {
        (m1) -- 
        [\SSpropgluon]
        (w);
    };
\end{feynman}
\end{tikzpicture}$
which we use to represent the sum over
all possibilities of
connecting the baubles to each other and finally to the external field.
Much of what follows is about the bookkeeping of these diagrams.
The lowest order is given by \Eref{stationarity_mechanism_app},
\begin{equation}\elabel{densities_order0}
    \densityRight[a]^{(0)}
    =\densityLeft[a]^{(0)}
    = \half \delta_{a,0} \ ,
\end{equation}
after the inverse Fourier transform that simply removes $\deltabar(\omega)$.

To construct the $n$ order diagrams of both particle species we need to attach a potential bauble to the $n-1$th order diagrams. Any such potential term \Eref{def_actionPert_Fourier} comes with a summation over all three indices associated with its three terms, namely the potential, a creator field with a derivative and an annihilator field. By adding a further perturbative term in this way, both $\densityRight[a]^{(n)}$ and $\densityLeft[a]^{(n)}$ each acquire two contributions due to  $\densityRight[b]^{(n-1)}$ and $\densityLeft[b]^{(n-1)}$,
both connecting to $\densityRight[b]^{(n)}$ and $\densityLeft[b]^{(n1)}$ using \Eref{bauble_vertices} and a suitable propagator \Eref{all_bare_propagators}. The resulting terms are
\begin{subequations}
\elabel{densityRight_orderN}
\begin{align}
    \nu^{n}\densityRight[a]^{(n)}&\corresponds
    \begin{tikzpicture}
    \begin{feynman}
    \vertex (m1) ;
    \vertex [left=2 em of m1] (w) ;
    \vertex [left=2 em of w] (w1) ;
    \vertex [below=3em of m1] (rc) ;
     \vertex [right=2 em of m1] (w5) ;
    \vertex [right=2 em of w5] (w0) ;
    \vertex [below=3 em of w0] (a0);
    \vertex [circ, below=3 em of m1];
    \vertex [circ, below=3 em of w0] ;
    \vertex [below=4 em of w0] (c);
    \vertex [dashhorizontallarge, below=1.5em of m1];
    \vertex [dashverticallarge, left=1em of m1];
    \vertex [dashhorizontallarge, below=1.5em of m1];
    \vertex [above=0.7em of w1] (b);
    \vertex[right=0.2em of b] {$b$};
     \vertex [dashhorizontallarge, below=1.5em of w0];
     \vertex [dashhorizontallarge, below=1.5em of w0];
     \vertex [below=1.5em of w5]{$...$};
    \vertex [xshift=0.3cm, yshift=-0.3cm] ;
    \vertex [circ, below=3 em of w1];
    \vertex [below=3 em of w1](l0);
     \vertex [left=2 em of w1] (l1);
     \vertex [left=2 em of l1] (l2);
     \vertex [dashhorizontallarge, below=1.5em of w1];
     \vertex [dashverticallarge, right=1em of l1];
     \vertex[above=0.005em of l2] {$a$};
     \vertex [right=1.5 em of m1] (z3);
    \vertex [left=1.5 em of w0] (z4);
    \vertex [above=-0.4em of w5]{...};
    \vertex [dashrotate, above=0 em of z3];
    \vertex [dashrotate, above=0 em of z4];
     \diagram* {
        (w1) -- [\SSpropline] (w)--[\SSpropgluon] (z1),
        (z2)-- [\SSpropgluon] (z3),
        (z4)-- [\SSpropgluon] (w0),
        (m1) -- [black, line width=0.4mm, -, dashed] (rc),
        (w0) -- [black, line width=0.4mm, -, dashed] (a0),
        (w1)--[black, line width=0.4mm, -, dashed](l0),
        (l1)--[\SSpropline](w1),
        (l1)--[\SSpropline](l2),
    };
    \end{feynman}
    \end{tikzpicture}+\begin{tikzpicture}
    \begin{feynman}
    \vertex (m1) ;
    \vertex [left=2 em of m1] (w) ;
    \vertex [left=2 em of w] (w1) ;
    \vertex [below=3em of m1] (rc) ;
     \vertex [right=2 em of m1] (w5) ;
    \vertex [right=2 em of w5] (w0) ;
    \vertex [below=3 em of w0] (a0);
    \vertex [circ, below=3 em of m1];
    \vertex [circ, below=3 em of w0] ;
    \vertex [below=4 em of w0] (c);
    \vertex [dashhorizontallarge, below=1.5em of m1];
    \vertex [dashverticallarge, left=1em of m1];
    \vertex [dashhorizontallarge, below=1.5em of m1];
    \vertex [above=0.7em of w1] (b);
    \vertex[right=0.2em of b] {$b$};
     \vertex [dashhorizontallarge, below=1.5em of w0];
     \vertex [dashhorizontallarge, below=1.5em of w0];
     \vertex [below=1.5em of w5]{$...$};
    \vertex [xshift=0.3cm, yshift=-0.3cm] ;
    \vertex [circ, below=3 em of w1];
    \vertex [below=3 em of w1](l0);
     \vertex [left=2 em of w1] (l1);
     \vertex [left=2 em of l1] (l2);
     \vertex [dashhorizontallarge, below=1.5em of w1];
     \vertex [dashverticallarge, right=1em of l1];
     \vertex[above=0.005em of l2] {$a$};
     \vertex [right=1.5 em of m1] (z3);
    \vertex [left=1.5 em of w0] (z4);
    \vertex [above=-0.4em of w5]{...};
    \vertex [dashrotate, above=0 em of z3];
    \vertex [dashrotate, above=0 em of z4];
     \diagram* {
       (w1) -- [\SSpropdashed] (w)--[\SSpropgluon] (z1),
        (z2)-- [\SSpropgluon] (z3),
        (z4)-- [\SSpropgluon] (w0),
        (m1) -- [black, line width=0.4mm, -, dashed] (rc),
        (w0) -- [black, line width=0.4mm, -, dashed] (a0),
        (w1)--[black, line width=0.4mm, -, dashed](l0),
        (l1)--[\SSpropdashed](w1),
        (l1)--[\SSpropline](l2),
    };
    \end{feynman}
    \end{tikzpicture}\\
    &\corresponds\frac{1}{L}\sum_b k_{a-b}\nu\extPot_{a-b}
    \Big(
    X_a \nu^{n-1}\densityRight[b]^{(n-1)}
    +
    Z_a \nu^{n-1}\densityLeft[b]^{(n-1)}
    \Big)
\end{align}
\end{subequations}
and
\begin{subequations}
\elabel{densityLeft_orderN}
\begin{align}
    \nu^{n}\densityLeft[a]^{(n)}&\corresponds
    \begin{tikzpicture}
    \begin{feynman}
    \vertex (m1) ;
    \vertex [left=2 em of m1] (w) ;
    \vertex [left=2 em of w] (w1) ;
    \vertex [below=3em of m1] (rc) ;
     \vertex [right=2 em of m1] (w5) ;
    \vertex [right=2 em of w5] (w0) ;
    \vertex [below=3 em of w0] (a0);
    \vertex [circ, below=3 em of m1];
    \vertex [circ, below=3 em of w0] ;
    \vertex [below=4 em of w0] (c);
    \vertex [dashhorizontallarge, below=1.5em of m1];
    \vertex [dashverticallarge, left=1em of m1];
    \vertex [dashhorizontallarge, below=1.5em of m1];
    \vertex [above=0.7em of w1] (b);
    \vertex[right=0.2em of b] {$b$};
     \vertex [dashhorizontallarge, below=1.5em of w0];
     \vertex [dashhorizontallarge, below=1.5em of w0];
     \vertex [below=1.5em of w5]{$...$};
    \vertex [xshift=0.3cm, yshift=-0.3cm] ;
    \vertex [circ, below=3 em of w1];
    \vertex [below=3 em of w1](l0);
     \vertex [left=2 em of w1] (l1);
     \vertex [left=2 em of l1] (l2);
     \vertex [dashhorizontallarge, below=1.5em of w1];
     \vertex [dashverticallarge, right=1em of l1];
     \vertex[above=0.005em of l2] {$a$};
       \vertex [right=1.5 em of m1] (z3);
    \vertex [left=1.5 em of w0] (z4);
    \vertex [above=-0.4em of w5]{...};
    \vertex [dashrotate, above=0 em of z3];
    \vertex [dashrotate, above=0 em of z4];
     \diagram* {
       (w1) -- [\SSpropdashed] (w)--[\SSpropgluon] (z1),
        (z2)-- [\SSpropgluon] (z3),
        (z4)-- [\SSpropgluon] (w0),
        (m1) -- [black, line width=0.4mm, -, dashed] (rc),
        (w0) -- [black, line width=0.4mm, -, dashed] (a0),
        (w1)--[black, line width=0.4mm, -, dashed](l0),
        (l1)--[\SSpropdashed](w1),
        (l1)--[\SSpropdashed](l2),
    };
    \end{feynman}
    \end{tikzpicture}+\begin{tikzpicture}
    \begin{feynman}
    \vertex (m1) ;
    \vertex [left=2 em of m1] (w) ;
    \vertex [left=2 em of w] (w1) ;
    \vertex [below=3em of m1] (rc) ;
     \vertex [right=2 em of m1] (w5) ;
    \vertex [right=2 em of w5] (w0) ;
    \vertex [below=3 em of w0] (a0);
    \vertex [circ, below=3 em of m1];
    \vertex [circ, below=3 em of w0] ;
    \vertex [below=4 em of w0] (c);
    \vertex [dashhorizontallarge, below=1.5em of m1];
    \vertex [dashverticallarge, left=1em of m1];
    \vertex [dashhorizontallarge, below=1.5em of m1];
    \vertex [above=0.7em of w1] (b);
    \vertex[right=0.2em of b] {$b$};
     \vertex [dashhorizontallarge, below=1.5em of w0];
     \vertex [dashhorizontallarge, below=1.5em of w0];
     \vertex [below=1.5em of w5]{$...$};
    \vertex [xshift=0.3cm, yshift=-0.3cm] ;
    \vertex [circ, below=3 em of w1];
    \vertex [below=3 em of w1](l0);
     \vertex [left=2 em of w1] (l1);
     \vertex [left=2 em of l1] (l2);
     \vertex [dashhorizontallarge, below=1.5em of w1];
     \vertex [dashverticallarge, right=1em of l1];
     \vertex[above=0.005em of l2] {$a$};
       \vertex [right=1.5 em of m1] (z3);
    \vertex [left=1.5 em of w0] (z4);
    \vertex [above=-0.4em of w5]{...};
    \vertex [dashrotate, above=0 em of z3];
    \vertex [dashrotate, above=0 em of z4];
     \diagram* {
        (w1) -- [\SSpropline] (w)--[\SSpropgluon] (z1),
        (z2)-- [\SSpropgluon] (z3),
        (z4)-- [\SSpropgluon] (w0),
        (m1) -- [black, line width=0.4mm, -, dashed] (rc),
        (w0) -- [black, line width=0.4mm, -, dashed] (a0),
        (w1)--[black, line width=0.4mm, -, dashed](l0),
        (l1)--[\SSpropline](w1),
        (l1)--[\SSpropdashed](l2),
    };
    \end{feynman}
    \end{tikzpicture}\\
    &\corresponds\frac{1}{L}\sum_b k_{a-b}\nu\extPot_{a-b}
    \Big(
    Z_a \nu^{n-1}\densityRight[b]^{(n-1)}
    +
    Y_a \nu^{n-1}\densityLeft[b]^{(n-1)}
    \Big) \ .
\end{align}
\end{subequations}
Using \Erefs{densityRight_orderN} and \eref{densities_order0} with $n=1$ produces for example \Erefs{first_order_correction_diagrams_all},
\begin{equation}
\nu\densityRight[a]^{(1)}=\frac{1}{L}k_a\nu\extPot_a(X_a+Z_a)\half
\end{equation}
with \Erefs{def_X} and \eref{def_Z}.

In the following, it is more instructive to work with the density
\begin{subequations}
\elabel{densPol_from_modes}
\begin{equation}
\density^{(n)}_a=\densityRight[a]^{(n)}+\densityLeft[a]^{(n)}
\quad\text{ so that }\quad
\density(x) = \frac{1}{L} \sum_{a=-\infty}^\infty \exp{\imag k_a x} \sum_{n=0}^\infty \nu^n \density^{(n)}_a
\end{equation}
and the polarity
\begin{equation}
\polarity^{(n)}_a=\densityRight[a]^{(n)}-\densityLeft[a]^{(n)} \quad\text{ so that }\quad
\polarity(x) = \frac{1}{L} \sum_{a=-\infty}^\infty \exp{\imag k_a x} \sum_{n=0}^\infty \nu^n \polarity^{(n)}_a
\ ,
\end{equation}
\end{subequations}
and write the convolutions \Erefs{densityRight_orderN} and \eref{densityLeft_orderN} in terms of a single matrix equation
\begin{equation}\elabel{densPol_from_matrixM}
    \begin{pmatrix}
    \density^{(n)}_a\\
    \polarity^{(n)}_a
    \end{pmatrix}
=\matrixM{a}\sum_{b=-\infty}^\infty W_{a-b}\begin{pmatrix}
    \density^{(n-1)}_b\\
    \polarity^{(n-1)}_b
    \end{pmatrix}
\end{equation}
with
\begin{equation}\elabel{def_W}
    W_{a-b}=\frac{k_{a-b}\extPot_{a-b}}{L}
\end{equation}
and the matrix
\begin{equation}\elabel{def_matrixM}
\matrixM{a}=
\half
\begin{pmatrix}
 X_a+Y_a+2Z_a &  X_a-Y_a\\
 X_a-Y_a &       X_a+Y_a-2Z_a
\end{pmatrix}
=
\frac{(-k_a)\indicator_a}{k_a^2(D^2k_a^2+2D\gamma+w^2)}
\begin{pmatrix}
Dk_a^2+2\gamma & -\imag wk_a \\
-\imag wk_a & Dk_a^2
\end{pmatrix} \ ,
\end{equation}
as a function of the mode $a$, closely related to \Eref{inverse_matrixHarm_a}. \Erefs{densPol_from_matrixM} and \eref{def_matrixM} are the main result of the present section as they determine density and polarity to arbitrary order in $\nu$. As a sanity check, we may verify that both density $\density(x)$ and polarity $\polarity(x)$, \Erefs{densPol_from_modes}, are real, or equivalently that complex conjugates of the modes obey
$\Big(\density^{(n)}_a\Big)^*=\density^{(n)}_{-a}$
and
$\Big(\polarity^{(n)}_a\Big)^*=\polarity^{(n)}_{-a}$.
This can be done by induction, assuming that
$\Big(\density^{(n-1)}_a\Big)^*=\density^{(n-1)}_{-a}$
and
$\Big(\polarity^{(n-1)}_a\Big)^*=\polarity^{(n-1)}_{-a}$,
and further observing that
$\Big(\matrixM{a}\Big)^*=-\matrixM{-a}$
from \Eref{def_matrixM}
and that a real external potential implies $\Big(W_{a-b}\Big)^*=-W_{b-a}$.
From \Eref{densities_order0} we have
\begin{equation}\elabel{induction_basis}
    \begin{pmatrix}
    \density^{(0)}_a\\
    \polarity^{(0)}_a
    \end{pmatrix}
    =
    \begin{pmatrix}
    1\\
    0
    \end{pmatrix}
    \delta_{a,0}
\end{equation}
as induction basis and, more importantly, as the staring point for the systematic calculation \Eref{densPol_from_matrixM} of the modes of the density and polarity \Eref{densPol_from_modes}. With that established, we may write \Eref{def_denPol}
\begin{equation}\elabel{densPol_from_matrixM_iterated}
    \begin{pmatrix}
    \density^{(n)}_{a_1}\\
    \polarity^{(n)}_{a_1}
    \end{pmatrix}
=\matrixM{a_1}\sum_{a_2,a_3,\ldots,a_n=-\infty}^\infty
W_{a_1-a_2}
\matrixM{a_2}
W_{a_2-a_3}
\matrixM{a_3}
\cdots
W_{a_{n-1}-a_n}
\matrixM{a_n}
W_{a_n}
\begin{pmatrix}
    1\\
    0
    \end{pmatrix}
    \ ,
\end{equation}
with each of the $n-1$ indeces running from $-\infty$ to $\infty$.
We will now
turn our attention to the steady-state current.

\subsection{Steady-state current}
\label{current}
From Fick's law of diffusion and the coupled Fokker-Planck equation, the steady-state probability current $J(x)$ reads
\begin{equation}\elabel{J_from_densPol}
\current(x)=w\polarity(x)-\nu\rho(x) \extPot'(x)-D\partial_x\rho(x)
\ ,
\end{equation}
which in the steady state must in fact be constant in $x$. We will demonstrate this property perturbatively, order by order in $\nu$. To this end, we introduce the inverse Fourier-transforms
\begin{subequations}
\elabel{densPol_n_from_modes}
\begin{align}
    \density^{(n)}(x) &= \frac{1}{L} \sum_{a=-\infty}^\infty \exp{\imag k_a x} \density[a]^{(n)} \\
    \polarity^{(n)}(x) &= \frac{1}{L} \sum_{a=-\infty}^\infty \exp{\imag k_a x} \polarity[a]^{(n)}
\end{align}
\end{subequations}
of the $n$th order Fourier coefficients
$\density[a]^{(n)}$
and
$\polarity[a]^{(n)}$, similar to \Eref{densPol_from_modes}. The steady-state current $\current(x)$ in \Eref{J_from_densPol} may then be written order by order in $\nu$,
\begin{equation}\elabel{expansion_current}
    \current(x)=\sum^\infty_{n=0}\nu^n\current^{(n)}(x)
\end{equation}
with $\current^{(n)}(x)$ for $n>0$ extracted from \Eref{J_from_densPol} by identifying terms of order $n$ in $\nu$,
\begin{equation}
    \current^{(n)}(x)=w\polarity^{(n)}(x)-\rho^{(n-1)}(x) \extPot'(x)-D\partial_x\rho^{(n)}(x) \ .
\elabel{def_Jn}
\end{equation}
We will proceed by showing that $\current^{(n)}(x)$ is constant in $x$ and can easily be expressed in terms of the Fourier coefficients $\density[a]^{(n)}$ and $\polarity[a]^{(n)}$ using \Erefs{densPol_n_from_modes} and \eref{densPol_from_matrixM},
\begin{equation}\elabel{Jn_indep_x_step1}
    \current^{(n)}(x)=
    -\density^{(n-1)}(x) \extPot'(x)
    \frac{1}{L}
    \sum_{a,b=-\infty}^\infty \exp{\imag k_a x}
    \begin{pmatrix}
    -D \imag k_a \\
    w
    \end{pmatrix}^\transpose
    \matrixM{a} W_{a-b}
    \begin{pmatrix}
    \density^{(n-1)}_b\\
    \polarity^{(n-1)}_b
    \end{pmatrix} \ .
\end{equation}
By direct calculation the linear algebra involving $\matrixM{a}$, \Eref{def_matrixM}, simplifies drastically,
\begin{equation}\elabel{Jn_indep_x_step2_ingredient1}
        \begin{pmatrix}
    -D \imag k_a \\
    w
    \end{pmatrix}^\transpose
    \matrixM{a}
    =
    \imag \indicator_a
        \begin{pmatrix}
    1 \\
    0
    \end{pmatrix}^\transpose
    \ .
\end{equation}
The first term in \Eref{Jn_indep_x_step1} can be written in modes as
\begin{equation}\elabel{Jn_indep_x_step2_ingredient2}
    \density^{(n-1)}(x) \extPot'(x)
    =
    \frac{1}{L^2}\sum_{a,b=-\infty}^\infty
    \exp{\imag k_a x}
    \exp{\imag k_b x}
    \density[b]^{(n-1)}
    \imag k_a \extPot_a
    =
    \frac{\imag}{L}\sum_{a,b=-\infty}^\infty
    \exp{\imag k_a x}
    \density[b]^{(n-1)}
    W_{a-b}
\end{equation}
with \Erefs{extPot_from_modes} and \eref{def_W} and
using $k_{a-b}+k_b=k_a$ after
shifting the dummy index $a$ by $-b$, so that the final expression is the inverse Fourier-transform of a convolution.
Using \Erefs{Jn_indep_x_step2_ingredient1} and \eref{Jn_indep_x_step2_ingredient2} in \Eref{Jn_indep_x_step1} finally gives
\begin{equation}\elabel{Jn_from_denPol_n-1}
    \current^{(n)}(x)=
    \frac{\imag}{L}\sum_{a,b=-\infty}^\infty
    \exp{\imag k_a x}
    \density[b]^{(n-1)}
    W_{a-b}
    (-1 + \indicator_a)
    =
    -\frac{\imag}{L}\sum_{b=-\infty}^\infty
    \density[b]^{(n-1)}
    W_{-b}
\end{equation}
using the definition of the indicator function, 
$\indicator_a=1-\delta_{a,0}$. \Eref{Jn_from_denPol_n-1} not only shows that every perturbative order of the steady-state current is independent of the position $x$, as it should be, but also expresses the current to order $n$ in terms of the density to order $n-1$. The $0$th order of current follows immediately from \Eref{J_from_densPol} as
\begin{equation}
    \current^{(0)}(x)=w\polarity^{(0)}(x)-D\partial_x\rho^{(0)}(x)=0
\end{equation}
because $\polarity^{(0)}(x)=0$ and $\density^{(0)}(x)=1/L$ from
the inverse Fourier transform of \Eref{induction_basis}.

\Eref{Jn_from_denPol_n-1} is the main result of the present section.
With \Eref{densPol_from_matrixM_iterated} we can write it as \Eref{J_from_M},
\begin{equation}\elabel{J_iterated}
    \current^{(n)}(x)=
    - \frac{\imag}{L}
    \sum_{a_1,a_2,\ldots,a_{n-1}=-\infty}^\infty
    W_{-a_1} W_{a_1-a_2}\cdots  W_{a_{n-2}-a_{n-1}}W_{a_{n-1}}
    \begin{pmatrix}
    1\\
    0
    \end{pmatrix}^\transpose
    \matrixM{a_1} \matrixM{a_2}\cdots \matrixM{a_{n-1}}
    \begin{pmatrix}
    1\\
    0
    \end{pmatrix}
\end{equation}
which contains $n$ factors of $W$ and $n-1$ factors of the matrix $\matrixM{a}$.
It will prove useful to relabel the $n-1$ dummy variables $a_i$ by $n$ dummy variables
$b_1=-a_1$,
$b_2=a_1-a_2$,
$b_3=a_2-a_3$,
\ldots,
$b_{n-1}=a_{n-2}-a_{n-1}$,
$b_n=a_{n-1}$, imposing that $\sum_{i=1}^nb_i=0$ by a Kronecker $\delta$-function,
\begin{multline}\elabel{J_iterated_relabelled} 
    \current^{(n)}(x)=
    - \frac{\imag}{L}
    \sum_{b_1,b_2,\ldots,b_{n}=-\infty}^\infty
    \delta_{b_1+b_2+\ldots+b_n,0}
    W_{b_1} W_{b_2}\cdots  W_{b_{n-1}} W_{b_n}\\
    \times \begin{pmatrix}
    1\\
    0
    \end{pmatrix}^\transpose
    \matrixM{-b_1} \matrixM{-(b_1+b_2)}\matrixM{-(b_1+b_2+b_3)}\cdots \matrixM{-(b_1+b_2+\ldots+b_{n-1})}
    \begin{pmatrix}
    1\\
    0
    \end{pmatrix} \ .
\end{multline}
In \SMref{sym_prop_curr} we consider properties of the steady-state current generally and in relation to the potential.

\section{Comparison between the field theory and the exact result \cite{Astumian1994Mar} for a piece-wise linear potential}
\label{AstumianBier}
In this section, we compare our field-theoretic result of the steady-state current $J$ to the exact result \cite{Astumian1994Mar}. We consider an RnT particle in  a periodic, piece-wise linear ratchet potential $\extPot(x)=\extPot(x+L)$ of the form
\begin{equation}
\extPot(x)=\begin{cases}
\extPot_0 \frac{x}{\alpha L}
& \text{for}\ x \in[0,\alpha L)\\
\extPot_0 \frac{L-x}{(1-\alpha) L}
& \text{for}\ x \in[\alpha L,L)
\end{cases}
\elabel{piecewise_linear_potential}
\end{equation}
as shown in \fref{AB_pot}.

\begin{figure}[t]
\input{Fig_ABplot}
\caption{The periodic, piece-wise linear ratchet described by  \Eref{piecewise_linear_potential} with $\alpha=0.9$. The ratchet has $\extPot(0)=0$ rises up to $\extPot(\alpha L)=U_0$ and descends again to $\extPot(L)=0$.
The point $x^*=\alpha L/2$ marks the point where
$\extPot(x^*+x) - \extPot(x^*) = \extPot(x^*) - \extPot(x^*-x)$
with $\extPot(x^*)=U_0/2$,
\Eref{generalised_odd_pot}, \SMref{sym_prop_curr}.
}
\flabel{AB_pot}
\end{figure}

The Fokker-Planck equations for RnT particles in a piece-wise linear potential are most easily written down by distinguishing the two intervals $x\in[0,\alpha L)$, by, say, superscript $\ABindex{i}=\ABindex{1}$ and $x\in[\alpha L, L)$ by, say, superscript $\ABindex{i}=\ABindex{2}$, so that
\begin{subequations}
\elabel{AB_FP}
\begin{align}
\partial_t\densityRight^{\ABindex{i}}(x,t)&=-\partial_x[(w-\nu
\extPot'^{\ABindex{i}}
)\densityRight^{\ABindex{i}}(x,t)]-\gamma (\densityRight^{\ABindex{i}}(x,t)-\densityLeft^{\ABindex{i}}(x,t))+D\partial_x^2 \densityRight^{\ABindex{i}}(x,t)\nonumber\\
\partial_t\densityLeft^{\ABindex{i}}(x,t)&=-\partial_x[(-w-\nu
\extPot'^{\ABindex{i}}
)\densityLeft^{\ABindex{i}}(x,t)]-\gamma (\densityLeft^{\ABindex{i}}(x,t)-\densityRight^{\ABindex{i}}(x,t))+D\partial_x^2 \densityLeft^{\ABindex{i}}(x,t)
\end{align}
\end{subequations}
with constant slopes $\extPot'^{\ABindex{1}}=\extPot_0/(\alpha L)$ and $\extPot'^{\ABindex{2}}=-\extPot_0/((1-\alpha)L)$. Similar to the main text, the densities $\densityRight^{\ABindex{i}}(x,t)$ and $\densityLeft^{\ABindex{i}}(x,t)$ refer to the right-moving and the left-moving particles respectively. Solving this system of equations is greatly facilitated by the slopes being piece-wise constant.

Following the procedure outlined in \cite{Astumian1994Mar}, the steady-state densities $\densityLR^{\ABindex{i}}(x)$ are determined by making the ansatz
$\density_\pm^{\ABindex{i}}=Z^{\ABindex{i}}_\pm e^{\lambda^{\ABindex{i}}x}$, to find $4$ linearly independent solutions of \Erefs{AB_FP} for $i=1,2$ and both particle species $\density_\pm$, and solving two coupled quadratic characteristic equations for the eigenvalues. One of those always vanishes, while the others can be relabelled as $\alpha^{\ABindex{i}}$, $\beta^{\ABindex{i}}$ and $\gamma^{\ABindex{i}}$, producing the ansatz
\begin{equation}
\rho_\pm^{\ABindex{i}}(x)=
 A_{\pm}^{\ABindex{i}}e^{\alpha^{\ABindex{i}}x}
+B_{\pm}^{\ABindex{i}}e^{\beta^{\ABindex{i}}x}
+C_{\pm}^{\ABindex{i}}e^{\gamma^{\ABindex{i}}x}
+D_{\pm}
\elabel{AB_rho} \ ,
\end{equation}
which leaves $16$ amplitudes to be determined: $A_+^{\ABindex{1}}$, $A_-^{\ABindex{1}}$, $A_+^{\ABindex{2}}$, \ldots, $D_-^{\ABindex{2}}$.  

The $16$ amplitudes are determined by substituting \Erefs{AB_rho} into \Erefs{AB_FP} at steady state, when the left hand side of the latter vanishes. As a result, the pre-factor of every linearly independent exponential on the right has to vanish, fixing the ratios $A_+^{\ABindex{i}}/A_-^{\ABindex{i}}$, $B_+^{\ABindex{i}}/B_-^{\ABindex{i}}$, $C_+^{\ABindex{i}}/C_-^{\ABindex{i}}$ and  $D_+^{\ABindex{i}}/D_-^{\ABindex{i}}$
in terms of system parameters
for
both $i=1,2$, which amounts to $8$ equations for $16$ unknowns. A further $8$ equations are obtained by matching conditions at the boundaries of intervals $i=1$ and $i=2$, more specifically for the density,
$\densityLR^{\ABindex{1}}(0)=\densityLR^{\ABindex{2}}(L)$
and
$\densityLR^{\ABindex{1}}(\alpha L)=\densityLR^{\ABindex{2}}(\alpha L)$,
and correspondingly for the current
$\current_{\pm}^{\ABindex{i}}(x)=(\pm w-\nu \extPot'^{\ABindex{i}}-\plaind/\plaind x)\rho_\pm^{\ABindex{i}}(x)$. It turns out that only $15$ of those $16$ equations are linearly independent, as expected in a homogeneous, linear system of equations. The final equation is in fact obtained by demanding in addition overall normalisation of the density. This procedure, best done in a computational algebra system \cite{Mathematica}, results in the density and the current being determined in closed form.

To compare to our field-theory, we choose the height of the piece-wise linear potential $\extPot(x)$ in \Eref{piecewise_linear_potential} by setting $\extPot_0=1$ and vary the coupling $\nu$ in \Erefs{AB_FP}. Density and current are then determined through the procedure outlined above.

The field-theoretic approach on the other hand is based on the Fourier coefficients of the potential \Eref{piecewise_linear_potential}. Taking the first $200$ such modes,
\begin{equation}
    U_a=\frac{U_0 \left(1-e^{-2 i \pi  a \alpha }\right)}{4 \pi ^2 a^2 \alpha  (\alpha -1)} \ ,
\end{equation}
for $a=-200,-199,\cdots,200$,
we have calculated the steady-state current $\current$ to order $75$ in $\nu$.
A comparison between this field-theoretic result and the exact result obtained above is shown in \fref{comparisonAB}. It shows perfect agreement for $\nu$ up to the radius of convergence
$\nuRocSmall=4.272\ldots$
estimated via \Eref{def_nuRoc}.

\begin{figure}
\input{Fig_comparisonAB_75}
\caption{
Comparison of the present field-theoretic result (filled circles) for the steady-state current and the exact solution \cite{Astumian1994Mar}
(solid line) for a piece-wise linear ratchet potential \Eref{piecewise_linear_potential} with $\alpha=0.9$
and $\extPot_0=1$
as shown in \fref{AB_pot}.
The other parameters are $w=1,\gamma=5$, $D=1$ and $L=1$.
The field theoretic results are based on $200$ Fourier coefficients and up to $75$
orders in $\nu$.
The agreement is perfect (deviation of less than $0.15\%$
up until
$\nuRocSmall=4.272\ldots$
the radius of convergence
estimated via \Eref{def_nuRoc} and shown as a dotted vertical line 
Beyond that, the current due to the field theory diverges sharply.
The coupling $\nu$ that produces the best current
$\currentOptim=0.015739\ldots$
in this potential is $\nuOptim=3.90\ldots$, where even the shallow part of the potential has a slope greater than the self-propulsion speed, $\nuOptim \extPot_0/(\alpha L)=3.9\cdot1/(0.9\cdot1)>w=1$, indicating that transport is dominated by diffusion.
}
\flabel{comparisonAB}
\end{figure}
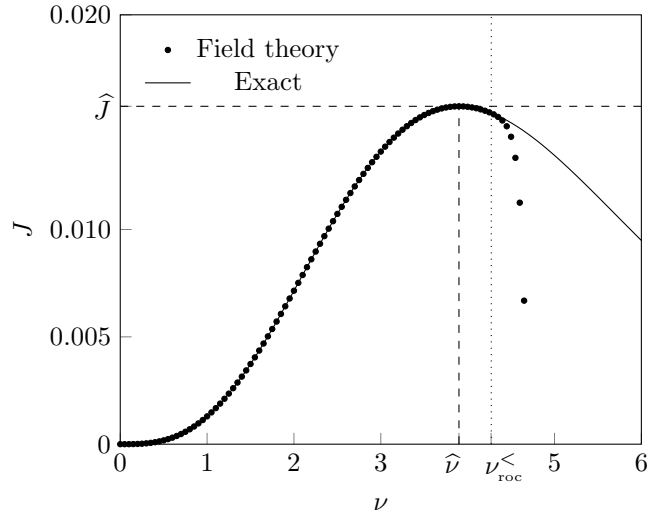

\section{Symmetry properties of the steady-state current}
\label{sym_prop_curr}
In the following we derive some general properties of the steady-state current on the basis of \Erefs{Jn_from_denPol_n-1} and \eref{densPol_from_matrixM}, that hold to all orders of the perturbation theory. We may briefly consider an odd potential, $\extPot(-x)=-\extPot(x)$,
or, given periodicity, more generally
a potential with
\begin{equation}
\elabel{generalised_odd_pot}
    \extPot(x^*+x) - \extPot(x^*) = \extPot(x^*) - \extPot(x^*-x)
\end{equation}
allowing for some offset $\extPot(x^*)$ and a more general point of inversion $x^*$. An example is shown in \fref{AB_pot}.
If such a potential \Eref{generalised_odd_pot} sustains a finite current, it must revert if the coupling changes sign,
$\nu\mapsto-\nu$ because by \Eref{generalised_odd_pot} this is equivalent to mirroring space $x^*+x \mapsto x^*-x$. We conclude that the expansion of the steady-state current \Eref{expansion_current} of RnT particles \emph{in such a potential} \Eref{generalised_odd_pot} must be odd in $\nu$, in other words, $\current^{(2n)}=0$ for $n\in\Nset$.

It is far from obvious why this should be the case for a general potential.
In \cite{Reimann2001May} an argument is sketched that time reversal $z(t)=x(T-t)$ of trajectories $x(t)$ obeying, say,
\begin{equation}\elabel{orig_ReimannLangevin}
    \dot{x} = -\nu \extPot'(x) + \xi_x(t)
\end{equation}
with white noise $\xi_x$
are generally solutions of the Langevin equation
\begin{equation}\elabel{reverse_ReimannLangevin}
    \dot{z} = \nu \extPot'(z) + \xi_z(t)
\end{equation}
with white noise $\xi_z$ distributed identically to $\xi_x$. The time reversal from \Eref{orig_ReimannLangevin} to \eref{reverse_ReimannLangevin} apparently amounts to a change of sign of the coupling $\nu$ to $-\nu$ and
clearly inverts the total displacement $z(T)-z(0)=-(x(T)-x(0))$. However, in order to conclude that time reversal reverts the \emph{steady-state} current, one has to show that the probabilistic weight of trajectories is unchanged under time reversal. To put it more succinctly, one has to show that typical steady-state trajectories $x(t)$ of \Eref{orig_ReimannLangevin} are typical steady-state trajectories $z(t)$ of \Eref{reverse_ReimannLangevin}. That this is generally not the case can be seen in the counter example of a harmonic $\extPot(x)=x^2/2$, where only one of the two \Erefs{orig_ReimannLangevin} and \eref{reverse_ReimannLangevin} has a stationary state at all. We conclude that there is no such simple argument as to why the steady-state current is generally odd in $\nu$.

\begin{figure}[t]
\resizebox{.9\linewidth}{!}{
\subfloat[\flabel{examplePotForInversionPlus}
$\nu=1$, $\current=-0.001873\ldots$]
{
\input{density_U}
}
\subfloat[\flabel{examplePotForInversionMinus}
$\nu=-1$, $\current=0.001873\ldots$]
{
\input{density_minus_U}
}
}
\caption{Particle density $\density(x)$ for an RnT particle in the periodic ``hurdle" potential  $\extPot(x)$ \Eref{hurdle_potential} shown in \fref{hurdle_potential}, based on $100$ modes calculated
perturbatively to $75$th order in $\nu$. Even when the density profiles cannot be related by a simple geometrical transformation, the steady-state currents are simple the negative of each other.
Parameters: $D=1$, $w=1$, $L=1$ and $\gamma=1$, so that $\peclet=\qeclet=\pqRatio=1$.
\flabel{examplePotForInversion}
}
\end{figure}
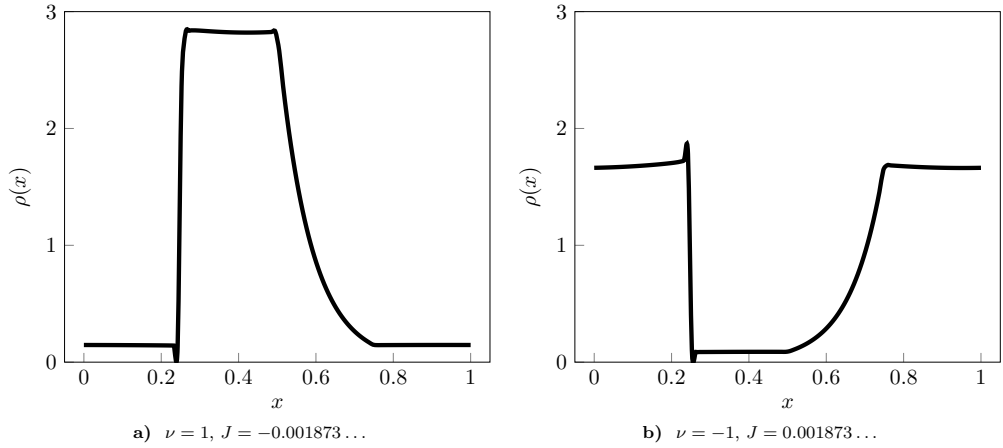

\begin{figure}
\input{Fig_reversalplot}
    \caption{The ``hurdle" potential \Eref{hurdle_potential}, whose resulting RnT particle density is shown in \fref{examplePotForInversionPlus}.
    }
    \flabel{hurdle_potential}
\end{figure}

To illustrate the subtleties further, we show in \fref{examplePotForInversionPlus} the particle density $\density(x)$ of RnT particles in the ``hurdle" potential $\extPot(x)$
\begin{equation}
\extPot(x)=\begin{cases}
0& x\in[0,\frac{L}{4})\\
3D & x \in[\frac{L}{4} ,\frac{L}{2})\\
(-12\frac{x}{L}+9)D &x  \in [\frac{L}{2},\frac{3L}{4})\\
0 &x \in [\frac{3L}{4},L)
\end{cases}
\elabel{hurdle_potential}
\end{equation}
shown in \fref{hurdle_potential} for $\nu=1$. This is contrasted by \fref{examplePotForInversionMinus}, which shows the particle density for the inverted potential, \ie for the potential \fref{hurdle_potential} but with coupling $\nu=-1$.
The two density profiles are clearly and obviously very distinct, even when they both seem to suffer slightly from the Gibbs phenomenon.
While they share some features,
there is also no simple geometrical transformation that maps one density profile to the other, and surely not one that is its own inverse as is $\nu\mapsto-\nu$.
For example,
both densities being convex from below for the sloped region $x\in[L/2,3L/4)$ makes it impossible to, say, mirror and overlay the two.
The steady-state trajectories in the potential $\extPot(x)$ of \Eref{hurdle_potential} produce therefore demonstrably a very different density $\density(x)$ compared to $-\extPot(x)$. And yet, we find that the steady-state currents in the two setups are indeed simply the negative of each other. The following section derives this property perturbatively.

\subsection{The steady-state current 
\texorpdfstring{$\current$}{J} is odd in 
\texorpdfstring{$\nu$}{n}}
\label{curr_odd_in_nu}
In the present section we show that changing the sign of $\nu$ of the potential simply reverses the steady-state current of RnT particles, despite the density $\density(x)$ showing none such simple relationship as outlined above. We will do so by showing $\current^{(n)}=(-1)^n \current^{(n)}$, \ie $\current^{(n)}=0$ for all even $n$.

The starting point for the derivation is \Eref{J_iterated_relabelled}, where we will rearrange the products of $\matrixM{b}$ to arrive at the desired expression. The key observation is that the complex matrix $\matrixM{b}$, \Eref{def_matrixM}, is symmetric (rather than Hermitian), with purely real entries along the diagonal and purely imaginary off-diagonal elements. Firstly, the symmetry implies
\begin{equation}\elabel{first_feature}
\dvec^\transpose \matrixM{c_1} \matrixM{c_2} \cdots \matrixM{c_{n-1}} \evec
=
\evec^\transpose \matrixM{c_{n-1}} \matrixM{c_{n-2}} \cdots \matrixM{c_1} \dvec \ ,
\end{equation}
for arbitrary vectors $\evec$ and $\dvec$,
so that \Eref{J_iterated_relabelled}, that has
$\evec=\dvec=(1,0)^\transpose$,
is invariant under the inversion of
the sequence of Fourier-indices in the product of $\matrixM{b}$.

Secondly, the make-up of the elements of $\matrixM{b}$, namely purely real diagonal elements and purely imaginary off-diagonal elements, means that any product of $\matrixM{b}$ has the same property, so that
\begin{equation}
    \begin{pmatrix}
    1\\
    0
    \end{pmatrix}^\transpose
    \matrixM{c_1} \matrixM{c_2}\cdots \matrixM{c_{n-1}}
    \begin{pmatrix}
    1\\
    0
    \end{pmatrix}
    \in \Rset \ .
\end{equation}
Taking the complex conjugate of this expression thus leaves it unchanged. By the definition \Eref{def_matrixM}, the complex conjugate can be written as $(\matrixM{a})^*=-\matrixM{-a}$, so that
\begin{equation}\elabel{second_feature}
    \begin{pmatrix}
    1\\
    0
    \end{pmatrix}^\transpose
    \matrixM{c_1} \matrixM{c_2}\cdots \matrixM{c_{n-1}}
    \begin{pmatrix}
    1\\
    0
    \end{pmatrix}
    =
    (-1)^{n-1}
    \begin{pmatrix}
    1\\
    0
    \end{pmatrix}^\transpose
    \matrixM{-c_1} \matrixM{-c_2}\cdots \matrixM{-c_{n-1}}
    \begin{pmatrix}
    1\\
    0
    \end{pmatrix} \ .
\end{equation}

Applying then \Eref{first_feature} followed by \Eref{second_feature}
to \Eref{J_iterated_relabelled} gives
\begin{multline}\elabel{J_iterated_relabelled2}
    \current^{(n)}(x)=
    - (-1)^{n-1} \frac{\imag}{L}
    \sum_{b_1,b_2,\ldots,b_{n}=-\infty}
    \delta_{b_1+b_2+\ldots+b_n,0}
    W_{b_1} W_{b_2}\cdots  W_{b_{n-1}} W_{b_n}\\
    \times \begin{pmatrix}
    1\\
    0
    \end{pmatrix}^\transpose
    \matrixM{b_1+b_2+\ldots+b_{n-1}}
    \matrixM{b_1+b_2+\ldots+b_{n-2}}
\cdots
\matrixM{b_1+b_2}
    \matrixM{b_1}
    \begin{pmatrix}
    1\\
    0
    \end{pmatrix} \ .
\end{multline}
Because $\sum_{i=1}^nb_i=0$, each of the matrices on the right hand side of \Eref{J_iterated_relabelled2} may be
indexed alternatively by
$b_1+b_2+\ldots+b_{n-1}=-b_n$,
$b_1+b_2+\ldots+b_{n-2}=-(b_n+b_{n-1})$,
\ldots,
$b_1+b_2=-(b_n+b_{n-1}+\ldots+b_3)$
and
$b_1=-(b_n+b_{n-1}+\ldots+b_2)$, resulting in
\begin{multline}\elabel{J_iterated_relabelled3}
    \current^{(n)}(x)=
    - (-1)^{n-1} \frac{\imag}{L}
    \sum_{b_1,b_2,\ldots,b_{n}=-\infty}
    \delta_{b_1+b_2+\ldots+b_n,0}
    W_{b_1} W_{b_2}\cdots  W_{b_{n-1}} W_{b_n}\\
    \times \begin{pmatrix}
    1\\
    0
    \end{pmatrix}^\transpose
    \matrixM{-b_n}
    \matrixM{-(b_n+b_{n-1})}
\cdots
\matrixM{-(b_n+b_{n-1}+\ldots+b_3)}
    \matrixM{-(b_n+b_{n-1}+\ldots+b_2)}
    \begin{pmatrix}
    1\\
    0
    \end{pmatrix} \ ,
\end{multline}
before ``mirroring" the indices' indices,
$c_i=b_{n+1-i}$, so that
\begin{multline}\elabel{J_iterated_relabelled4}
    \current^{(n)}(x)=
    - (-1)^{n-1} \frac{\imag}{L}
    \sum_{c_1,c_2,\ldots,c_{n}=-\infty}
    \delta_{c_1+c_2+\ldots+c_n,0}
    W_{c_n} W_{c_{n-1}}\cdots  W_{c_2} W_{c_1}\\
    \times \begin{pmatrix}
    1\\
    0
    \end{pmatrix}^\transpose
    \matrixM{-c_1}
    \matrixM{-(c_1+c_2)}
\cdots
\matrixM{-(c_1+b_2+\ldots+c_{n-2})}
    \matrixM{-(c_1+b_2+\ldots+c_{n-1})}
    \begin{pmatrix}
    1\\
    0
    \end{pmatrix} \ .
\end{multline}
As the factors $W_{c_i}$ are scalars, \Eref{J_iterated_relabelled4} is identical to \Eref{J_iterated_relabelled} up to a factor of $(-1)^{n-1}$, which produces the desired identity
\begin{equation}
    \current^{(n)}(x) = (-1)^{n-1} \current^{(n)}(x) \ ,
\end{equation}
which implies that $\current^{(n)}(x)$ vanishes for even $n$. This is the main result of the present section. It follows immediately, that $\current$, \Eref{J_from_sum}, is odd in $\nu$ and thus changes sign when $\nu$ does.

\subsection{The steady-state current vanishes in potentials with only odd Fourier coefficients}
\label{superSym_pot}
We first show that
potentials whose even Fourier coefficients all vanish obey
\begin{equation}
    \extPot(x)=-\extPot(x+L/2)
\end{equation}
and vice versa and
are thus identical to those that have been dubbed ``supersymmetric" \cite{Reimann2001May}.

Firstly, if $\extPot_a=0$ for all even $a$, then \Eref{extPot_from_modes} gives
\begin{equation}
    \extPot(x+L/2)
    = \frac{1}{L} \sum_{a=-\infty}^\infty \exp{\imag k_a (x+L/2)} \extPot_a
    = - \extPot(x)
\end{equation}
using that $\exp{\imag k_a (x+L/2)}=(-1)^a\exp{\imag k_a x}$ and only odd $a$ enters.

Secondly, if $U(x)=U(x+L/2)$, then \Eref{extPot_from_modes} gives
\begin{equation}
    \extPot_a
    =- \int_0^L \dint{x} \exp{-\imag k_a x} \extPot(x+L/2)
    =- (-1)^a \int_{L/2}^{3L/2} \dint{x} \exp{-\imag k_a x} \extPot(x)
    = - (-1)^{a} \extPot_a \ ,
\end{equation}
which means that $\extPot_a$ must vanish for even $a$.

In the following we demonstrate that the steady-state current vanishes if $\extPot_a=0$ for all even $a$.

The potential enters into the perturbative expansion \Eref{J_iterated_relabelled} of the steady-state current only via $W_a$ \Eref{def_W}. It features in \Eref{J_iterated_relabelled} as a product involving all $n$ indices $b_1$, $b_2$, \ldots, $b_n$. If $W_a$ vanishes for all even $a$, then all indices $b_1$, $b_2$, \ldots, $b_n$ need to be odd in order for the product $W_{b_1}W_{b_1}\ldots W_{b_n}$ to be non-zero. However, if all indices $b_i$ are odd, then the sum
of an odd number of them
cannot possible vanish, $\sum_{i=1}^n b_i\ne0$, so that $\delta_{b_1+b_2+\ldots+b_n}=0$ for odd $n$. It follows that $\current^{(n)}$ vanishes for all odd $n$ if $\extPot_a$ vanishes for all even $a$.

In \SMref{curr_odd_in_nu} we have demonstrated that $\current^{(n)}$ vanishes generally for even $n$. If $\current^{(n)}$ also vanishes for all odd $n$, then it must vanish overall. In summary, if $\extPot_a$ vanishes for all even $a$, which is equivalent to saying that the potential is supersymmetric, then $\current$ vanishes.

\subsection{The steady-state current vanishes in potentials even about some 
\texorpdfstring{$x^*$}{x}}
\label{even_pot}
The steady-state current of RnT particles in
a potential even about $x^*$,
\begin{equation}\elabel{even_pot}
    \extPot(x^*+x)=\extPot(x^*-x)
\end{equation}
vanishes trivially because there is no preferred direction.
In the following, we rederive this property
perturbatively from the expression for the current \Eref{J_iterated_relabelled}.

From \Eref{even_pot}, we derive for the
Fourier coefficients using \Eref{extPot_from_modes},
\begin{multline}\elabel{even_pot_a}
    \exp{\imag k_a x^*} \extPot_a
    =\exp{\imag k_a x^*} \int_0^L \dint{x} \exp{-\imag k_a (x+x^*)} \extPot(x^*+x)
    =\int_0^L \dint{x} \exp{-\imag k_a x} \extPot(x^*-x)\\
    =\exp{-\imag k_a x^*} \int_0^L \dint{x} \exp{\imag k_a (x^*-x)} \extPot(x^*-x)
    = \exp{-\imag k_a x^*}  \extPot_{-a} \ .
\end{multline}

Considering now the product of $W_b$, \Eref{def_W}, in \Eref{J_iterated_relabelled}, it generally has the property
\begin{multline}
    \big(W_{b_1}W_{b_2}\cdots W_{b_n}\big) +
    \big(W_{-b_1}W_{-b_2}\cdots W_{-b_n}\big) \\
    =
    L^{-n} k_{b_1} k_{b_2}\cdots  k_{b_n}
    \Big(
    \extPot_{b_1}\extPot_{b_2}\cdots\extPot_{b_n}
    + (-1)^n
    \extPot_{-b_1}\extPot_{-b_2}\cdots\extPot_{-b_n}
    \Big)    \ ,
\end{multline}
With \Eref{even_pot_a}, the product of modes of the potential obeys
\begin{equation}
    \extPot_{b_1}\extPot_{b_2}\cdots\extPot_{b_n}
    = \exp{-2\imag \sum_{a=1}^n k_a x^*}
    \extPot_{-b_1}\extPot_{-b_2}\cdots\extPot_{-b_n}
    =
    \extPot_{-b_1}\extPot_{-b_2}\cdots\extPot_{-b_n}
    \ ,
\end{equation}
provided $\sum_{i=1}^n k_{b_i}=(2\pi/L) \sum_{i=1}^n b_i = 0$, as is enforced by the Kronecker $\delta$-function in \Eref{J_iterated_relabelled}. For a potential even about some $x^*$, we thus have
\begin{multline}
\delta_{b_1+b_2+\ldots+b_n,0}
\big(
    W_{b_1}W_{b_2}\cdots W_{b_n} +
    W_{-b_1}W_{-b_2}\cdots W_{-b_n}
\big) \\
=
\delta_{b_1+b_2+\ldots+b_n,0}
    L^{-n} k_{b_1} k_{b_2}\ldots k_{b_n}
\times
\begin{cases}
2 \extPot_{b_1}\extPot_{b_2}\ldots\extPot_{b_n} &
  \text{for $n$ even}\\
0 &
  \text{for $n$ odd} \ .
\end{cases}
\elabel{E19}
\end{multline}
This identity can be used in the expression for the steady-state current \Eref{J_iterated_relabelled} once it has been suitably rewritten. To this end, we double up all terms in \Eref{J_iterated_relabelled},
\begin{align}\elabel{J_iterated_relabelled_doubled}
    2\current^{(n)}(x)&=
    - \frac{\imag}{L}
    \sum_{b_1,b_2,\ldots,b_{n}=-\infty}^\infty
    \delta_{b_1+b_2+\ldots+b_n,0}
    \\
    \nonumber&
    \quad\times\Bigg\{
    W_{b_1} W_{b_2}\cdots  W_{b_{n-1}} W_{b_n}
    \begin{pmatrix}
    1\\
    0
    \end{pmatrix}^\transpose
    \matrixM{-b_1} \matrixM{-(b_1+b_2)}
    \cdots \matrixM{-(b_1+b_2+\ldots+b_{n-1})}
    \begin{pmatrix}
    1\\
    0
    \end{pmatrix} \\
    \nonumber&
    \quad+
    W_{-b_1} W_{-b_2}\cdots  W_{-b_{n-1}} W_{-b_n}
    (-1)^{n-1}
    \begin{pmatrix}
    1\\
    0
    \end{pmatrix}^\transpose
    \matrixM{b_1} \matrixM{b_1+b_2}
    \cdots \matrixM{b_1+b_2+\ldots+b_{n-1}}
    \begin{pmatrix}
    1\\
    0
    \end{pmatrix} \Bigg\} \ ,
\end{align}
which by means of \Eref{second_feature} simplifies to
\begin{align}
    \elabel{J_iterated_relabelled_doubled_simplified}
    2\current^{(n)}(x)&=
    - \frac{\imag}{L}
    \sum_{b_1,b_2,\ldots,b_{n}=-\infty}^\infty
    \delta_{b_1+b_2+\ldots+b_n,0}
    \begin{pmatrix}
    1\\
    0
    \end{pmatrix}^\transpose
    \matrixM{-b_1} \matrixM{-(b_1+b_2)}
\cdots \matrixM{-(b_1+b_2+\ldots+b_{n-1})}
    \begin{pmatrix}
    1\\
    0
    \end{pmatrix} \\
    \nonumber&\quad\times
    \big\{
    W_{b_1} W_{b_2}\cdots  W_{b_{n-1}} W_{b_n}
+ (-1)^{n-1}
    W_{-b_1} W_{-b_2}\cdots  W_{-b_{n-1}} W_{-b_n}
    \big\} \ .
\end{align}
If the potential is even, then \Eref{E19} can be used on the right hand side for odd $n$, where $(-1)^{n-1}=1$, so that
\begin{equation}
    2\current^{(n)}(x) = 0 \quad\text{ for odd $n$ if the potential is even, \Eref{even_pot} } \ .
\end{equation}
As $\current^{(n)}(x)=0$ for all even $n$, it implies that the steady-state current vanishes altogether. In other words, even potentials, \Eref{even_pot}, have no steady-state current, $\current(x)=0$.
\end{document}

%% file: preamble.tex
\usepackage{physics}

%% file: gp_macros.tex
\usepackage{amsmath}
\usepackage{amsfonts}
\usepackage{cases}
\usepackage{xspace}
\usepackage{color}
\usepackage{xcolor}

\newcommand{\ave}[1]{\mathchoice{\left\langle #1 \right\rangle}{\langle #1 \rangle}{\langle #1 \rangle}{\langle #1 \rangle}}

\newcommand{\imag}{\mathring{\imath}}

\newcommand{\plaind}{\mathrm{d}}
\newcommand{\dint}[1]{\mathchoice{\!\plaind#1\,}{\!\plaind#1\,}{\!\plaind#1\,}{\!\plaind#1\,}}
\newcommand{\dbar}{\plaind\mkern-7mu\mathchar'26}
\newcommand{\deltabar}{\delta\mkern-8mu\mathchar'26}
\newcommand{\dintbar}[1]{\mathchoice{\!\dbar#1\,}{\!\dbar#1\,}{\!\dbar#1\,}{\!\dbar#1\,}}

\usepackage{dsfont}

\newcommand{\gpset}[1]{\mathds{#1}}

\newcommand{\canetset}[1]{{\mathchoice {\hbox{$\sf\textstyle #1\kern-0.4em #1$}}
{\hbox{$\sf\textstyle #1\kern-0.4em #1$}}
{\hbox{$\sf\scriptstyle #1\kern-0.3em #1$}}
{\hbox{$\sf\scriptscriptstyle #1\kern-0.2em #1$}}}}

\newcommand{\Nset}{\gpset{N}}
\newcommand{\Zset}{\gpset{Z}}

\newcommand{\Rset}{\gpset{R}}

\def\nbZ{{\mathchoice {\hbox{$\sf\textstyle Z\kern-0.4em Z$}}
{\hbox{$\sf\textstyle Z\kern-0.4em Z$}}
{\hbox{$\sf\scriptstyle Z\kern-0.3em Z$}}
{\hbox{$\sf\scriptscriptstyle Z\kern-0.2em Z$}}}}

\newcommand{\gpvec}[1]{\mathbf{#1}}

\newcommand{\dvec}{\gpvec{d}}
\newcommand{\evec}{\gpvec{e}}

\newcommand{\transpose}{\mathsf{T}}

\newcommand{\FC}{\mathcal{F}}

\newcommand{\NC}{\mathcal{N}}

\newcommand{\phitilde}{\tilde{\phi}}
\newcommand{\psitilde}{\tilde{\psi}}

\newcommand{\half}{\mathchoice{\frac{1}{2}}{(1/2)}{\frac{1}{2}}{(1/2)}}

\renewcommand{\exp}[1]{\mathchoice{\mathrm{e}^{#1}}{\operatorname{exp}\left(#1\right)}{\operatorname{exp}\left(#1\right)}{\operatorname{exp}\left(#1\right)}}

\newcommand{\elabel}[1]{\label{eq:#1}}
\newcommand{\eref}[1]{(\ref{eq:#1})}
\newcommand{\Eref}[1]{Eq.~(\ref{eq:#1})}
\newcommand{\Erefs}[1]{Eqs.~(\ref{eq:#1})}

\newcommand{\flabel}[1]{\label{fig:#1}}
\newcommand{\fref}[1]{Fig.~\ref{fig:#1}}
\newcommand{\frefs}[1]{Figs.~\ref{fig:#1}}

\newcommand{\latin}[1]{{\it #1}}
\newcommand{\ie}{\latin{i.e.}\@\xspace}
\newcommand{\eg}{\latin{e.g.}\@\xspace}

\newlength \standardfigwidth
\DeclareMathAlphabet{\matheub}{U}{eur}{m}{n}

\newcounter{exercise}
{\addtocounter{exercise}{1}\begin{center}\begin{minipage}{0.8\linewidth}\textbf{Exercise
\arabic{exercise}:}\begin{itshape}}
{\end{itshape}\end{minipage}\end{center}}

\makeatletter
\newcommand{\creat}[3][]{\@ifempty{#1}{#2^{\dagger}}{\left(#2^{\dagger}\right)^{#1}}\@ifempty{#3}{}{\!(#3)}}

\newcommand{\creatDoi}[3][]{\@ifempty{#1}{\tilde{#2}}{\left(\tilde{#2}\right)^{#1}}\@ifempty{#3}{}{(#3)}}

\newcommand{\annih}[3][]{#2\@ifempty{#1}{}{^{#1}}\@ifempty{#3}{}{(#3)}}

\makeatother

% ...

\newlength{\bibmarkkeyAleft}

\newlength{\bibmarkkeyBleft}

\newlength{\bibmarkkeyCleft}

\newlength{\bibmarkkeyDleft}

\usepackage{wasysym}

%% file: tikz_macros.tex
\usepackage[colorlinks=true,linkcolor=black, urlcolor = black, citecolor = blue, anchorcolor = blue]{hyperref}

\usepackage{graphicx}
\usepackage{multirow}
\usepackage{subeqnarray}
\usepackage{wasysym}
\usepackage{eufrak}
\usepackage{amsopn}

\usepackage{amssymb}

\usepackage{framed}

\usepackage{vmargin}
\usepackage{epsfig}
\usepackage{color}
\usepackage{alltt}
\usepackage{pstricks,pst-node,pst-tree,pst-grad,pst-text,graphics}
\usepackage{dcolumn}
\usepackage{url}

\usepackage{cases}

\usepackage{natbib}

\usepackage{ifthen}

\usepackage{afterpage}

\usepackage{comment}

\usepackage{supertabular}

\usepackage{xspace}

\usepackage{setspace}

\usepackage[bf,footnotesize]{caption}

\usepackage{textcomp}

\usepackage{tikz}
\usetikzlibrary{decorations.pathmorphing}
\usetikzlibrary{decorations.markings}
\usetikzlibrary{arrows,shapes,snakes,automata,backgrounds,petri}
\usetikzlibrary{calc}
\usetikzlibrary{patterns}
\usetikzlibrary{decorations.text}
\tikzset{
xxtsubstrate/.style={decorate, 
line width=1pt,
draw=olive, 
decoration=snake, 
segment amplitude=0.75mm, 
line after snake=0.25mm,
line before snake=0.25mm
},
tsubstrate/.style={decorate, 
line width=1pt,
draw=olive, 
decoration=snake, 
segment amplitude=0.5mm, 
segment length=5pt,
segment amplitude=0.2mm, 
line after snake=1mm,
line before snake=1mm
},
Bsubstrate/.style={decorate, 
line width=1pt,
draw=orange, 
decoration=snake,
segment length=5pt,
segment aspect=0,
segment amplitude=0.5mm, 
line after snake=0mm,
line before snake=0mm
},
substrate/.style={decorate, 
line width=1pt,
draw=orange,
decoration=snake, 
segment length=5pt,
segment amplitude=0.5mm, 
line after snake=0.5mm,
line before snake=0.5mm
},
activity/.style={very thick,draw=red,postaction={decorate},
decoration={markings,mark=at position .5 with
{\arrow[draw=red]{>}}}},
tactivity/.style={thick,draw=red,postaction={decorate},
decoration={markings,mark=at position .5 with
{\arrow[draw=red]{>}}}},
tEPSactivity/.style={thick,draw=red,postaction={decorate},
decoration={markings,mark=at position .55 with
{\arrow[draw=red]{>}}}},
tAactivity/.style={thick,draw=red},
Aactivity/.style={thick,draw=black},
tSactivity/.style={thick,draw=red,postaction={decorate},
decoration={markings,mark=at position .7 with
{\arrow[draw=red]{>}}}},
Sactivity/.style={very thick,draw=red,postaction={decorate},
decoration={markings,mark=at position .7 with
{\arrow[draw=red]{>}}}},
polarity/.style={decorate, 
line width=1pt,
draw=black,
decoration={markings,mark=between positions 0 and 1 step 1.6mm with {\draw[black,thick]  (0,0) circle (0.02)} },
segment length=5pt,
segment amplitude=0.5mm, 
},
Bpolarity/.style={decorate, 
line width=1.5pt,
draw=red,
segment length=5pt,
segment amplitude=0.5mm, 
},
density/.style={ 
line width=1.5pt,
draw=magenta,
densely dashed,
segment length=5pt,
segment amplitude=0.5mm, 
},
}

%% file: Fig_p=1_q=1.tex
\begin{tikzpicture}[scale=0.9]
\begin{axis}[ymin=-2.5, ymax=2.5, xmin=-0.15, xmax=1.15,
  ytick={-2,-1,...,2}, ytick pos=left,
  xtick={0,0.2,...,1}, xtick pos=left,
  xlabel={$x/L$},
  ylabel={$\extPotOptim(x)/D$},
  legend pos=north west,
  legend style={draw=none}]
\addplot+[
  black, mark options={black, scale=0.75},
  only marks, 
  error bars/.cd, 
    y fixed,
    y dir=both, 
    y explicit
] table [x=x, y=y,y error=error, col sep=comma] {
    x,  y,       error
    0.000000,-0.060946,0.033595
0.010309,-1.748414,0.474121
0.020619,-1.362555,0.041874
0.030928,-1.315469,0.106960
0.041237,-1.175563,0.028847
0.051546,-1.117221,0.031694
0.061856,-1.046794,0.019382
0.072165,-0.987162,0.014755
0.082474,-0.946271,0.008715
0.092784,-0.886876,0.016980
0.103093,-0.858714,0.005218
0.113402,-0.804486,0.019638
0.123711,-0.779557,0.012164
0.134021,-0.735001,0.017985
0.144330,-0.708502,0.016899
0.154639,-0.675257,0.016911
0.164948,-0.646122,0.015593
0.175258,-0.621993,0.013043
0.185567,-0.591540,0.011934
0.195876,-0.572403,0.006655
0.206186,-0.542131,0.008930
0.216495,-0.524328,0.002754
0.226804,-0.496314,0.006413
0.237113,-0.476885,0.003506
0.247423,-0.452109,0.004548
0.257732,-0.430281,0.004596
0.268041,-0.409438,0.005386
0.278351,-0.385791,0.003665
0.288660,-0.368285,0.007326
0.298969,-0.344623,0.002302
0.309278,-0.329312,0.007106
0.319588,-0.307020,0.001163
0.329897,-0.292124,0.004604
0.340206,-0.272594,0.000815
0.350515,-0.256285,0.000877
0.360825,-0.239607,0.000568
0.371134,-0.221158,0.003275
0.381443,-0.206393,0.000567
0.391753,-0.185885,0.005318
0.402062,-0.171575,0.002045
0.412371,-0.150303,0.006159
0.422680,-0.135184,0.004149
0.432990,-0.114751,0.005280
0.443299,-0.097964,0.005400
0.453608,-0.079623,0.005431
0.463918,-0.060950,0.004809
0.474227,-0.044972,0.003921
0.484536,-0.024718,0.003597
0.494845,-0.010439,0.003103
0.505155,0.010421,0.003063
0.515464,0.024842,0.003493
0.525773,0.044938,0.004117
0.536082,0.060922,0.004999
0.546392,0.079549,0.005512
0.556701,0.097996,0.005328
0.567010,0.114689,0.005373
0.577320,0.135152,0.004155
0.587629,0.150255,0.006207
0.597938,0.171486,0.001938
0.608247,0.185785,0.005247
0.618557,0.206108,0.000533
0.628866,0.221039,0.003096
0.639175,0.239414,0.000670
0.649485,0.256224,0.001137
0.659794,0.272406,0.000848
0.670103,0.292100,0.004780
0.680412,0.307089,0.001437
0.690722,0.329355,0.007202
0.701031,0.344459,0.002139
0.711340,0.368407,0.007383
0.721649,0.386030,0.003922
0.731959,0.409558,0.005162
0.742268,0.430666,0.004711
0.752577,0.452330,0.004584
0.762887,0.477140,0.003336
0.773196,0.496506,0.006486
0.783505,0.524545,0.002900
0.793814,0.542373,0.009084
0.804124,0.572518,0.007026
0.814433,0.591724,0.012014
0.824742,0.621941,0.013315
0.835052,0.646380,0.015628
0.845361,0.675267,0.016856
0.855670,0.708847,0.016648
0.865979,0.735308,0.018041
0.876289,0.779908,0.011515
0.886598,0.804749,0.019434
0.896907,0.858906,0.004771
0.907216,0.887077,0.017031
0.917526,0.945789,0.009192
0.927835,0.986898,0.014393
0.938144,1.045798,0.019279
0.948454,1.116780,0.033022
0.958763,1.174270,0.028199
0.969072,1.315116,0.108008
0.979381,1.362170,0.040352
0.989691,1.751750,0.471374
1.000000,-0.060946,0.033595
};
\end{axis}
\node (A) at (0.3, 0.65){};
\node (B) at (0.3, 2.2) {};
\node [rotate=90](C) at (0.3, 2.8) {\figOlab{barrier}};
\node (D) at (0.3, 3.5){} ;
\node (E) at (0.3, 5.0) {};
\node (F) at (2.2, 1.919) {} ;
\node (G) at (3.2, 2.3) {} ;
\node[rotate=20] (H) at (3.6, 2.45) {\figOlab{plane}};
\node (H) at (4, 2.61) {} ;
\node (I) at (5, 2.985) {} ;
\draw [->,gray](B) -- (A);
\draw [->,gray](D) -- (E);
\draw [->,gray](G) -- (F);
\draw [->,gray](H) -- (I);
\node (1) at (0.9, 1.7) {} ;
\node (2) at (2, 3.2) {} ;
\node (3) at (2, 3.3) {\figOlab{crevasse}} ;
\node (4) at (4.7, 3.8) {} ;
\node (5) at (5.9, 4.8) {} ;
\node (6) at (4.3, 3.8) {\figOlab{peak}} ;
\draw[->,gray] (2) to[bend right] (1);
\draw[->,gray] (4) to[bend right] (5);
\node (7) at (0.1,4.85) {};
\node (8) at (6, 4.85){};
\node (9) at (2.2, 1) {};
\node (10) at (5, 1) {};
\draw[->,gray] (2.2, 1) -- (5, 1) node[midway,above] {\figOlab{current}};
 \draw [dotted,gray] (7) to (8);
\end{tikzpicture}

%% file: p=2_q=100_with_error_bar.tex
\begin{tikzpicture}[scale=0.9]
\begin{axis}[ymin=-2.5, ymax=2.5,
  ytick={-2,-1,...,2}, ytick pos=left,
  xtick={0,0.2,...,1}, xtick pos=left,
  xlabel={$x/L$},
  ylabel={$\extPotOptim(x)/D$},
  legend pos=north west,
  legend style={draw=none}]
\addplot+[
  black, mark options={black, scale=0.75},
  only marks, 
  error bars/.cd, 
    y fixed,
    y dir=both, 
    y explicit
] table [x=x, y=y,y error=error, col sep=comma] {
    x,  y,       error
0.000000,-0.012008,0.011398
0.010309,-1.816719,0.483611
0.020619,-1.412844,0.044798
0.030928,-1.353842,0.105431
0.041237,-1.209878,0.028056
0.051546,-1.146263,0.031032
0.061856,-1.069367,0.020238
0.072165,-1.005398,0.014207
0.082474,-0.959673,0.008635
0.092784,-0.895589,0.017449
0.103093,-0.864327,0.005261
0.113402,-0.806505,0.020091
0.123711,-0.779110,0.012841
0.134021,-0.731996,0.018063
0.144330,-0.703312,0.017539
0.154639,-0.667934,0.017601
0.164948,-0.637127,0.016134
0.175258,-0.611239,0.013578
0.185567,-0.579216,0.012177
0.195876,-0.559067,0.006879
0.206186,-0.527539,0.009070
0.216495,-0.509301,0.002573
0.226804,-0.480083,0.006337
0.237113,-0.460664,0.003704
0.247423,-0.435171,0.004874
0.257732,-0.413327,0.004574
0.268041,-0.392282,0.006106
0.278351,-0.368526,0.003380
0.288660,-0.351510,0.008313
0.298969,-0.327537,0.001716
0.309278,-0.313031,0.007876
0.319588,-0.290738,0.000659
0.329897,-0.276678,0.005159
0.340206,-0.257518,0.000549
0.350515,-0.241957,0.001133
0.360825,-0.226077,0.000601
0.371134,-0.208077,0.003099
0.381443,-0.194604,0.000135
0.391753,-0.174452,0.005765
0.402062,-0.161668,0.001840
0.412371,-0.140871,0.006778
0.422680,-0.127184,0.004055
0.432990,-0.107443,0.005857
0.443299,-0.091847,0.005527
0.453608,-0.074649,0.005671
0.463918,-0.056820,0.005100
0.474227,-0.042359,0.004030
0.484536,-0.022748,0.003721
0.494845,-0.010111,0.002872
0.505155,0.010098,0.002947
0.515464,0.022760,0.003735
0.525773,0.042295,0.004081
0.536082,0.056777,0.005158
0.546392,0.074627,0.005685
0.556701,0.091856,0.005487
0.567010,0.107483,0.005902
0.577320,0.127224,0.004011
0.587629,0.140810,0.006818
0.597938,0.161627,0.001843
0.608247,0.174465,0.005719
0.618557,0.194574,0.000127
0.628866,0.208137,0.003096
0.639175,0.226065,0.000621
0.649485,0.242025,0.001171
0.659794,0.257497,0.000562
0.670103,0.276740,0.005148
0.680412,0.290719,0.000658
0.690722,0.313044,0.007909
0.701031,0.327555,0.001750
0.711340,0.351495,0.008291
0.721649,0.368587,0.003432
0.731959,0.392306,0.006096
0.742268,0.413420,0.004675
0.752577,0.435142,0.004819
0.762887,0.460696,0.003683
0.773196,0.480061,0.006356
0.783505,0.509235,0.002609
0.793814,0.527562,0.009130
0.804124,0.559115,0.006859
0.814433,0.579181,0.012142
0.824742,0.611254,0.013555
0.835052,0.637200,0.016110
0.845361,0.667969,0.017508
0.855670,0.703474,0.017415
0.865979,0.732076,0.018106
0.876289,0.779226,0.012681
0.886598,0.806597,0.020093
0.896907,0.864448,0.005090
0.907216,0.895718,0.017501
0.917526,0.959662,0.008730
0.927835,1.005439,0.014112
0.938144,1.069215,0.020108
0.948454,1.146200,0.031271
0.958763,1.209674,0.027852
0.969072,1.353862,0.105678
0.979381,1.412750,0.044650
0.989691,1.817112,0.483803
1.000000,-0.012008,0.011398
};
\end{axis}
\end{tikzpicture}

%% file: p=2_q=point_zero_1_with_error_bar.tex
\begin{tikzpicture}[scale=0.9]
\begin{axis}[ymin=-2.5, ymax=2.5,
  ytick={-2,-1,...,2}, ytick pos=left,
  xtick={0,0.2,...,1}, xtick pos=left,
  xlabel={$x/L$},
  ylabel={$\extPotOptim(x)/D$},
  legend pos=north west,
  legend style={draw=none}]
\addplot+[
  black, mark options={black, scale=0.75},
  only marks, 
  error bars/.cd, 
    y fixed,
    y dir=both, 
    y explicit
] table [x=x, y=y,y error=error, col sep=comma] {
    x,  y,       error
0.000000,-0.006111,0.005619
0.010309,-1.486671,0.425999
0.020619,-1.319487,0.076444
0.030928,-1.307279,0.093267
0.041237,-1.191682,0.033628
0.051546,-1.112160,0.023469
0.061856,-1.021518,0.015586
0.072165,-0.923022,0.016390
0.082474,-0.846828,0.003755
0.092784,-0.747413,0.021159
0.103093,-0.678926,0.012679
0.113402,-0.589924,0.022851
0.123711,-0.525724,0.020886
0.134021,-0.453533,0.022198
0.144330,-0.392903,0.021793
0.154639,-0.338859,0.019555
0.164948,-0.282920,0.016111
0.175258,-0.244525,0.011494
0.185567,-0.195517,0.012266
0.195876,-0.167947,0.005047
0.206186,-0.128373,0.008076
0.216495,-0.106416,0.005928
0.226804,-0.078238,0.008027
0.237113,-0.058368,0.009037
0.247423,-0.042194,0.009842
0.257732,-0.023278,0.008141
0.268041,-0.017835,0.014362
0.278351,-0.001136,0.005097
0.288660,-0.003321,0.014851
0.298969,0.008965,0.002204
0.309278,0.003293,0.011106
0.319588,0.009200,0.000941
0.329897,0.004439,0.004264
0.340206,0.003437,0.000842
0.350515,0.002674,0.004411
0.360825,-0.003828,0.000342
0.371134,0.000350,0.010891
0.381443,-0.008949,0.002194
0.391753,-0.001041,0.015181
0.402062,-0.010049,0.006344
0.412371,-0.001472,0.015678
0.422680,-0.007239,0.010847
0.432990,-0.001539,0.012514
0.443299,-0.002620,0.013068
0.453608,-0.002091,0.012767
0.463918,0.001706,0.011480
0.474227,-0.003264,0.008940
0.484536,0.004056,0.009007
0.494845,-0.004217,0.007494
0.505155,0.004280,0.007477
0.515464,-0.004074,0.009010
0.525773,0.003170,0.009108
0.536082,-0.001687,0.011524
0.546392,0.002043,0.012790
0.556701,0.002630,0.013104
0.567010,0.001522,0.012532
0.577320,0.007349,0.010700
0.587629,0.001438,0.015710
0.597938,0.009997,0.006411
0.608247,0.001008,0.015156
0.618557,0.008950,0.002163
0.628866,-0.000385,0.010863
0.639175,0.003795,0.000326
0.649485,-0.002719,0.004301
0.659794,-0.003511,0.000770
0.670103,-0.004415,0.004330
0.680412,-0.009203,0.000985
0.690722,-0.003287,0.011142
0.701031,-0.008903,0.002306
0.711340,0.003356,0.014873
0.721649,0.001167,0.005123
0.731959,0.017805,0.014335
0.742268,0.023370,0.008248
0.752577,0.042225,0.009790
0.762887,0.058429,0.009109
0.773196,0.078272,0.008053
0.783505,0.106493,0.005965
0.793814,0.128362,0.008048
0.804124,0.167957,0.005120
0.814433,0.195587,0.012176
0.824742,0.244482,0.011539
0.835052,0.282958,0.016118
0.845361,0.338842,0.019575
0.855670,0.392956,0.021771
0.865979,0.453556,0.022171
0.876289,0.525813,0.020793
0.886598,0.590010,0.022834
0.896907,0.678995,0.012584
0.907216,0.747477,0.021114
0.917526,0.846823,0.003747
0.927835,0.923071,0.016354
0.938144,1.021475,0.015610
0.948454,1.112168,0.023390
0.958763,1.191528,0.033487
0.969072,1.307292,0.093439
0.979381,1.319329,0.076294
0.989691,1.486742,0.426071
1.000000,-0.006111,0.005619
};
\end{axis}
\end{tikzpicture}

%% file: Fig_modes_convergence.tex
\centering
\begin{tikzpicture}[]
\begin{axis}[ymin=-0.02, ymax=0.1, xmin=0,xmax=220,
  ytick={-0.02,0,0.02,0.04,0.06,0.08,0.10,0.12}, ytick pos=left,
  xtick={0,20,40,...,200}, xtick pos=left,
  xlabel={$a$},
  yticklabels={$-0.02$,$0.00$,$0.02$,$0.04$,$0.06$,$0.08$,$0.10$,$0.12$},
  ylabel={$\operatorname{Im}\left(\extPotOptim_a\right)$},
  legend pos=north east,
  legend cell align={left},
  legend style={draw=none},
  scaled y ticks=false,
  yticklabel style={
            /pgf/number format/fixed,
            /pgf/number format/precision=3,
            /pgf/number format/fixed zerofill
        },]
\addplot+[
  black, mark options={black,mark= square*,scale=0.75},
  only marks, 
  error bars/.cd, 
    y fixed,
    y dir=both, 
    y explicit
] table [x=x, y=y, col sep=comma] {
    x,  y,      
7.000000,0.097258
8.000000,0.091344
9.000000,0.086785
10.000000,0.083179
11.000000,0.080259
12.000000,0.077843
13.000000,0.075809
14.000000,0.074069
15.000000,0.072556
16.000000,0.071225
17.000000,0.070036
18.000000,0.068963
19.000000,0.067975
20.000000,0.067056
21.000000,0.066185
22.000000,0.065347
23.000000,0.064530
24.000000,0.063721
25.000000,0.062910
26.000000,0.062087
27.000000,0.061245
28.000000,0.060376
29.000000,0.059475
30.000000,0.058534
31.000000,0.057548
32.000000,0.056512
33.000000,0.055422
34.000000,0.054272
35.000000,0.053058
36.000000,0.051776
37.000000,0.050420
38.000000,0.048985
39.000000,0.047464
40.000000,0.045852
41.000000,0.044141
42.000000,0.042320
43.000000,0.040377
44.000000,0.038294
45.000000,0.036040
46.000000,0.033573
47.000000,0.030820
48.000000,0.027658
49.000000,0.023807
50.000000,0.018423
};
\label{p1}
\addplot+[
  black, mark options={black,mark=diamond*,scale=1},
  only marks, 
  error bars/.cd, 
    y fixed,
    y dir=both, 
    y explicit
] table [x=x, y=y, col sep=comma] {
    x,  y,      
6.000000,0.093416
7.000000,0.084714
8.000000,0.077978
9.000000,0.072611
10.000000,0.068239
11.000000,0.064611
12.000000,0.061560
13.000000,0.058960
14.000000,0.056728
15.000000,0.054786
16.000000,0.053089
17.000000,0.051595
18.000000,0.050272
19.000000,0.049096
20.000000,0.048042
21.000000,0.047096
22.000000,0.046241
23.000000,0.045469
24.000000,0.044766
25.000000,0.044125
26.000000,0.043539
27.000000,0.043001
28.000000,0.042506
29.000000,0.042046
30.000000,0.041619
31.000000,0.041222
32.000000,0.040854
33.000000,0.040503
34.000000,0.040174
35.000000,0.039865
36.000000,0.039570
37.000000,0.039289
38.000000,0.039020
39.000000,0.038763
40.000000,0.038512
41.000000,0.038270
42.000000,0.038034
43.000000,0.037801
44.000000,0.037575
45.000000,0.037350
46.000000,0.037129
47.000000,0.036908
48.000000,0.036686
49.000000,0.036467
50.000000,0.036247
51.000000,0.036023
52.000000,0.035798
53.000000,0.035571
54.000000,0.035341
55.000000,0.035105
56.000000,0.034868
57.000000,0.034625
58.000000,0.034377
59.000000,0.034125
60.000000,0.033864
61.000000,0.033598
62.000000,0.033327
63.000000,0.033046
64.000000,0.032759
65.000000,0.032464
66.000000,0.032160
67.000000,0.031849
68.000000,0.031528
69.000000,0.031195
70.000000,0.030854
71.000000,0.030505
72.000000,0.030142
73.000000,0.029770
74.000000,0.029386
75.000000,0.028987
76.000000,0.028578
77.000000,0.028156
78.000000,0.027722
79.000000,0.027275
80.000000,0.026811
81.000000,0.026334
82.000000,0.025840
83.000000,0.025328
84.000000,0.024796
85.000000,0.024242
86.000000,0.023667
87.000000,0.023072
88.000000,0.022448
89.000000,0.021797
90.000000,0.021115
91.000000,0.020396
92.000000,0.019633
93.000000,0.018821
94.000000,0.017951
95.000000,0.017006
96.000000,0.015961
97.000000,0.014784
98.000000,0.013408
99.000000,0.011689
100.000000,0.009188
};
\label{p2}
\addplot+[
  black, mark options={black,mark=triangle*,scale=1},
  only marks, 
  error bars/.cd, 
    y fixed,
    y dir=both, 
    y explicit
] table [x=x, y=y, col sep=comma] {
    x,  y,      
6.000000,0.090056
7.000000,0.081125
8.000000,0.074167
9.000000,0.068585
10.000000,0.064006
11.000000,0.060175
12.000000,0.056924
13.000000,0.054137
14.000000,0.051705
15.000000,0.049584
16.000000,0.047713
17.000000,0.046048
18.000000,0.044556
19.000000,0.043221
20.000000,0.042006
21.000000,0.040913
22.000000,0.039922
23.000000,0.039017
24.000000,0.038187
25.000000,0.037417
26.000000,0.036712
27.000000,0.036067
28.000000,0.035462
29.000000,0.034911
30.000000,0.034391
31.000000,0.033914
32.000000,0.033461
33.000000,0.033052
34.000000,0.032665
35.000000,0.032300
36.000000,0.031962
37.000000,0.031643
38.000000,0.031335
39.000000,0.031054
40.000000,0.030794
41.000000,0.030538
42.000000,0.030303
43.000000,0.030071
44.000000,0.029865
45.000000,0.029665
46.000000,0.029480
47.000000,0.029292
48.000000,0.029118
49.000000,0.028953
50.000000,0.028788
51.000000,0.028651
52.000000,0.028502
53.000000,0.028365
54.000000,0.028230
55.000000,0.028104
56.000000,0.027989
57.000000,0.027871
58.000000,0.027767
59.000000,0.027660
60.000000,0.027537
61.000000,0.027440
62.000000,0.027340
63.000000,0.027237
64.000000,0.027142
65.000000,0.027047
66.000000,0.026954
67.000000,0.026857
68.000000,0.026769
69.000000,0.026674
70.000000,0.026579
71.000000,0.026498
72.000000,0.026400
73.000000,0.026308
74.000000,0.026216
75.000000,0.026113
76.000000,0.026025
77.000000,0.025917
78.000000,0.025819
79.000000,0.025714
80.000000,0.025605
81.000000,0.025490
82.000000,0.025379
83.000000,0.025255
84.000000,0.025129
85.000000,0.025002
86.000000,0.024870
87.000000,0.024736
88.000000,0.024589
89.000000,0.024457
90.000000,0.024322
91.000000,0.024190
92.000000,0.024080
93.000000,0.023984
94.000000,0.023919
95.000000,0.023867
96.000000,0.023800
97.000000,0.023668
98.000000,0.023397
99.000000,0.023177
100.000000,0.023213
101.000000,0.022964
102.000000,0.022975
103.000000,0.022686
104.000000,0.022441
105.000000,0.022345
106.000000,0.022287
107.000000,0.022191
108.000000,0.022050
109.000000,0.021857
110.000000,0.021660
111.000000,0.021447
112.000000,0.021231
113.000000,0.021007
114.000000,0.020797
115.000000,0.020588
116.000000,0.020375
117.000000,0.020169
118.000000,0.019960
119.000000,0.019750
120.000000,0.019545
121.000000,0.019324
122.000000,0.019099
123.000000,0.018886
124.000000,0.018658
125.000000,0.018421
126.000000,0.018181
127.000000,0.017926
128.000000,0.017672
129.000000,0.017410
130.000000,0.017144
131.000000,0.016851
132.000000,0.016561
133.000000,0.016249
134.000000,0.015935
135.000000,0.015602
136.000000,0.015263
137.000000,0.014904
138.000000,0.014531
139.000000,0.014140
140.000000,0.013727
141.000000,0.013288
142.000000,0.012823
143.000000,0.012337
144.000000,0.011802
145.000000,0.011210
146.000000,0.010554
147.000000,0.009810
148.000000,0.008928
149.000000,0.007827
150.000000,0.006196
};
\label{p3}
\addplot+[
  black, mark options={black,mark=*,scale=0.75},
  only marks, 
  error bars/.cd, 
    y fixed,
    y dir=both, 
    y explicit
] table [x=x, y=y, col sep=comma] {
    x,  y,      
6.000000,0.088567
7.000000,0.079531
8.000000,0.072484
9.000000,0.066819
10.000000,0.062150
11.000000,0.058224
12.000000,0.054903
13.000000,0.052030
14.000000,0.049531
15.000000,0.047321
16.000000,0.045397
17.000000,0.043657
18.000000,0.042110
19.000000,0.040692
20.000000,0.039417
21.000000,0.038266
22.000000,0.037208
23.000000,0.036251
24.000000,0.035346
25.000000,0.034532
26.000000,0.033779
27.000000,0.033068
28.000000,0.032413
29.000000,0.031809
30.000000,0.031248
31.000000,0.030717
32.000000,0.030224
33.000000,0.029759
34.000000,0.029326
35.000000,0.028912
36.000000,0.028534
37.000000,0.028150
38.000000,0.027815
39.000000,0.027503
40.000000,0.027198
41.000000,0.026904
42.000000,0.026631
43.000000,0.026364
44.000000,0.026115
45.000000,0.025880
46.000000,0.025660
47.000000,0.025453
48.000000,0.025242
49.000000,0.025051
50.000000,0.024865
51.000000,0.024698
52.000000,0.024531
53.000000,0.024360
54.000000,0.024205
55.000000,0.024050
56.000000,0.023911
57.000000,0.023786
58.000000,0.023648
59.000000,0.023524
60.000000,0.023401
61.000000,0.023283
62.000000,0.023160
63.000000,0.023060
64.000000,0.022942
65.000000,0.022841
66.000000,0.022758
67.000000,0.022665
68.000000,0.022573
69.000000,0.022473
70.000000,0.022383
71.000000,0.022313
72.000000,0.022214
73.000000,0.022138
74.000000,0.022056
75.000000,0.021982
76.000000,0.021913
77.000000,0.021826
78.000000,0.021758
79.000000,0.021698
80.000000,0.021623
81.000000,0.021549
82.000000,0.021494
83.000000,0.021433
84.000000,0.021368
85.000000,0.021304
86.000000,0.021231
87.000000,0.021185
88.000000,0.021130
89.000000,0.021053
90.000000,0.020991
91.000000,0.020930
92.000000,0.020883
93.000000,0.020818
94.000000,0.020764
95.000000,0.020712
96.000000,0.020653
97.000000,0.020588
98.000000,0.020540
99.000000,0.020493
100.000000,0.020433
101.000000,0.020392
102.000000,0.020322
103.000000,0.020265
104.000000,0.020214
105.000000,0.020160
106.000000,0.020097
107.000000,0.020047
108.000000,0.020004
109.000000,0.019944
110.000000,0.019890
111.000000,0.019831
112.000000,0.019778
113.000000,0.019723
114.000000,0.019685
115.000000,0.019617
116.000000,0.019576
117.000000,0.019518
118.000000,0.019460
119.000000,0.019403
120.000000,0.019356
121.000000,0.019285
122.000000,0.019222
123.000000,0.019174
124.000000,0.019112
125.000000,0.019058
126.000000,0.018989
127.000000,0.018923
128.000000,0.018850
129.000000,0.018787
130.000000,0.018725
131.000000,0.018658
132.000000,0.018586
133.000000,0.018493
134.000000,0.018418
135.000000,0.018326
136.000000,0.018234
137.000000,0.018128
138.000000,0.018027
139.000000,0.017918
140.000000,0.017789
141.000000,0.017673
142.000000,0.017531
143.000000,0.017400
144.000000,0.017267
145.000000,0.017189
146.000000,0.017130
147.000000,0.017120
148.000000,0.017063
149.000000,0.016831
150.000000,0.016660
151.000000,0.016736
152.000000,0.016513
153.000000,0.016296
154.000000,0.016283
155.000000,0.016250
156.000000,0.016145
157.000000,0.015977
158.000000,0.015811
159.000000,0.015653
160.000000,0.015480
161.000000,0.015323
162.000000,0.015181
163.000000,0.015038
164.000000,0.014920
165.000000,0.014784
166.000000,0.014662
167.000000,0.014539
168.000000,0.014400
169.000000,0.014270
170.000000,0.014129
171.000000,0.013994
172.000000,0.013850
173.000000,0.013702
174.000000,0.013555
175.000000,0.013386
176.000000,0.013222
177.000000,0.013049
178.000000,0.012875
179.000000,0.012687
180.000000,0.012498
181.000000,0.012287
182.000000,0.012095
183.000000,0.011873
184.000000,0.011646
185.000000,0.011409
186.000000,0.011172
187.000000,0.010920
188.000000,0.010661
189.000000,0.010383
190.000000,0.010088
191.000000,0.009788
192.000000,0.009464
193.000000,0.009115
194.000000,0.008740
195.000000,0.008321
196.000000,0.007854
197.000000,0.007319
198.000000,0.006667
199.000000,0.005855
200.000000,0.004648
};
\label{p4}
\addplot+[
  black, mark options={black,mark=square,scale=0.75},
  only marks, 
  error bars/.cd, 
    y fixed,
    y dir=both, 
    y explicit
] table [x=x, y=y, col sep=comma] {
    x,  y,      
5.000000,0.092089
6.000000,0.079590
7.000000,0.070133
8.000000,0.062688
9.000000,0.056652
10.000000,0.051635
11.000000,0.047367
12.000000,0.043721
13.000000,0.040542
14.000000,0.037744
15.000000,0.035252
16.000000,0.033046
17.000000,0.031036
18.000000,0.029224
19.000000,0.027572
20.000000,0.026045
21.000000,0.024665
22.000000,0.023380
23.000000,0.022210
24.000000,0.021096
25.000000,0.020081
26.000000,0.019124
27.000000,0.018232
28.000000,0.017396
29.000000,0.016617
30.000000,0.015889
31.000000,0.015188
32.000000,0.014535
33.000000,0.013905
34.000000,0.013322
35.000000,0.012760
36.000000,0.012224
37.000000,0.011700
38.000000,0.011203
39.000000,0.010730
40.000000,0.010255
41.000000,0.009779
42.000000,0.009338
43.000000,0.008900
44.000000,0.008480
45.000000,0.008130
46.000000,0.007854
47.000000,0.007610
48.000000,0.007195
49.000000,0.006797
50.000000,0.006651
51.000000,0.006321
52.000000,0.005948
53.000000,0.005715
54.000000,0.005533
55.000000,0.005315
56.000000,0.005097
57.000000,0.004852
58.000000,0.004601
59.000000,0.004356
60.000000,0.004135
61.000000,0.003912
62.000000,0.003702
63.000000,0.003518
64.000000,0.003321
65.000000,0.003149
66.000000,0.002992
67.000000,0.002842
68.000000,0.002684
69.000000,0.002536
70.000000,0.002402
71.000000,0.002286
72.000000,0.002154
73.000000,0.002031
74.000000,0.001911
75.000000,0.001798
76.000000,0.001699
77.000000,0.001587
78.000000,0.001496
79.000000,0.001403
80.000000,0.001299
81.000000,0.001211
82.000000,0.001130
83.000000,0.001048
84.000000,0.000966
85.000000,0.000894
86.000000,0.000807
87.000000,0.000751
88.000000,0.000683
89.000000,0.000606
90.000000,0.000537
91.000000,0.000473
92.000000,0.000414
93.000000,0.000357
94.000000,0.000303
95.000000,0.000246
96.000000,0.000192
97.000000,0.000139
98.000000,0.000092
99.000000,0.000042
100.000000,0.000000
101.000000,-0.000040
102.000000,-0.000085
103.000000,-0.000123
104.000000,-0.000164
105.000000,-0.000202
106.000000,-0.000238
107.000000,-0.000270
108.000000,-0.000301
109.000000,-0.000330
110.000000,-0.000360
111.000000,-0.000390
112.000000,-0.000422
113.000000,-0.000445
114.000000,-0.000459
115.000000,-0.000489
116.000000,-0.000506
117.000000,-0.000527
118.000000,-0.000546
119.000000,-0.000561
120.000000,-0.000577
121.000000,-0.000598
122.000000,-0.000611
123.000000,-0.000622
124.000000,-0.000638
125.000000,-0.000647
126.000000,-0.000659
127.000000,-0.000671
128.000000,-0.000681
129.000000,-0.000693
130.000000,-0.000696
131.000000,-0.000704
132.000000,-0.000712
133.000000,-0.000721
134.000000,-0.000726
135.000000,-0.000730
136.000000,-0.000735
137.000000,-0.000744
138.000000,-0.000747
139.000000,-0.000753
140.000000,-0.000760
141.000000,-0.000763
142.000000,-0.000768
143.000000,-0.000771
144.000000,-0.000771
145.000000,-0.000765
146.000000,-0.000757
147.000000,-0.000743
148.000000,-0.000734
149.000000,-0.000741
150.000000,-0.000739
151.000000,-0.000716
152.000000,-0.000717
153.000000,-0.000718
154.000000,-0.000701
155.000000,-0.000685
156.000000,-0.000675
157.000000,-0.000668
158.000000,-0.000660
159.000000,-0.000650
160.000000,-0.000641
161.000000,-0.000630
162.000000,-0.000616
163.000000,-0.000603
164.000000,-0.000589
165.000000,-0.000574
166.000000,-0.000559
167.000000,-0.000543
168.000000,-0.000527
169.000000,-0.000511
170.000000,-0.000495
171.000000,-0.000478
172.000000,-0.000461
173.000000,-0.000444
174.000000,-0.000427
175.000000,-0.000410
176.000000,-0.000392
177.000000,-0.000375
178.000000,-0.000357
179.000000,-0.000339
180.000000,-0.000322
181.000000,-0.000304
182.000000,-0.000286
183.000000,-0.000268
184.000000,-0.000250
185.000000,-0.000232
186.000000,-0.000214
187.000000,-0.000196
188.000000,-0.000178
189.000000,-0.000160
190.000000,-0.000143
191.000000,-0.000126
192.000000,-0.000109
193.000000,-0.000092
194.000000,-0.000076
195.000000,-0.000061
196.000000,-0.000046
197.000000,-0.000032
198.000000,-0.000019
199.000000,-0.000008
200.000000,0.000000
};
\label{p5}

\node (A) at (60, 0.09) {\ref{p1} $A=50$};
\node (A1) at (60, 0.085) {};
\node (A2) at (30, 0.065){}; 
\draw [->](A1) to (A2);
\node (B) at (70, 0.06) {\ref{p2} $A=100$};
\node (B1) at (70, 0.055) {};
\node (B2) at (60, 0.035){}; 
\draw [->](B1) to (B2);
\node (C) at (100, 0.045) {\ref{p3} $A=150$};
\node (C1) at (100, 0.040) {};
\node (C2) at (95, 0.025){}; 
\draw [->](C1) to (C2);
\node (D) at (180, 0.04) {\ref{p4} $A=200$};
\node (D1) at (180, 0.035) {};
\node (D2) at (170, 0.02){}; 
\draw [->](D1) to (D2);
\node (E) at (60, -0.01) {\ref{p5} Linear interpolation};
\node (E1) at (60, -0.005) {};
\node (E2) at (50, 0.005){}; 
\draw [->](E1) to (E2);
\end{axis}
\end{tikzpicture}

%% file: Fig_p=1_q=1_dense_A=50.tex
\centering
\begin{tikzpicture}[]
\begin{axis}[ymin=-6.2, ymax=6.2, xmin=-0.05, xmax=1.05,
  ytick={-6,-4,...,6}, ytick pos=left,
  xtick={0,0.2,...,1}, xtick pos=left,
  xlabel={$x/L$},
  ylabel={$\extPotOptim(x)/D$},
  legend pos=north west,
  legend style={draw=none}]
\addplot+[line width=2pt,
  black, smooth,mark=empty,
] table [x=x, y=y, col sep=comma] {
    x,  y,      
0.000000,-0.000220
0.000500,-0.416760
0.001001,-0.829109
0.001501,-1.233137
0.002001,-1.624840
0.002501,-2.000394
0.003002,-2.356215
0.003502,-2.689007
0.004002,-2.995810
0.004502,-3.274042
0.005003,-3.521533
0.005503,-3.736551
0.006003,-3.917825
0.006503,-4.064555
0.007004,-4.176418
0.007504,-4.253562
0.008004,-4.296600
0.008504,-4.306587
0.009005,-4.284996
0.009505,-4.233688
0.010005,-4.154870
0.010505,-4.051057
0.011006,-3.925020
0.011506,-3.779739
0.012006,-3.618349
0.012506,-3.444087
0.013007,-3.260231
0.013507,-3.070056
0.014007,-2.876773
0.014507,-2.683481
0.015008,-2.493121
0.015508,-2.308433
0.016008,-2.131917
0.016508,-1.965799
0.017009,-1.812007
0.017509,-1.672148
0.018009,-1.547495
0.018509,-1.438976
0.019010,-1.347175
0.019510,-1.272339
0.020010,-1.214384
0.020510,-1.172916
0.021011,-1.147250
0.021511,-1.136437
0.022011,-1.139298
0.022511,-1.154452
0.023012,-1.180359
0.023512,-1.215352
0.024012,-1.257680
0.024512,-1.305546
0.025013,-1.357146
0.025513,-1.410706
0.026013,-1.464518
0.026513,-1.516970
0.027014,-1.566579
0.027514,-1.612012
0.028014,-1.652111
0.028514,-1.685908
0.029015,-1.712637
0.029515,-1.731741
0.030015,-1.742876
0.030515,-1.745907
0.031016,-1.740902
0.031516,-1.728124
0.032016,-1.708012
0.032516,-1.681166
0.033017,-1.648325
0.033517,-1.610345
0.034017,-1.568174
0.034517,-1.522825
0.035018,-1.475353
0.035518,-1.426824
0.036018,-1.378291
0.036518,-1.330774
0.037019,-1.285229
0.037519,-1.242534
0.038019,-1.203470
0.038519,-1.168702
0.039020,-1.138772
0.039520,-1.114089
0.040020,-1.094920
0.040520,-1.081393
0.041021,-1.073497
0.041521,-1.071086
0.042021,-1.073886
0.042521,-1.081510
0.043022,-1.093463
0.043522,-1.109167
0.044022,-1.127968
0.044522,-1.149161
0.045023,-1.172005
0.045523,-1.195741
0.046023,-1.219616
0.046523,-1.242895
0.047024,-1.264881
0.047524,-1.284933
0.048024,-1.302474
0.048524,-1.317010
0.049025,-1.328135
0.049525,-1.335539
0.050025,-1.339017
0.050525,-1.338466
0.051026,-1.333886
0.051526,-1.325380
0.052026,-1.313146
0.052526,-1.297472
0.053027,-1.278722
0.053527,-1.257333
0.054027,-1.233794
0.054527,-1.208639
0.055028,-1.182429
0.055528,-1.155741
0.056028,-1.129148
0.056528,-1.103211
0.057029,-1.078462
0.057529,-1.055393
0.058029,-1.034443
0.058529,-1.015990
0.059030,-1.000342
0.059530,-0.987733
0.060030,-0.978315
0.060530,-0.972161
0.061031,-0.969259
0.061531,-0.969521
0.062031,-0.972780
0.062531,-0.978801
0.063032,-0.987283
0.063532,-0.997874
0.064032,-1.010176
0.064532,-1.023758
0.065033,-1.038166
0.065533,-1.052935
0.066033,-1.067602
0.066533,-1.081716
0.067034,-1.094848
0.067534,-1.106606
0.068034,-1.116636
0.068534,-1.124637
0.069035,-1.130366
0.069535,-1.133639
0.070035,-1.134341
0.070535,-1.132420
0.071036,-1.127895
0.071536,-1.120846
0.072036,-1.111419
0.072536,-1.099816
0.073037,-1.086289
0.073537,-1.071136
0.074037,-1.054691
0.074537,-1.037313
0.075038,-1.019382
0.075538,-1.001282
0.076038,-0.983398
0.076538,-0.966103
0.077039,-0.949749
0.077539,-0.934660
0.078039,-0.921122
0.078539,-0.909379
0.079040,-0.899628
0.079540,-0.892010
0.080040,-0.886614
0.080540,-0.883471
0.081041,-0.882557
0.081541,-0.883790
0.082041,-0.887039
0.082541,-0.892124
0.083042,-0.898821
0.083542,-0.906872
0.084042,-0.915988
0.084542,-0.925857
0.085043,-0.936156
0.085543,-0.946553
0.086043,-0.956723
0.086543,-0.966347
0.087044,-0.975130
0.087544,-0.982797
0.088044,-0.989111
0.088544,-0.993869
0.089045,-0.996910
0.089545,-0.998122
0.090045,-0.997437
0.090545,-0.994838
0.091046,-0.990356
0.091546,-0.984070
0.092046,-0.976101
0.092546,-0.966613
0.093047,-0.955806
0.093547,-0.943909
0.094047,-0.931179
0.094547,-0.917887
0.095048,-0.904316
0.095548,-0.890755
0.096048,-0.877485
0.096548,-0.864779
0.097049,-0.852894
0.097549,-0.842059
0.098049,-0.832478
0.098549,-0.824319
0.099050,-0.817714
0.099550,-0.812753
0.100050,-0.809485
0.100550,-0.807916
0.101051,-0.808011
0.101551,-0.809690
0.102051,-0.812838
0.102551,-0.817304
0.103052,-0.822905
0.103552,-0.829434
0.104052,-0.836661
0.104552,-0.844344
0.105053,-0.852230
0.105553,-0.860066
0.106053,-0.867603
0.106553,-0.874603
0.107054,-0.880843
0.107554,-0.886125
0.108054,-0.890274
0.108554,-0.893148
0.109055,-0.894638
0.109555,-0.894671
0.110055,-0.893213
0.110555,-0.890266
0.111056,-0.885872
0.111556,-0.880107
0.112056,-0.873081
0.112556,-0.864936
0.113057,-0.855840
0.113557,-0.845984
0.114057,-0.835574
0.114557,-0.824830
0.115058,-0.813978
0.115558,-0.803244
0.116058,-0.792850
0.116558,-0.783005
0.117059,-0.773903
0.117559,-0.765720
0.118059,-0.758604
0.118559,-0.752676
0.119060,-0.748027
0.119560,-0.744714
0.120060,-0.742760
0.120560,-0.742154
0.121061,-0.742853
0.121561,-0.744779
0.122061,-0.747825
0.122561,-0.751860
0.123062,-0.756725
0.123562,-0.762244
0.124062,-0.768229
0.124562,-0.774477
0.125063,-0.780785
0.125563,-0.786949
0.126063,-0.792771
0.126563,-0.798063
0.127064,-0.802653
0.127564,-0.806388
0.128064,-0.809138
0.128564,-0.810801
0.129065,-0.811301
0.129565,-0.810593
0.130065,-0.808664
0.130565,-0.805530
0.131066,-0.801238
0.131566,-0.795866
0.132066,-0.789515
0.132566,-0.782313
0.133067,-0.774406
0.133567,-0.765958
0.134067,-0.757146
0.134567,-0.748152
0.135068,-0.739163
0.135568,-0.730365
0.136068,-0.721936
0.136568,-0.714044
0.137069,-0.706842
0.137569,-0.700466
0.138069,-0.695027
0.138569,-0.690615
0.139070,-0.687291
0.139570,-0.685089
0.140070,-0.684015
0.140570,-0.684047
0.141071,-0.685134
0.141571,-0.687201
0.142071,-0.690148
0.142571,-0.693853
0.143072,-0.698178
0.143572,-0.702969
0.144072,-0.708062
0.144572,-0.713288
0.145073,-0.718475
0.145573,-0.723453
0.146073,-0.728061
0.146573,-0.732147
0.147074,-0.735573
0.147574,-0.738220
0.148074,-0.739990
0.148574,-0.740807
0.149075,-0.740619
0.149575,-0.739402
0.150075,-0.737157
0.150575,-0.733910
0.151076,-0.729714
0.151576,-0.724645
0.152076,-0.718799
0.152576,-0.712293
0.153077,-0.705258
0.153577,-0.697840
0.154077,-0.690190
0.154577,-0.682468
0.155078,-0.674830
0.155578,-0.667434
0.156078,-0.660426
0.156578,-0.653944
0.157079,-0.648112
0.157579,-0.643035
0.158079,-0.638799
0.158579,-0.635470
0.159080,-0.633088
0.159580,-0.631672
0.160080,-0.631213
0.160580,-0.631681
0.161081,-0.633022
0.161581,-0.635160
0.162081,-0.638001
0.162581,-0.641433
0.163082,-0.645329
0.163582,-0.649554
0.164082,-0.653963
0.164582,-0.658408
0.165083,-0.662744
0.165583,-0.666826
0.166083,-0.670520
0.166583,-0.673698
0.167084,-0.676252
0.167584,-0.678086
0.168084,-0.679124
0.168584,-0.679312
0.169085,-0.678616
0.169585,-0.677027
0.170085,-0.674555
0.170585,-0.671235
0.171086,-0.667123
0.171586,-0.662294
0.172086,-0.656841
0.172586,-0.650871
0.173087,-0.644506
0.173587,-0.637874
0.174087,-0.631111
0.174587,-0.624355
0.175088,-0.617744
0.175588,-0.611409
0.176088,-0.605476
0.176588,-0.600059
0.177089,-0.595257
0.177589,-0.591156
0.178089,-0.587821
0.178589,-0.585298
0.179090,-0.583614
0.179590,-0.582773
0.180090,-0.582759
0.180590,-0.583533
0.181091,-0.585040
0.181591,-0.587205
0.182091,-0.589935
0.182591,-0.593128
0.183092,-0.596666
0.183592,-0.600428
0.184092,-0.604284
0.184592,-0.608106
0.185093,-0.611764
0.185593,-0.615135
0.186093,-0.618105
0.186593,-0.620570
0.187094,-0.622438
0.187594,-0.623634
0.188094,-0.624101
0.188594,-0.623799
0.189095,-0.622710
0.189595,-0.620833
0.190095,-0.618190
0.190595,-0.614821
0.191096,-0.610783
0.191596,-0.606151
0.192096,-0.601013
0.192596,-0.595473
0.193097,-0.589639
0.193597,-0.583631
0.194097,-0.577570
0.194597,-0.571578
0.195098,-0.565776
0.195598,-0.560277
0.196098,-0.555188
0.196598,-0.550605
0.197099,-0.546609
0.197599,-0.543269
0.198099,-0.540633
0.198599,-0.538735
0.199100,-0.537587
0.199600,-0.537184
0.200100,-0.537503
0.200600,-0.538500
0.201101,-0.540118
0.201601,-0.542281
0.202101,-0.544902
0.202601,-0.547884
0.203102,-0.551117
0.203602,-0.554491
0.204102,-0.557889
0.204602,-0.561195
0.205103,-0.564298
0.205603,-0.567089
0.206103,-0.569471
0.206603,-0.571356
0.207104,-0.572669
0.207604,-0.573352
0.208104,-0.573359
0.208604,-0.572666
0.209105,-0.571264
0.209605,-0.569162
0.210105,-0.566388
0.210605,-0.562986
0.211106,-0.559016
0.211606,-0.554551
0.212106,-0.549678
0.212606,-0.544492
0.213107,-0.539097
0.213607,-0.533600
0.214107,-0.528113
0.214607,-0.522743
0.215108,-0.517598
0.215608,-0.512778
0.216108,-0.508372
0.216608,-0.504463
0.217109,-0.501116
0.217609,-0.498386
0.218109,-0.496309
0.218609,-0.494907
0.219110,-0.494182
0.219610,-0.494121
0.220110,-0.494694
0.220610,-0.495856
0.221111,-0.497545
0.221611,-0.499688
0.222111,-0.502201
0.222611,-0.504990
0.223112,-0.507956
0.223612,-0.510994
0.224112,-0.514000
0.224612,-0.516869
0.225113,-0.519502
0.225613,-0.521806
0.226113,-0.523695
0.226613,-0.525096
0.227114,-0.525948
0.227614,-0.526204
0.228114,-0.525832
0.228614,-0.524817
0.229115,-0.523159
0.229615,-0.520876
0.230115,-0.518000
0.230615,-0.514578
0.231116,-0.510671
0.231616,-0.506353
0.232116,-0.501707
0.232616,-0.496823
0.233117,-0.491799
0.233617,-0.486734
0.234117,-0.481729
0.234617,-0.476882
0.235118,-0.472288
0.235618,-0.468033
0.236118,-0.464197
0.236618,-0.460847
0.237119,-0.458038
0.237619,-0.455812
0.238119,-0.454195
0.238619,-0.453198
0.239120,-0.452816
0.239620,-0.453030
0.240120,-0.453804
0.240620,-0.455089
0.241121,-0.456824
0.241621,-0.458937
0.242121,-0.461345
0.242621,-0.463959
0.243122,-0.466686
0.243622,-0.469430
0.244122,-0.472095
0.244622,-0.474586
0.245123,-0.476815
0.245623,-0.478700
0.246123,-0.480168
0.246623,-0.481157
0.247124,-0.481618
0.247624,-0.481513
0.248124,-0.480822
0.248624,-0.479538
0.249125,-0.477668
0.249625,-0.475237
0.250125,-0.472280
0.250625,-0.468848
0.251126,-0.465003
0.251626,-0.460817
0.252126,-0.456371
0.252626,-0.451752
0.253127,-0.447050
0.253627,-0.442359
0.254127,-0.437769
0.254627,-0.433371
0.255128,-0.429248
0.255628,-0.425477
0.256128,-0.422126
0.256628,-0.419252
0.257129,-0.416899
0.257629,-0.415098
0.258129,-0.413868
0.258629,-0.413210
0.259130,-0.413113
0.259630,-0.413552
0.260130,-0.414487
0.260630,-0.415868
0.261131,-0.417631
0.261631,-0.419706
0.262131,-0.422012
0.262631,-0.424465
0.263132,-0.426977
0.263632,-0.429459
0.264132,-0.431821
0.264632,-0.433979
0.265133,-0.435855
0.265633,-0.437375
0.266133,-0.438477
0.266633,-0.439109
0.267134,-0.439231
0.267634,-0.438816
0.268134,-0.437852
0.268634,-0.436338
0.269135,-0.434290
0.269635,-0.431735
0.270135,-0.428715
0.270635,-0.425281
0.271136,-0.421497
0.271636,-0.417433
0.272136,-0.413167
0.272636,-0.408784
0.273137,-0.404368
0.273637,-0.400005
0.274137,-0.395781
0.274637,-0.391775
0.275138,-0.388065
0.275638,-0.384715
0.276138,-0.381786
0.276638,-0.379323
0.277139,-0.377364
0.277639,-0.375929
0.278139,-0.375029
0.278639,-0.374660
0.279140,-0.374804
0.279640,-0.375430
0.280140,-0.376497
0.280640,-0.377951
0.281141,-0.379728
0.281641,-0.381760
0.282141,-0.383967
0.282641,-0.386269
0.283142,-0.388583
0.283642,-0.390826
0.284142,-0.392916
0.284642,-0.394776
0.285143,-0.396335
0.285643,-0.397530
0.286143,-0.398307
0.286643,-0.398624
0.287144,-0.398448
0.287644,-0.397762
0.288144,-0.396558
0.288644,-0.394846
0.289145,-0.392644
0.289645,-0.389985
0.290145,-0.386913
0.290645,-0.383483
0.291146,-0.379756
0.291646,-0.375805
0.292146,-0.371704
0.292646,-0.367533
0.293147,-0.363373
0.293647,-0.359304
0.294147,-0.355405
0.294647,-0.351748
0.295148,-0.348402
0.295648,-0.345425
0.296148,-0.342866
0.296648,-0.340765
0.297149,-0.339149
0.297649,-0.338034
0.298149,-0.337421
0.298649,-0.337300
0.299150,-0.337648
0.299650,-0.338432
0.300150,-0.339606
0.300650,-0.341114
0.301151,-0.342894
0.301651,-0.344876
0.302151,-0.346986
0.302651,-0.349144
0.303152,-0.351273
0.303652,-0.353294
0.304152,-0.355134
0.304652,-0.356721
0.305153,-0.357993
0.305653,-0.358894
0.306153,-0.359378
0.306653,-0.359411
0.307154,-0.358968
0.307654,-0.358038
0.308154,-0.356622
0.308654,-0.354734
0.309155,-0.352397
0.309655,-0.349648
0.310155,-0.346535
0.310655,-0.343112
0.311156,-0.339442
0.311656,-0.335596
0.312156,-0.331647
0.312656,-0.327671
0.313157,-0.323745
0.313657,-0.319943
0.314157,-0.316338
0.314657,-0.312997
0.315158,-0.309979
0.315658,-0.307336
0.316158,-0.305110
0.316658,-0.303332
0.317159,-0.302023
0.317659,-0.301190
0.318159,-0.300830
0.318659,-0.300927
0.319160,-0.301454
0.319660,-0.302373
0.320160,-0.303636
0.320660,-0.305187
0.321161,-0.306962
0.321661,-0.308892
0.322161,-0.310905
0.322661,-0.312926
0.323162,-0.314881
0.323662,-0.316697
0.324162,-0.318304
0.324662,-0.319640
0.325163,-0.320648
0.325663,-0.321280
0.326163,-0.321496
0.326663,-0.321270
0.327164,-0.320585
0.327664,-0.319435
0.328164,-0.317827
0.328664,-0.315781
0.329165,-0.313324
0.329665,-0.310496
0.330165,-0.307347
0.330665,-0.303933
0.331166,-0.300318
0.331666,-0.296571
0.332166,-0.292762
0.332666,-0.288964
0.333167,-0.285252
0.333667,-0.281694
0.334167,-0.278357
0.334667,-0.275301
0.335168,-0.272580
0.335668,-0.270239
0.336168,-0.268312
0.336668,-0.266824
0.337169,-0.265789
0.337669,-0.265208
0.338169,-0.265073
0.338669,-0.265362
0.339170,-0.266044
0.339670,-0.267078
0.340170,-0.268414
0.340670,-0.269994
0.341171,-0.271756
0.341671,-0.273631
0.342171,-0.275548
0.342671,-0.277437
0.343172,-0.279226
0.343672,-0.280847
0.344172,-0.282237
0.344672,-0.283339
0.345173,-0.284102
0.345673,-0.284484
0.346173,-0.284454
0.346673,-0.283990
0.347174,-0.283083
0.347674,-0.281733
0.348174,-0.279951
0.348674,-0.277762
0.349175,-0.275198
0.349675,-0.272302
0.350175,-0.269124
0.350675,-0.265723
0.351176,-0.262163
0.351676,-0.258509
0.352176,-0.254833
0.352676,-0.251204
0.353177,-0.247690
0.353677,-0.244358
0.354177,-0.241269
0.354677,-0.238477
0.355178,-0.236029
0.355678,-0.233963
0.356178,-0.232310
0.356678,-0.231086
0.357179,-0.230299
0.357679,-0.229945
0.358179,-0.230010
0.358679,-0.230469
0.359180,-0.231286
0.359680,-0.232418
0.360180,-0.233812
0.360680,-0.235412
0.361181,-0.237152
0.361681,-0.238966
0.362181,-0.240787
0.362681,-0.242544
0.363182,-0.244171
0.363682,-0.245605
0.364182,-0.246788
0.364682,-0.247666
0.365183,-0.248197
0.365683,-0.248344
0.366183,-0.248082
0.366683,-0.247397
0.367184,-0.246282
0.367684,-0.244745
0.368184,-0.242803
0.368684,-0.240483
0.369185,-0.237821
0.369685,-0.234863
0.370185,-0.231662
0.370685,-0.228276
0.371186,-0.224768
0.371686,-0.221206
0.372186,-0.217656
0.372686,-0.214185
0.373187,-0.210860
0.373687,-0.207740
0.374187,-0.204882
0.374687,-0.202336
0.375188,-0.200142
0.375688,-0.198333
0.376188,-0.196933
0.376688,-0.195952
0.377189,-0.195393
0.377689,-0.195247
0.378189,-0.195494
0.378689,-0.196106
0.379190,-0.197044
0.379690,-0.198261
0.380190,-0.199705
0.380690,-0.201315
0.381191,-0.203029
0.381691,-0.204781
0.382191,-0.206504
0.382691,-0.208132
0.383192,-0.209603
0.383692,-0.210858
0.384192,-0.211841
0.384692,-0.212508
0.385193,-0.212819
0.385693,-0.212744
0.386193,-0.212264
0.386693,-0.211370
0.387194,-0.210062
0.387694,-0.208352
0.388194,-0.206261
0.388694,-0.203820
0.389195,-0.201070
0.389695,-0.198057
0.390195,-0.194837
0.390695,-0.191469
0.391196,-0.188016
0.391696,-0.184542
0.392196,-0.181114
0.392696,-0.177796
0.393197,-0.174649
0.393697,-0.171731
0.394197,-0.169093
0.394697,-0.166778
0.395198,-0.164823
0.395698,-0.163254
0.396198,-0.162089
0.396698,-0.161334
0.397199,-0.160986
0.397699,-0.161031
0.398199,-0.161445
0.398699,-0.162194
0.399200,-0.163239
0.399700,-0.164529
0.400200,-0.166011
0.400700,-0.167623
0.401201,-0.169304
0.401701,-0.170988
0.402201,-0.172612
0.402701,-0.174111
0.403202,-0.175427
0.403702,-0.176504
0.404202,-0.177294
0.404702,-0.177754
0.405203,-0.177852
0.405703,-0.177563
0.406203,-0.176874
0.406703,-0.175780
0.407204,-0.174287
0.407704,-0.172412
0.408204,-0.170179
0.408704,-0.167625
0.409205,-0.164792
0.409705,-0.161730
0.410205,-0.158494
0.410705,-0.155145
0.411206,-0.151745
0.411706,-0.148359
0.412206,-0.145049
0.412706,-0.141878
0.413207,-0.138903
0.413707,-0.136177
0.414207,-0.133748
0.414707,-0.131653
0.415208,-0.129925
0.415708,-0.128584
0.416208,-0.127642
0.416708,-0.127100
0.417209,-0.126949
0.417709,-0.127172
0.418209,-0.127740
0.418709,-0.128617
0.419210,-0.129758
0.419710,-0.131113
0.420210,-0.132625
0.420710,-0.134234
0.421211,-0.135878
0.421711,-0.137492
0.422211,-0.139015
0.422711,-0.140386
0.423212,-0.141549
0.423712,-0.142453
0.424212,-0.143054
0.424712,-0.143314
0.425213,-0.143206
0.425713,-0.142711
0.426213,-0.141820
0.426713,-0.140535
0.427214,-0.138865
0.427714,-0.136832
0.428214,-0.134466
0.428714,-0.131805
0.429215,-0.128894
0.429715,-0.125786
0.430215,-0.122538
0.430715,-0.119209
0.431216,-0.115863
0.431716,-0.112561
0.432216,-0.109367
0.432716,-0.106337
0.433217,-0.103528
0.433717,-0.100988
0.434217,-0.098759
0.434717,-0.096875
0.435218,-0.095363
0.435718,-0.094238
0.436218,-0.093507
0.436718,-0.093165
0.437219,-0.093201
0.437719,-0.093590
0.438219,-0.094302
0.438719,-0.095295
0.439220,-0.096524
0.439720,-0.097934
0.440220,-0.099470
0.440720,-0.101069
0.441221,-0.102671
0.441721,-0.104212
0.442221,-0.105633
0.442721,-0.106876
0.443222,-0.107887
0.443722,-0.108621
0.444222,-0.109036
0.444722,-0.109101
0.445223,-0.108793
0.445723,-0.108099
0.446223,-0.107013
0.446723,-0.105543
0.447224,-0.103704
0.447724,-0.101521
0.448224,-0.099028
0.448724,-0.096267
0.449225,-0.093285
0.449725,-0.090136
0.450225,-0.086879
0.450725,-0.083574
0.451226,-0.080283
0.451726,-0.077067
0.452226,-0.073986
0.452726,-0.071097
0.453227,-0.068450
0.453727,-0.066090
0.454227,-0.064055
0.454727,-0.062376
0.455228,-0.061071
0.455728,-0.060154
0.456228,-0.059624
0.456728,-0.059473
0.457229,-0.059684
0.457729,-0.060230
0.458229,-0.061074
0.458729,-0.062173
0.459230,-0.063479
0.459730,-0.064937
0.460230,-0.066487
0.460730,-0.068069
0.461231,-0.069623
0.461731,-0.071085
0.462231,-0.072399
0.462731,-0.073510
0.463232,-0.074368
0.463732,-0.074929
0.464232,-0.075159
0.464732,-0.075030
0.465233,-0.074523
0.465733,-0.073631
0.466233,-0.072354
0.466733,-0.070704
0.467234,-0.068700
0.467734,-0.066371
0.468234,-0.063755
0.468734,-0.060898
0.469235,-0.057848
0.469735,-0.054662
0.470235,-0.051400
0.470735,-0.048121
0.471236,-0.044887
0.471736,-0.041758
0.472236,-0.038792
0.472736,-0.036041
0.473237,-0.033554
0.473737,-0.031373
0.474237,-0.029529
0.474737,-0.028049
0.475238,-0.026948
0.475738,-0.026232
0.476238,-0.025898
0.476738,-0.025931
0.477239,-0.026311
0.477739,-0.027005
0.478239,-0.027975
0.478739,-0.029174
0.479240,-0.030550
0.479740,-0.032048
0.480240,-0.033607
0.480740,-0.035166
0.481241,-0.036666
0.481741,-0.038045
0.482241,-0.039249
0.482741,-0.040225
0.483242,-0.040926
0.483742,-0.041315
0.484242,-0.041358
0.484742,-0.041035
0.485243,-0.040331
0.485743,-0.039243
0.486243,-0.037777
0.486743,-0.035948
0.487244,-0.033782
0.487744,-0.031311
0.488244,-0.028576
0.488744,-0.025626
0.489245,-0.022512
0.489745,-0.019293
0.490245,-0.016028
0.490745,-0.012778
0.491246,-0.009603
0.491746,-0.006563
0.492246,-0.003712
0.492746,-0.001102
0.493247,0.001224
0.493747,0.003229
0.494247,0.004882
0.494747,0.006165
0.495248,0.007065
0.495748,0.007583
0.496248,0.007725
0.496748,0.007511
0.497249,0.006968
0.497749,0.006129
0.498249,0.005039
0.498749,0.003746
0.499250,0.002304
0.499750,0.000772
0.500250,-0.000791
0.500750,-0.002323
0.501251,-0.003764
0.501751,-0.005057
0.502251,-0.006147
0.502751,-0.006984
0.503252,-0.007527
0.503752,-0.007741
0.504252,-0.007597
0.504752,-0.007079
0.505253,-0.006177
0.505753,-0.004893
0.506253,-0.003239
0.506753,-0.001233
0.507254,0.001094
0.507754,0.003706
0.508254,0.006557
0.508754,0.009599
0.509255,0.012774
0.509755,0.016025
0.510255,0.019291
0.510755,0.022511
0.511256,0.025625
0.511756,0.028577
0.512256,0.031312
0.512756,0.033783
0.513257,0.035950
0.513757,0.037779
0.514257,0.039245
0.514757,0.040333
0.515258,0.041036
0.515758,0.041360
0.516258,0.041316
0.516758,0.040927
0.517259,0.040225
0.517759,0.039249
0.518259,0.038044
0.518759,0.036664
0.519260,0.035164
0.519760,0.033604
0.520260,0.032044
0.520760,0.030546
0.521261,0.029169
0.521761,0.027969
0.522261,0.026998
0.522761,0.026303
0.523262,0.025923
0.523762,0.025888
0.524262,0.026222
0.524762,0.026938
0.525263,0.028038
0.525763,0.029518
0.526263,0.031361
0.526763,0.033542
0.527264,0.036028
0.527764,0.038778
0.528264,0.041744
0.528764,0.044873
0.529265,0.048107
0.529765,0.051386
0.530265,0.054649
0.530765,0.057834
0.531266,0.060884
0.531766,0.063742
0.532266,0.066358
0.532766,0.068687
0.533267,0.070692
0.533767,0.072343
0.534267,0.073621
0.534767,0.074513
0.535268,0.075021
0.535768,0.075150
0.536268,0.074921
0.536768,0.074361
0.537269,0.073504
0.537769,0.072394
0.538269,0.071081
0.538769,0.069619
0.539270,0.068066
0.539770,0.066485
0.540270,0.064935
0.540770,0.063478
0.541271,0.062173
0.541771,0.061073
0.542271,0.060230
0.542771,0.059684
0.543272,0.059473
0.543772,0.059624
0.544272,0.060154
0.544772,0.061071
0.545273,0.062375
0.545773,0.064054
0.546273,0.066088
0.546773,0.068447
0.547274,0.071094
0.547774,0.073982
0.548274,0.077062
0.548774,0.080277
0.549275,0.083567
0.549775,0.086871
0.550275,0.090127
0.550775,0.093274
0.551276,0.096255
0.551776,0.099016
0.552276,0.101508
0.552776,0.103690
0.553277,0.105528
0.553777,0.106997
0.554277,0.108082
0.554777,0.108776
0.555278,0.109084
0.555778,0.109019
0.556278,0.108603
0.556778,0.107870
0.557279,0.106858
0.557779,0.105616
0.558279,0.104196
0.558779,0.102655
0.559280,0.101054
0.559780,0.099455
0.560280,0.097921
0.560780,0.096511
0.561281,0.095284
0.561781,0.094292
0.562281,0.093581
0.562781,0.093193
0.563282,0.093159
0.563782,0.093501
0.564282,0.094234
0.564782,0.095360
0.565283,0.096873
0.565783,0.098757
0.566283,0.100987
0.566783,0.103528
0.567284,0.106337
0.567784,0.109367
0.568284,0.112562
0.568784,0.115864
0.569285,0.119210
0.569785,0.122538
0.570285,0.125787
0.570785,0.128894
0.571286,0.131804
0.571786,0.134465
0.572286,0.136830
0.572786,0.138862
0.573287,0.140531
0.573787,0.141816
0.574287,0.142706
0.574787,0.143201
0.575288,0.143308
0.575788,0.143047
0.576288,0.142446
0.576788,0.141541
0.577289,0.140377
0.577789,0.139006
0.578289,0.137483
0.578789,0.135868
0.579290,0.134225
0.579790,0.132616
0.580290,0.131104
0.580790,0.129750
0.581291,0.128609
0.581791,0.127733
0.582291,0.127165
0.582791,0.126943
0.583292,0.127094
0.583792,0.127637
0.584292,0.128581
0.584792,0.129923
0.585293,0.131652
0.585793,0.133748
0.586293,0.136178
0.586793,0.138904
0.587294,0.141880
0.587794,0.145053
0.588294,0.148363
0.588794,0.151751
0.589295,0.155151
0.589795,0.158501
0.590295,0.161737
0.590795,0.164799
0.591296,0.167633
0.591796,0.170187
0.592296,0.172419
0.592796,0.174295
0.593297,0.175787
0.593797,0.176881
0.594297,0.177570
0.594797,0.177858
0.595298,0.177760
0.595798,0.177299
0.596298,0.176509
0.596798,0.175431
0.597299,0.174115
0.597799,0.172615
0.598299,0.170991
0.598799,0.169306
0.599300,0.167624
0.599800,0.166011
0.600300,0.164529
0.600800,0.163238
0.601301,0.162193
0.601801,0.161443
0.602301,0.161028
0.602801,0.160983
0.603302,0.161331
0.603802,0.162085
0.604302,0.163250
0.604802,0.164818
0.605303,0.166773
0.605803,0.169088
0.606303,0.171726
0.606803,0.174644
0.607304,0.177791
0.607804,0.181108
0.608304,0.184536
0.608804,0.188010
0.609305,0.191463
0.609805,0.194831
0.610305,0.198051
0.610805,0.201063
0.611306,0.203814
0.611806,0.206254
0.612306,0.208346
0.612806,0.210056
0.613307,0.211364
0.613807,0.212259
0.614307,0.212739
0.614807,0.212814
0.615308,0.212504
0.615808,0.211838
0.616308,0.210855
0.616808,0.209601
0.617309,0.208131
0.617809,0.206503
0.618309,0.204781
0.618809,0.203030
0.619310,0.201317
0.619810,0.199708
0.620310,0.198265
0.620810,0.197049
0.621311,0.196112
0.621811,0.195501
0.622311,0.195254
0.622811,0.195401
0.623312,0.195960
0.623812,0.196942
0.624312,0.198343
0.624812,0.200152
0.625313,0.202346
0.625813,0.204892
0.626313,0.207750
0.626813,0.210870
0.627314,0.214195
0.627814,0.217665
0.628314,0.221215
0.628814,0.224777
0.629315,0.228283
0.629815,0.231669
0.630315,0.234869
0.630815,0.237826
0.631316,0.240487
0.631816,0.242807
0.632316,0.244748
0.632816,0.246284
0.633317,0.247397
0.633817,0.248082
0.634317,0.248343
0.634817,0.248195
0.635318,0.247664
0.635818,0.246785
0.636318,0.245602
0.636818,0.244168
0.637319,0.242540
0.637819,0.240783
0.638319,0.238963
0.638819,0.237148
0.639320,0.235408
0.639820,0.233810
0.640320,0.232415
0.640820,0.231284
0.641321,0.230467
0.641821,0.230009
0.642321,0.229945
0.642821,0.230299
0.643322,0.231087
0.643822,0.232311
0.644322,0.233966
0.644822,0.236031
0.645323,0.238479
0.645823,0.241272
0.646323,0.244362
0.646823,0.247694
0.647324,0.251207
0.647824,0.254836
0.648324,0.258512
0.648824,0.262166
0.649325,0.265726
0.649825,0.269126
0.650325,0.272303
0.650825,0.275199
0.651326,0.277762
0.651826,0.279950
0.652326,0.281731
0.652826,0.283080
0.653327,0.283987
0.653827,0.284449
0.654327,0.284478
0.654827,0.284095
0.655328,0.283332
0.655828,0.282229
0.656328,0.280839
0.656828,0.279217
0.657329,0.277427
0.657829,0.275539
0.658329,0.273621
0.658829,0.271746
0.659330,0.269985
0.659830,0.268405
0.660330,0.267069
0.660830,0.266035
0.661331,0.265354
0.661831,0.265066
0.662331,0.265202
0.662831,0.265783
0.663332,0.266819
0.663832,0.268307
0.664332,0.270235
0.664832,0.272577
0.665333,0.275299
0.665833,0.278355
0.666333,0.281693
0.666833,0.285251
0.667334,0.288964
0.667834,0.292762
0.668334,0.296571
0.668834,0.300318
0.669335,0.303933
0.669835,0.307347
0.670335,0.310496
0.670835,0.313323
0.671336,0.315780
0.671836,0.317826
0.672336,0.319433
0.672836,0.320582
0.673337,0.321267
0.673837,0.321492
0.674337,0.321275
0.674837,0.320643
0.675338,0.319635
0.675838,0.318298
0.676338,0.316690
0.676838,0.314874
0.677339,0.312919
0.677839,0.310898
0.678339,0.308885
0.678839,0.306955
0.679340,0.305181
0.679840,0.303631
0.680340,0.302369
0.680840,0.301451
0.681341,0.300925
0.681841,0.300828
0.682341,0.301189
0.682841,0.302023
0.683342,0.303334
0.683842,0.305113
0.684342,0.307340
0.684842,0.309985
0.685343,0.313004
0.685843,0.316347
0.686343,0.319953
0.686843,0.323756
0.687344,0.327683
0.687844,0.331660
0.688344,0.335610
0.688844,0.339457
0.689345,0.343127
0.689845,0.346550
0.690345,0.349664
0.690845,0.352413
0.691346,0.354749
0.691846,0.356638
0.692346,0.358053
0.692846,0.358983
0.693347,0.359425
0.693847,0.359391
0.694347,0.358906
0.694847,0.358005
0.695348,0.356732
0.695848,0.355144
0.696348,0.353303
0.696848,0.351281
0.697349,0.349151
0.697849,0.346991
0.698349,0.344881
0.698849,0.342899
0.699350,0.341118
0.699850,0.339608
0.700350,0.338434
0.700850,0.337650
0.701351,0.337301
0.701851,0.337421
0.702351,0.338034
0.702851,0.339150
0.703352,0.340766
0.703852,0.342867
0.704352,0.345425
0.704852,0.348403
0.705353,0.351750
0.705853,0.355407
0.706353,0.359306
0.706853,0.363375
0.707354,0.367536
0.707854,0.371707
0.708354,0.375808
0.708854,0.379760
0.709355,0.383487
0.709855,0.386918
0.710355,0.389990
0.710855,0.392649
0.711356,0.394851
0.711856,0.396564
0.712356,0.397767
0.712856,0.398454
0.713357,0.398630
0.713857,0.398313
0.714357,0.397536
0.714857,0.396341
0.715358,0.394782
0.715858,0.392921
0.716358,0.390832
0.716858,0.388589
0.717359,0.386275
0.717859,0.383972
0.718359,0.381765
0.718859,0.379734
0.719360,0.377956
0.719860,0.376503
0.720360,0.375436
0.720860,0.374810
0.721361,0.374666
0.721861,0.375035
0.722361,0.375935
0.722861,0.377370
0.723362,0.379330
0.723862,0.381793
0.724362,0.384722
0.724862,0.388072
0.725363,0.391783
0.725863,0.395788
0.726363,0.400013
0.726863,0.404375
0.727364,0.408792
0.727864,0.413175
0.728364,0.417440
0.728864,0.421504
0.729365,0.425289
0.729865,0.428722
0.730365,0.431743
0.730865,0.434297
0.731366,0.436345
0.731866,0.437858
0.732366,0.438822
0.732866,0.439237
0.733367,0.439114
0.733867,0.438482
0.734367,0.437379
0.734867,0.435859
0.735368,0.433984
0.735868,0.431825
0.736368,0.429463
0.736868,0.426981
0.737369,0.424469
0.737869,0.422016
0.738369,0.419710
0.738869,0.417635
0.739370,0.415872
0.739870,0.414491
0.740370,0.413556
0.740870,0.413118
0.741371,0.413215
0.741871,0.413873
0.742371,0.415105
0.742871,0.416906
0.743372,0.419259
0.743872,0.422134
0.744372,0.425485
0.744872,0.429256
0.745373,0.433380
0.745873,0.437778
0.746373,0.442368
0.746873,0.447060
0.747374,0.451762
0.747874,0.456382
0.748374,0.460828
0.748874,0.465014
0.749375,0.468859
0.749875,0.472291
0.750375,0.475247
0.750875,0.477679
0.751376,0.479548
0.751876,0.480833
0.752376,0.481523
0.752876,0.481627
0.753377,0.481167
0.753877,0.480177
0.754377,0.478709
0.754877,0.476823
0.755378,0.474594
0.755878,0.472102
0.756378,0.469437
0.756878,0.466693
0.757379,0.463965
0.757879,0.461350
0.758379,0.458942
0.758879,0.456829
0.759380,0.455093
0.759880,0.453807
0.760380,0.453033
0.760880,0.452819
0.761381,0.453200
0.761881,0.454197
0.762381,0.455814
0.762881,0.458040
0.763382,0.460848
0.763882,0.464198
0.764382,0.468034
0.764882,0.472288
0.765383,0.476882
0.765883,0.481729
0.766383,0.486733
0.766883,0.491798
0.767384,0.496822
0.767884,0.501705
0.768384,0.506351
0.768884,0.510669
0.769385,0.514575
0.769885,0.517997
0.770385,0.520873
0.770885,0.523156
0.771386,0.524813
0.771886,0.525828
0.772386,0.526199
0.772886,0.525943
0.773387,0.525091
0.773887,0.523690
0.774387,0.521801
0.774887,0.519498
0.775388,0.516865
0.775888,0.513996
0.776388,0.510990
0.776888,0.507952
0.777389,0.504987
0.777889,0.502198
0.778389,0.499685
0.778889,0.497542
0.779390,0.495854
0.779890,0.494693
0.780390,0.494120
0.780890,0.494182
0.781391,0.494907
0.781891,0.496310
0.782391,0.498388
0.782891,0.501118
0.783392,0.504465
0.783892,0.508375
0.784392,0.512781
0.784892,0.517602
0.785393,0.522747
0.785893,0.528117
0.786393,0.533604
0.786893,0.539101
0.787394,0.544496
0.787894,0.549682
0.788394,0.554556
0.788894,0.559020
0.789395,0.562990
0.789895,0.566392
0.790395,0.569166
0.790895,0.571268
0.791396,0.572670
0.791896,0.573363
0.792396,0.573355
0.792896,0.572673
0.793397,0.571359
0.793897,0.569474
0.794397,0.567092
0.794897,0.564301
0.795398,0.561199
0.795898,0.557893
0.796398,0.554496
0.796898,0.551123
0.797399,0.547889
0.797899,0.544909
0.798399,0.542288
0.798899,0.540126
0.799400,0.538509
0.799900,0.537512
0.800400,0.537195
0.800900,0.537598
0.801401,0.538747
0.801901,0.540646
0.802401,0.543283
0.802901,0.546625
0.803402,0.550621
0.803902,0.555205
0.804402,0.560294
0.804902,0.565794
0.805403,0.571597
0.805903,0.577589
0.806403,0.583651
0.806903,0.589659
0.807404,0.595492
0.807904,0.601033
0.808404,0.606170
0.808904,0.610802
0.809405,0.614840
0.809905,0.618209
0.810405,0.620852
0.810905,0.622728
0.811406,0.623817
0.811906,0.624118
0.812406,0.623650
0.812906,0.622453
0.813407,0.620585
0.813907,0.618120
0.814407,0.615149
0.814907,0.611777
0.815408,0.608118
0.815908,0.604297
0.816408,0.600440
0.816908,0.596678
0.817409,0.593138
0.817909,0.589946
0.818409,0.587215
0.818909,0.585050
0.819410,0.583543
0.819910,0.582768
0.820410,0.582782
0.820910,0.583623
0.821411,0.585307
0.821911,0.587829
0.822411,0.591164
0.822911,0.595266
0.823412,0.600067
0.823912,0.605484
0.824412,0.611417
0.824912,0.617751
0.825413,0.624363
0.825913,0.631118
0.826413,0.637881
0.826913,0.644512
0.827414,0.650878
0.827914,0.656847
0.828414,0.662300
0.828914,0.667129
0.829415,0.671241
0.829915,0.674560
0.830415,0.677032
0.830915,0.678622
0.831416,0.679318
0.831916,0.679130
0.832416,0.678092
0.832916,0.676258
0.833417,0.673705
0.833917,0.670526
0.834417,0.666834
0.834917,0.662752
0.835418,0.658417
0.835918,0.653972
0.836418,0.649564
0.836918,0.645340
0.837419,0.641445
0.837919,0.638014
0.838419,0.635174
0.838919,0.633036
0.839420,0.631696
0.839920,0.631229
0.840420,0.631688
0.840920,0.633105
0.841421,0.635488
0.841921,0.638817
0.842421,0.643053
0.842921,0.648131
0.843422,0.653963
0.843922,0.660445
0.844422,0.667452
0.844922,0.674848
0.845423,0.682485
0.845923,0.690207
0.846423,0.697856
0.846923,0.705273
0.847424,0.712307
0.847924,0.718812
0.848424,0.724657
0.848924,0.729725
0.849425,0.733920
0.849925,0.737166
0.850425,0.739410
0.850925,0.740626
0.851426,0.740813
0.851926,0.739995
0.852426,0.738225
0.852926,0.735577
0.853427,0.732150
0.853927,0.728064
0.854427,0.723456
0.854927,0.718478
0.855428,0.713292
0.855928,0.708066
0.856428,0.702974
0.856928,0.698184
0.857429,0.693860
0.857929,0.690156
0.858429,0.687210
0.858929,0.685145
0.859430,0.684058
0.859930,0.684028
0.860430,0.685103
0.860930,0.687306
0.861431,0.690631
0.861931,0.695045
0.862431,0.700485
0.862931,0.706862
0.863432,0.714064
0.863932,0.721957
0.864432,0.730386
0.864932,0.739185
0.865433,0.748174
0.865933,0.757167
0.866433,0.765980
0.866933,0.774427
0.867434,0.782333
0.867934,0.789534
0.868434,0.795884
0.868934,0.801255
0.869435,0.805546
0.869935,0.808679
0.870435,0.810607
0.870935,0.811314
0.871436,0.810812
0.871936,0.809149
0.872436,0.806397
0.872936,0.802661
0.873437,0.798070
0.873937,0.792778
0.874437,0.786956
0.874937,0.780791
0.875438,0.774483
0.875938,0.768234
0.876438,0.762250
0.876938,0.756731
0.877439,0.751867
0.877939,0.747833
0.878439,0.744787
0.878939,0.742862
0.879440,0.742165
0.879940,0.742771
0.880440,0.744726
0.880940,0.748040
0.881441,0.752690
0.881941,0.758619
0.882441,0.765736
0.882941,0.773920
0.883442,0.783021
0.883942,0.792866
0.884442,0.803261
0.884942,0.813995
0.885443,0.824846
0.885943,0.835589
0.886443,0.845998
0.886943,0.855853
0.887444,0.864948
0.887944,0.873091
0.888444,0.880116
0.888944,0.885879
0.889445,0.890272
0.889945,0.893217
0.890445,0.894673
0.890945,0.894638
0.891446,0.893146
0.891946,0.890270
0.892446,0.886120
0.892946,0.880837
0.893447,0.874595
0.893947,0.867595
0.894447,0.860057
0.894947,0.852221
0.895448,0.844335
0.895948,0.836653
0.896448,0.829426
0.896948,0.822898
0.897449,0.817298
0.897949,0.812834
0.898449,0.809687
0.898949,0.808010
0.899450,0.807918
0.899950,0.809489
0.900450,0.812759
0.900950,0.817722
0.901451,0.824330
0.901951,0.832491
0.902451,0.842075
0.902951,0.852912
0.903452,0.864800
0.903952,0.877507
0.904452,0.890778
0.904952,0.904342
0.905453,0.917913
0.905953,0.931206
0.906453,0.943937
0.906953,0.955834
0.907454,0.966641
0.907954,0.976129
0.908454,0.984097
0.908954,0.990383
0.909455,0.994864
0.909955,0.997461
0.910455,0.998145
0.910955,0.996932
0.911456,0.993889
0.911956,0.989130
0.912456,0.982814
0.912956,0.975145
0.913457,0.966361
0.913957,0.956736
0.914457,0.946565
0.914957,0.936167
0.915458,0.925867
0.915958,0.915997
0.916458,0.906881
0.916958,0.898830
0.917459,0.892133
0.917959,0.887048
0.918459,0.883800
0.918959,0.882567
0.919460,0.883483
0.919960,0.886627
0.920460,0.892024
0.920960,0.899643
0.921461,0.909396
0.921961,0.921140
0.922461,0.934679
0.922961,0.949769
0.923462,0.966124
0.923962,0.983420
0.924462,1.001305
0.924962,1.019405
0.925463,1.037337
0.925963,1.054715
0.926463,1.071160
0.926963,1.086313
0.927464,1.099839
0.927964,1.111442
0.928464,1.120868
0.928964,1.127915
0.929465,1.132440
0.929965,1.134359
0.930465,1.133657
0.930965,1.130382
0.931466,1.124652
0.931966,1.116649
0.932466,1.106618
0.932966,1.094860
0.933467,1.081727
0.933967,1.067612
0.934467,1.052945
0.934967,1.038175
0.935468,1.023768
0.935968,1.010186
0.936468,0.997884
0.936968,0.987294
0.937469,0.978812
0.937969,0.972793
0.938469,0.969536
0.938969,0.969275
0.939470,0.972178
0.939970,0.978334
0.940470,0.987753
0.940970,1.000364
0.941471,1.016013
0.941971,1.034468
0.942471,1.055419
0.942971,1.078489
0.943472,1.103239
0.943972,1.129176
0.944472,1.155769
0.944972,1.182458
0.945473,1.208667
0.945973,1.233822
0.946473,1.257359
0.946973,1.278748
0.947474,1.297496
0.947974,1.313169
0.948474,1.325401
0.948974,1.333905
0.949475,1.338483
0.949975,1.339032
0.950475,1.335552
0.950975,1.328145
0.951476,1.317019
0.951976,1.302482
0.952476,1.284939
0.952976,1.264887
0.953477,1.242900
0.953977,1.219621
0.954477,1.195747
0.954977,1.172011
0.955478,1.149168
0.955978,1.127977
0.956478,1.109178
0.956978,1.093476
0.957479,1.081525
0.957979,1.073904
0.958479,1.071107
0.958979,1.073521
0.959480,1.081420
0.959980,1.094950
0.960480,1.114122
0.960980,1.138809
0.961481,1.168741
0.961981,1.203511
0.962481,1.242577
0.962981,1.285273
0.963482,1.330818
0.963982,1.378336
0.964482,1.426868
0.964982,1.475395
0.965483,1.522866
0.965983,1.568211
0.966483,1.610379
0.966983,1.648355
0.967484,1.681192
0.967984,1.708033
0.968484,1.728139
0.968984,1.740912
0.969485,1.745910
0.969985,1.742873
0.970485,1.731733
0.970985,1.712623
0.971486,1.685889
0.971986,1.652087
0.972486,1.611984
0.972986,1.566547
0.973487,1.516936
0.973987,1.464482
0.974487,1.410670
0.974987,1.357110
0.975488,1.305511
0.975988,1.257648
0.976488,1.215324
0.976988,1.180336
0.977489,1.154436
0.977989,1.139288
0.978489,1.136435
0.978989,1.147256
0.979490,1.172932
0.979990,1.214409
0.980490,1.272374
0.980990,1.347219
0.981491,1.439029
0.981991,1.547557
0.982491,1.672218
0.982991,1.812084
0.983492,1.965881
0.983992,2.132003
0.984492,2.308522
0.984992,2.493211
0.985493,2.683570
0.985993,2.876858
0.986493,3.070136
0.986993,3.260303
0.987494,3.444148
0.987994,3.618398
0.988494,3.779773
0.988994,3.925036
0.989495,4.051054
0.989995,4.154846
0.990495,4.233640
0.990995,4.284924
0.991496,4.306488
0.991996,4.296474
0.992496,4.253408
0.992996,4.176236
0.993497,4.064345
0.993997,3.917587
0.994497,3.736285
0.994997,3.521240
0.995498,3.273724
0.995998,2.995468
0.996498,2.688644
0.996998,2.355833
0.997499,1.999995
0.997999,1.624427
0.998499,1.232713
0.998999,0.828676
0.999500,0.416323
1.000000,-0.000220
};
\end{axis}
\end{tikzpicture}

%% file: Fig_p=1_q=1_dense_A=200.tex
\centering
\begin{tikzpicture}[]
\begin{axis}[ymin=-6.2, ymax=6.2, xmin=-0.05, xmax=1.05,
  ytick={-6,-4,...,6}, ytick pos=left,
  xtick={0,0.2,...,1}, xtick pos=left,
  xlabel={$x/L$},
  ylabel={$\extPotOptim(x)/D$},
  legend pos=north west,
  legend style={draw=none}]
\addplot+[
  line width=2pt,
  black, smooth,mark=empty,
] table [x=x, y=y, col sep=comma] {
    x,  y,      
0.000000,-0.102270
0.000500,-2.245302
0.001000,-4.038426
0.001500,-5.228009
0.002000,-5.695558
0.002500,-5.477403
0.003000,-4.744975
0.003500,-3.752665
0.004000,-2.768957
0.004500,-2.010328
0.005000,-1.595496
0.005500,-1.530623
0.006000,-1.726537
0.006500,-2.039759
0.007000,-2.323000
0.007500,-2.469382
0.008000,-2.438105
0.008500,-2.256150
0.009000,-1.998526
0.009500,-1.755947
0.010000,-1.601830
0.010500,-1.569399
0.011000,-1.645243
0.011500,-1.779582
0.012000,-1.907907
0.012500,-1.975387
0.013000,-1.955333
0.013500,-1.855906
0.014000,-1.713829
0.014500,-1.578510
0.015000,-1.493034
0.015500,-1.479049
0.016000,-1.530588
0.016500,-1.618353
0.017000,-1.702098
0.017500,-1.746071
0.018000,-1.731708
0.018500,-1.663208
0.019000,-1.564452
0.019500,-1.468967
0.020000,-1.407036
0.020500,-1.394819
0.021000,-1.429335
0.021500,-1.490744
0.022000,-1.550676
0.022500,-1.583182
0.023000,-1.574141
0.023500,-1.525772
0.024000,-1.454851
0.024500,-1.385664
0.025000,-1.340566
0.025500,-1.331769
0.026000,-1.357330
0.026500,-1.402606
0.027000,-1.446356
0.027500,-1.469006
0.028000,-1.459887
0.028500,-1.420809
0.029000,-1.364815
0.029500,-1.310797
0.030000,-1.276162
0.030500,-1.270379
0.031000,-1.291767
0.031500,-1.328592
0.032000,-1.363872
0.032500,-1.381949
0.033000,-1.374294
0.033500,-1.342401
0.034000,-1.296817
0.034500,-1.252814
0.035000,-1.224463
0.035500,-1.219402
0.036000,-1.236218
0.036500,-1.265350
0.037000,-1.293012
0.037500,-1.306568
0.038000,-1.299269
0.038500,-1.272591
0.039000,-1.235368
0.039500,-1.200179
0.040000,-1.178422
0.040500,-1.176006
0.041000,-1.191270
0.041500,-1.215838
0.042000,-1.238015
0.042500,-1.247348
0.043000,-1.238589
0.043500,-1.213572
0.044000,-1.180352
0.044500,-1.149984
0.045000,-1.132250
0.045500,-1.131961
0.046000,-1.147224
0.046500,-1.170299
0.047000,-1.190647
0.047500,-1.198968
0.048000,-1.190692
0.048500,-1.167600
0.049000,-1.137021
0.049500,-1.108952
0.050000,-1.092246
0.050500,-1.091313
0.051000,-1.104573
0.051500,-1.125201
0.052000,-1.143815
0.052500,-1.152065
0.053000,-1.145735
0.053500,-1.126208
0.054000,-1.099792
0.054500,-1.075199
0.055000,-1.060197
0.055500,-1.058709
0.056000,-1.069436
0.056500,-1.086507
0.057000,-1.101831
0.057500,-1.108237
0.058000,-1.102175
0.058500,-1.084974
0.059000,-1.062212
0.059500,-1.041481
0.060000,-1.029428
0.060500,-1.029239
0.061000,-1.039483
0.061500,-1.054753
0.062000,-1.067815
0.062500,-1.072405
0.063000,-1.065613
0.063500,-1.048937
0.064000,-1.027637
0.064500,-1.008669
0.065000,-0.998004
0.065500,-0.998371
0.066000,-1.008272
0.066500,-1.022626
0.067000,-1.034756
0.067500,-1.038959
0.068000,-1.032655
0.068500,-1.017309
0.069000,-0.997805
0.069500,-0.980526
0.070000,-0.970901
0.070500,-0.971356
0.071000,-0.980443
0.071500,-0.993464
0.072000,-1.004309
0.072500,-1.007814
0.073000,-1.001705
0.073500,-0.987421
0.074000,-0.969502
0.074500,-0.953806
0.075000,-0.945261
0.075500,-0.945996
0.076000,-0.954574
0.076500,-0.966595
0.077000,-0.976416
0.077500,-0.979314
0.078000,-0.973275
0.078500,-0.959717
0.079000,-0.942904
0.079500,-0.928279
0.080000,-0.920385
0.080500,-0.921153
0.081000,-0.929209
0.081500,-0.940451
0.082000,-0.949644
0.082500,-0.952428
0.083000,-0.946950
0.083500,-0.934523
0.084000,-0.919068
0.084500,-0.905578
0.085000,-0.898200
0.085500,-0.898679
0.086000,-0.905739
0.086500,-0.915633
0.087000,-0.923617
0.087500,-0.925787
0.088000,-0.920550
0.088500,-0.909200
0.089000,-0.895360
0.089500,-0.883555
0.090000,-0.877453
0.090500,-0.878464
0.091000,-0.885227
0.091500,-0.894173
0.092000,-0.900924
0.092500,-0.901999
0.093000,-0.896145
0.093500,-0.884807
0.094000,-0.871551
0.094500,-0.860677
0.095000,-0.855560
0.095500,-0.857358
0.096000,-0.864580
0.096500,-0.873664
0.097000,-0.880352
0.097500,-0.881312
0.098000,-0.875400
0.098500,-0.864085
0.099000,-0.850867
0.099500,-0.839943
0.100000,-0.834613
0.100500,-0.836047
0.101000,-0.842868
0.101500,-0.851706
0.102000,-0.858490
0.102500,-0.859995
0.103000,-0.855037
0.103500,-0.844883
0.104000,-0.832728
0.104500,-0.822447
0.105000,-0.817119
0.105500,-0.817877
0.106000,-0.823522
0.106500,-0.831026
0.107000,-0.836733
0.107500,-0.837765
0.108000,-0.833110
0.108500,-0.823977
0.109000,-0.813282
0.109500,-0.804496
0.110000,-0.800293
0.110500,-0.801518
0.111000,-0.806888
0.111500,-0.813514
0.112000,-0.818036
0.112500,-0.817941
0.113000,-0.812531
0.113500,-0.803192
0.114000,-0.792849
0.114500,-0.784826
0.115000,-0.781563
0.115500,-0.783669
0.116000,-0.789692
0.116500,-0.796667
0.117000,-0.801261
0.117500,-0.801042
0.118000,-0.795411
0.118500,-0.785822
0.119000,-0.775230
0.119500,-0.766965
0.120000,-0.763469
0.120500,-0.765384
0.121000,-0.771319
0.121500,-0.778396
0.122000,-0.783345
0.122500,-0.783739
0.123000,-0.778897
0.123500,-0.770109
0.124000,-0.760124
0.124500,-0.752083
0.125000,-0.748327
0.125500,-0.749526
0.126000,-0.754461
0.126500,-0.760517
0.127000,-0.764715
0.127500,-0.764854
0.128000,-0.760341
0.128500,-0.752393
0.129000,-0.743537
0.129500,-0.736613
0.130000,-0.733675
0.130500,-0.735202
0.131000,-0.739924
0.131500,-0.745337
0.132000,-0.748675
0.132500,-0.747991
0.133000,-0.742906
0.133500,-0.734753
0.134000,-0.726056
0.134500,-0.719559
0.135000,-0.717165
0.135500,-0.719208
0.136000,-0.724327
0.136500,-0.729990
0.137000,-0.733465
0.137500,-0.732859
0.138000,-0.727838
0.138500,-0.719745
0.139000,-0.711078
0.139500,-0.704534
0.140000,-0.701983
0.140500,-0.703766
0.141000,-0.708582
0.141500,-0.713994
0.142000,-0.717375
0.142500,-0.716902
0.143000,-0.712244
0.143500,-0.704667
0.144000,-0.696538
0.144500,-0.690401
0.145000,-0.688009
0.145500,-0.689661
0.146000,-0.694105
0.146500,-0.699035
0.147000,-0.701989
0.147500,-0.701290
0.148000,-0.696690
0.148500,-0.689443
0.149000,-0.681817
0.149500,-0.676207
0.150000,-0.674215
0.150500,-0.676042
0.151000,-0.680415
0.151500,-0.685091
0.152000,-0.687722
0.152500,-0.686760
0.153000,-0.682047
0.153500,-0.674866
0.154000,-0.667442
0.154500,-0.662082
0.155000,-0.660290
0.155500,-0.662194
0.156000,-0.666507
0.156500,-0.671025
0.157000,-0.673480
0.157500,-0.672408
0.158000,-0.667705
0.158500,-0.660660
0.159000,-0.653454
0.159500,-0.648319
0.160000,-0.646685
0.160500,-0.648637
0.161000,-0.652890
0.161500,-0.657287
0.162000,-0.659633
0.162500,-0.658528
0.163000,-0.653901
0.163500,-0.647025
0.164000,-0.640022
0.164500,-0.635048
0.165000,-0.633467
0.165500,-0.635337
0.166000,-0.639400
0.166500,-0.643571
0.167000,-0.645744
0.167500,-0.644599
0.168000,-0.640098
0.168500,-0.633493
0.169000,-0.626835
0.169500,-0.622185
0.170000,-0.620822
0.170500,-0.622760
0.171000,-0.626749
0.171500,-0.630764
0.172000,-0.632783
0.172500,-0.631559
0.173000,-0.627093
0.173500,-0.620622
0.174000,-0.614139
0.174500,-0.609623
0.175000,-0.608284
0.175500,-0.610109
0.176000,-0.613874
0.176500,-0.617631
0.177000,-0.619455
0.177500,-0.618186
0.178000,-0.613865
0.178500,-0.607711
0.179000,-0.601636
0.179500,-0.597508
0.180000,-0.596429
0.180500,-0.598315
0.181000,-0.601933
0.181500,-0.605388
0.182000,-0.606851
0.182500,-0.605273
0.183000,-0.600781
0.183500,-0.594633
0.184000,-0.588727
0.184500,-0.584871
0.185000,-0.584084
0.185500,-0.586205
0.186000,-0.589954
0.186500,-0.593418
0.187000,-0.594791
0.187500,-0.593059
0.188000,-0.588391
0.188500,-0.582071
0.189000,-0.576015
0.189500,-0.572039
0.190000,-0.571170
0.190500,-0.573265
0.191000,-0.577070
0.191500,-0.580701
0.192000,-0.582363
0.192500,-0.581022
0.193000,-0.576795
0.193500,-0.570882
0.194000,-0.565103
0.194500,-0.561195
0.195000,-0.560154
0.195500,-0.561865
0.196000,-0.565164
0.196500,-0.568296
0.197000,-0.569601
0.197500,-0.568146
0.198000,-0.564080
0.198500,-0.558563
0.199000,-0.553308
0.199500,-0.549911
0.200000,-0.549232
0.200500,-0.551069
0.201000,-0.554236
0.201500,-0.557032
0.202000,-0.557898
0.202500,-0.556027
0.203000,-0.551677
0.203500,-0.546069
0.204000,-0.540924
0.204500,-0.537790
0.205000,-0.537453
0.205500,-0.539624
0.206000,-0.543054
0.206500,-0.546002
0.207000,-0.546906
0.207500,-0.544978
0.208000,-0.540505
0.208500,-0.534743
0.209000,-0.529436
0.209500,-0.526160
0.210000,-0.525720
0.210500,-0.527856
0.211000,-0.531339
0.211500,-0.534445
0.212000,-0.535606
0.212500,-0.534000
0.213000,-0.529852
0.213500,-0.524335
0.214000,-0.519114
0.214500,-0.515711
0.215000,-0.514930
0.215500,-0.516567
0.216000,-0.519506
0.216500,-0.522160
0.217000,-0.523090
0.217500,-0.521555
0.218000,-0.517788
0.218500,-0.512885
0.219000,-0.508371
0.219500,-0.505597
0.220000,-0.505211
0.220500,-0.506913
0.221000,-0.509572
0.221500,-0.511673
0.222000,-0.511918
0.222500,-0.509732
0.223000,-0.505497
0.223500,-0.500410
0.224000,-0.496021
0.224500,-0.493635
0.225000,-0.493800
0.225500,-0.496090
0.226000,-0.499255
0.226500,-0.501696
0.227000,-0.502074
0.227500,-0.499820
0.228000,-0.495361
0.228500,-0.489961
0.229000,-0.485247
0.229500,-0.482597
0.230000,-0.482616
0.230500,-0.484915
0.231000,-0.488256
0.231500,-0.491017
0.232000,-0.491807
0.232500,-0.489977
0.233000,-0.485857
0.233500,-0.480619
0.234000,-0.475821
0.234500,-0.472819
0.235000,-0.472261
0.235500,-0.473857
0.236000,-0.476512
0.236500,-0.478759
0.237000,-0.479335
0.237500,-0.477660
0.238000,-0.474046
0.238500,-0.469568
0.239000,-0.465622
0.239500,-0.463381
0.240000,-0.463331
0.240500,-0.465081
0.241000,-0.467521
0.241500,-0.469256
0.242000,-0.469156
0.242500,-0.466803
0.243000,-0.462660
0.243500,-0.457898
0.244000,-0.453949
0.244500,-0.451950
0.245000,-0.452300
0.245500,-0.454496
0.246000,-0.457323
0.246500,-0.459306
0.247000,-0.459279
0.247500,-0.456827
0.248000,-0.452453
0.248500,-0.447391
0.249000,-0.443144
0.249500,-0.440915
0.250000,-0.441157
0.250500,-0.443404
0.251000,-0.446446
0.251500,-0.448788
0.252000,-0.449211
0.252500,-0.447223
0.253000,-0.443237
0.253500,-0.438399
0.254000,-0.434148
0.254500,-0.431669
0.255000,-0.431451
0.255500,-0.433119
0.256000,-0.435591
0.256500,-0.437508
0.257000,-0.437763
0.257500,-0.435918
0.258000,-0.432364
0.258500,-0.428153
0.259000,-0.424576
0.259500,-0.422657
0.260000,-0.422748
0.260500,-0.424397
0.261000,-0.426531
0.261500,-0.427877
0.262000,-0.427469
0.262500,-0.425030
0.263000,-0.421091
0.263500,-0.416786
0.264000,-0.413418
0.264500,-0.411949
0.265000,-0.412620
0.265500,-0.414845
0.266000,-0.417430
0.266500,-0.419018
0.267000,-0.418612
0.267500,-0.415957
0.268000,-0.411648
0.268500,-0.406907
0.269000,-0.403128
0.269500,-0.401349
0.270000,-0.401864
0.270500,-0.404108
0.271000,-0.406872
0.271500,-0.408756
0.272000,-0.408699
0.272500,-0.406373
0.273000,-0.402298
0.273500,-0.397641
0.274000,-0.393769
0.274500,-0.391736
0.275000,-0.391886
0.275500,-0.393738
0.276000,-0.396182
0.276500,-0.397909
0.277000,-0.397912
0.277500,-0.395869
0.278000,-0.392250
0.278500,-0.388126
0.279000,-0.384744
0.279500,-0.383042
0.280000,-0.383287
0.280500,-0.384979
0.281000,-0.387049
0.281500,-0.388286
0.282000,-0.387805
0.282500,-0.385404
0.283000,-0.381637
0.283500,-0.377603
0.284000,-0.374521
0.284500,-0.373249
0.285000,-0.373949
0.285500,-0.376016
0.286000,-0.378305
0.286500,-0.379567
0.287000,-0.378928
0.287500,-0.376232
0.288000,-0.372105
0.288500,-0.367724
0.289000,-0.364374
0.289500,-0.362958
0.290000,-0.363658
0.290500,-0.365862
0.291000,-0.368393
0.291500,-0.369956
0.292000,-0.369619
0.292500,-0.367165
0.293000,-0.363170
0.293500,-0.358775
0.294000,-0.355258
0.294500,-0.353551
0.295000,-0.353893
0.295500,-0.355758
0.296000,-0.358057
0.296500,-0.359569
0.297000,-0.359402
0.297500,-0.357327
0.298000,-0.353848
0.298500,-0.350000
0.299000,-0.346933
0.299500,-0.345463
0.300000,-0.345761
0.300500,-0.347293
0.301000,-0.349034
0.301500,-0.349882
0.302000,-0.349090
0.302500,-0.346566
0.303000,-0.342913
0.303500,-0.339200
0.304000,-0.336540
0.304500,-0.335657
0.305000,-0.336594
0.305500,-0.338686
0.306000,-0.340807
0.306500,-0.341794
0.307000,-0.340899
0.307500,-0.338073
0.308000,-0.333999
0.308500,-0.329836
0.309000,-0.326784
0.309500,-0.325636
0.310000,-0.326472
0.310500,-0.328633
0.311000,-0.330971
0.311500,-0.332276
0.312000,-0.331736
0.312500,-0.329235
0.313000,-0.325394
0.313500,-0.321323
0.314000,-0.318204
0.314500,-0.316840
0.315000,-0.317359
0.315500,-0.319173
0.316000,-0.321215
0.316500,-0.322352
0.317000,-0.321818
0.317500,-0.319503
0.318000,-0.315985
0.318500,-0.312297
0.319000,-0.309521
0.319500,-0.308365
0.320000,-0.308893
0.320500,-0.310503
0.321000,-0.312166
0.321500,-0.312830
0.322000,-0.311836
0.322500,-0.309180
0.323000,-0.305514
0.323500,-0.301902
0.324000,-0.299405
0.324500,-0.298668
0.325000,-0.299666
0.325500,-0.301705
0.326000,-0.303679
0.326500,-0.304495
0.327000,-0.303488
0.327500,-0.300677
0.328000,-0.296764
0.328500,-0.292872
0.329000,-0.290115
0.329500,-0.289185
0.330000,-0.290084
0.330500,-0.292128
0.331000,-0.294207
0.331500,-0.295203
0.332000,-0.294416
0.332500,-0.291823
0.333000,-0.288082
0.333500,-0.284277
0.334000,-0.281502
0.334500,-0.280448
0.335000,-0.281146
0.335500,-0.282959
0.336000,-0.284837
0.336500,-0.285721
0.337000,-0.284948
0.337500,-0.282501
0.338000,-0.279010
0.338500,-0.275499
0.339000,-0.272990
0.339500,-0.272102
0.340000,-0.272820
0.340500,-0.274498
0.341000,-0.276118
0.341500,-0.276683
0.342000,-0.275610
0.342500,-0.272957
0.343000,-0.269400
0.343500,-0.265975
0.344000,-0.263676
0.344500,-0.263070
0.345000,-0.264070
0.345500,-0.265968
0.346000,-0.267700
0.346500,-0.268256
0.347000,-0.267062
0.347500,-0.264209
0.348000,-0.260418
0.348500,-0.256769
0.349000,-0.254292
0.349500,-0.253579
0.350000,-0.254555
0.350500,-0.256512
0.351000,-0.258374
0.351500,-0.259108
0.352000,-0.258112
0.352500,-0.255441
0.353000,-0.251780
0.353500,-0.248177
0.354000,-0.245647
0.354500,-0.244784
0.355000,-0.245542
0.355500,-0.247263
0.356000,-0.248933
0.356500,-0.249575
0.357000,-0.248626
0.357500,-0.246147
0.358000,-0.242789
0.358500,-0.239539
0.359000,-0.237330
0.359500,-0.236679
0.360000,-0.237488
0.360500,-0.239084
0.361000,-0.240483
0.361500,-0.240774
0.362000,-0.239476
0.362500,-0.236729
0.363000,-0.233242
0.363500,-0.230020
0.364000,-0.227978
0.364500,-0.227586
0.365000,-0.228682
0.365500,-0.230535
0.366000,-0.232118
0.366500,-0.232503
0.367000,-0.231218
0.367500,-0.228423
0.368000,-0.224857
0.368500,-0.221551
0.369000,-0.219435
0.369500,-0.218986
0.370000,-0.220040
0.370500,-0.221861
0.371000,-0.223419
0.371500,-0.223785
0.372000,-0.222483
0.372500,-0.219674
0.373000,-0.216088
0.373500,-0.212746
0.374000,-0.210568
0.374500,-0.210024
0.375000,-0.210956
0.375500,-0.212648
0.376000,-0.214097
0.376500,-0.214407
0.377000,-0.213131
0.377500,-0.210436
0.378000,-0.207043
0.378500,-0.203937
0.379000,-0.201989
0.379500,-0.201620
0.380000,-0.202635
0.380500,-0.204301
0.381000,-0.205632
0.381500,-0.205772
0.382000,-0.204325
0.382500,-0.201514
0.383000,-0.198092
0.383500,-0.195055
0.384000,-0.193257
0.384500,-0.193075
0.385000,-0.194267
0.385500,-0.196053
0.386000,-0.197420
0.386500,-0.197505
0.387000,-0.195931
0.387500,-0.192952
0.388000,-0.189362
0.388500,-0.186192
0.389000,-0.184319
0.389500,-0.184129
0.390000,-0.185373
0.390500,-0.187256
0.391000,-0.188741
0.391500,-0.188942
0.392000,-0.187461
0.392500,-0.184537
0.393000,-0.180953
0.393500,-0.177738
0.394000,-0.175775
0.394500,-0.175468
0.395000,-0.176588
0.395500,-0.178370
0.396000,-0.179801
0.396500,-0.180015
0.397000,-0.178613
0.397500,-0.175820
0.398000,-0.172386
0.398500,-0.169300
0.399000,-0.167406
0.399500,-0.167080
0.400000,-0.168091
0.400500,-0.169696
0.401000,-0.170928
0.401500,-0.170973
0.402000,-0.169485
0.402500,-0.166715
0.403000,-0.163417
0.403500,-0.160548
0.404000,-0.158899
0.404500,-0.158784
0.405000,-0.159919
0.405500,-0.161527
0.406000,-0.162642
0.406500,-0.162478
0.407000,-0.160738
0.407500,-0.157733
0.408000,-0.154266
0.408500,-0.151327
0.409000,-0.149715
0.409500,-0.149730
0.410000,-0.151056
0.410500,-0.152882
0.411000,-0.154201
0.411500,-0.154202
0.412000,-0.152571
0.412500,-0.149610
0.413000,-0.146124
0.413500,-0.143111
0.414000,-0.141383
0.414500,-0.141259
0.415000,-0.142444
0.415500,-0.144152
0.416000,-0.145400
0.416500,-0.145392
0.417000,-0.143815
0.417500,-0.140961
0.418000,-0.137606
0.418500,-0.134711
0.419000,-0.133051
0.419500,-0.132920
0.420000,-0.134020
0.420500,-0.135579
0.421000,-0.136655
0.421500,-0.136498
0.422000,-0.134842
0.422500,-0.132005
0.423000,-0.128768
0.423500,-0.126067
0.424000,-0.124633
0.424500,-0.124708
0.425000,-0.125948
0.425500,-0.127556
0.426000,-0.128589
0.426500,-0.128318
0.427000,-0.126516
0.427500,-0.123541
0.428000,-0.120203
0.428500,-0.117454
0.429000,-0.116023
0.429500,-0.116133
0.430000,-0.117419
0.430500,-0.119064
0.431000,-0.120113
0.431500,-0.119839
0.432000,-0.118021
0.432500,-0.115026
0.433000,-0.111670
0.433500,-0.108905
0.434000,-0.107453
0.434500,-0.107529
0.435000,-0.108765
0.435500,-0.110347
0.436000,-0.111338
0.436500,-0.111029
0.437000,-0.109220
0.437500,-0.106290
0.438000,-0.103055
0.438500,-0.100447
0.439000,-0.099158
0.439500,-0.099369
0.440000,-0.100682
0.440500,-0.102272
0.441000,-0.103202
0.441500,-0.102786
0.442000,-0.100853
0.442500,-0.097814
0.443000,-0.094506
0.443500,-0.091871
0.444000,-0.090594
0.444500,-0.090840
0.445000,-0.092191
0.445500,-0.093807
0.446000,-0.094746
0.446500,-0.094328
0.447000,-0.092395
0.447500,-0.089372
0.448000,-0.086105
0.448500,-0.083534
0.449000,-0.082333
0.449500,-0.082648
0.450000,-0.084043
0.450500,-0.085661
0.451000,-0.086555
0.451500,-0.086051
0.452000,-0.084004
0.452500,-0.080857
0.453000,-0.077474
0.453500,-0.074811
0.454000,-0.073549
0.454500,-0.073842
0.455000,-0.075254
0.455500,-0.076928
0.456000,-0.077914
0.456500,-0.077531
0.457000,-0.075620
0.457500,-0.072605
0.458000,-0.069324
0.458500,-0.066706
0.459000,-0.065414
0.459500,-0.065596
0.460000,-0.066828
0.460500,-0.068285
0.461000,-0.069060
0.461500,-0.068519
0.462000,-0.066542
0.462500,-0.063569
0.463000,-0.060429
0.463500,-0.058020
0.464000,-0.056956
0.464500,-0.057332
0.465000,-0.058684
0.465500,-0.060168
0.466000,-0.060884
0.466500,-0.060226
0.467000,-0.058113
0.467500,-0.055027
0.468000,-0.051824
0.468500,-0.049409
0.469000,-0.048386
0.469500,-0.048825
0.470000,-0.050233
0.470500,-0.051745
0.471000,-0.052452
0.471500,-0.051753
0.472000,-0.049587
0.472500,-0.046455
0.473000,-0.043233
0.473500,-0.040836
0.474000,-0.039866
0.474500,-0.040383
0.475000,-0.041881
0.475500,-0.043480
0.476000,-0.044263
0.476500,-0.043625
0.477000,-0.041501
0.477500,-0.038393
0.478000,-0.035172
0.478500,-0.032745
0.479000,-0.031707
0.479500,-0.032115
0.480000,-0.033468
0.480500,-0.034904
0.481000,-0.035533
0.481500,-0.034780
0.482000,-0.032609
0.482500,-0.029534
0.483000,-0.026422
0.483500,-0.024156
0.484000,-0.023292
0.484500,-0.023845
0.485000,-0.025278
0.485500,-0.026711
0.486000,-0.027258
0.486500,-0.026368
0.487000,-0.024044
0.487500,-0.020843
0.488000,-0.017666
0.488500,-0.015412
0.489000,-0.014636
0.489500,-0.015333
0.490000,-0.016935
0.490500,-0.018528
0.491000,-0.019196
0.491500,-0.018370
0.492000,-0.016048
0.492500,-0.012792
0.493000,-0.009514
0.493500,-0.007133
0.494000,-0.006224
0.494500,-0.006798
0.495000,-0.008305
0.495500,-0.009843
0.496000,-0.010505
0.496500,-0.009725
0.497000,-0.007490
0.497500,-0.004347
0.498000,-0.001184
0.498500,0.001104
0.499000,0.001966
0.499500,0.001399
0.500000,-0.000047
0.500500,-0.001489
0.501000,-0.002046
0.501500,-0.001182
0.502000,0.001093
0.502500,0.004223
0.503000,0.007324
0.503500,0.009524
0.504000,0.010299
0.504500,0.009674
0.505000,0.008212
0.505500,0.006800
0.506000,0.006303
0.506500,0.007233
0.507000,0.009550
0.507500,0.012682
0.508000,0.015737
0.508500,0.017854
0.509000,0.018529
0.509500,0.017814
0.510000,0.016293
0.510500,0.014865
0.511000,0.014395
0.511500,0.015385
0.512000,0.017783
0.512500,0.021002
0.513000,0.024145
0.513500,0.026349
0.514000,0.027114
0.514500,0.026491
0.515000,0.025061
0.515500,0.023707
0.516000,0.023276
0.516500,0.024251
0.517000,0.026571
0.517500,0.029653
0.518000,0.032623
0.518500,0.034654
0.519000,0.035286
0.519500,0.034605
0.520000,0.033205
0.520500,0.031960
0.521000,0.031684
0.521500,0.032813
0.522000,0.035236
0.522500,0.038338
0.523000,0.041234
0.523500,0.043116
0.524000,0.043564
0.524500,0.042714
0.525000,0.041206
0.525500,0.039942
0.526000,0.039740
0.526500,0.041014
0.527000,0.043615
0.527500,0.046882
0.528000,0.049896
0.528500,0.051829
0.529000,0.052263
0.529500,0.051350
0.530000,0.049757
0.530500,0.048412
0.531000,0.048148
0.531500,0.049389
0.532000,0.051980
0.532500,0.055253
0.533000,0.058280
0.533500,0.060228
0.534000,0.060678
0.534500,0.059782
0.535000,0.058210
0.535500,0.056886
0.536000,0.056635
0.536500,0.057871
0.537000,0.060431
0.537500,0.063649
0.538000,0.066605
0.538500,0.068487
0.539000,0.068899
0.539500,0.068019
0.540000,0.066524
0.540500,0.065333
0.541000,0.065249
0.541500,0.066646
0.542000,0.069325
0.542500,0.072585
0.543000,0.075493
0.543500,0.077242
0.544000,0.077465
0.544500,0.076379
0.545000,0.074705
0.545500,0.073396
0.546000,0.073274
0.546500,0.074715
0.547000,0.077503
0.547500,0.080912
0.548000,0.083980
0.548500,0.085876
0.549000,0.086214
0.549500,0.085203
0.550000,0.083561
0.550500,0.082244
0.551000,0.082078
0.551500,0.083442
0.552000,0.086130
0.552500,0.089426
0.553000,0.092384
0.553500,0.094191
0.554000,0.094477
0.554500,0.093459
0.555000,0.091853
0.555500,0.090603
0.556000,0.090512
0.556500,0.091937
0.557000,0.094650
0.557500,0.097930
0.558000,0.100836
0.558500,0.102575
0.559000,0.102806
0.559500,0.101768
0.560000,0.100193
0.560500,0.099021
0.561000,0.099038
0.561500,0.100570
0.562000,0.103360
0.562500,0.106663
0.563000,0.109534
0.563500,0.111189
0.564000,0.111312
0.564500,0.110173
0.565000,0.108529
0.565500,0.107333
0.566000,0.107367
0.566500,0.108942
0.567000,0.111774
0.567500,0.115097
0.568000,0.117955
0.568500,0.119568
0.569000,0.119634
0.569500,0.118450
0.570000,0.116795
0.570500,0.115638
0.571000,0.115760
0.571500,0.117459
0.572000,0.120428
0.572500,0.123875
0.573000,0.126819
0.573500,0.128466
0.574000,0.128511
0.574500,0.127257
0.575000,0.125494
0.575500,0.124204
0.576000,0.124183
0.576500,0.125739
0.577000,0.128581
0.577500,0.131928
0.578000,0.134816
0.578500,0.136465
0.579000,0.136578
0.579500,0.135457
0.580000,0.133877
0.580500,0.132791
0.581000,0.132961
0.581500,0.134658
0.582000,0.137564
0.582500,0.140891
0.583000,0.143685
0.583500,0.145199
0.584000,0.145177
0.584500,0.143961
0.585000,0.142354
0.585500,0.141315
0.586000,0.141591
0.586500,0.143421
0.587000,0.146448
0.587500,0.149850
0.588000,0.152655
0.588500,0.154113
0.589000,0.153987
0.589500,0.152646
0.590000,0.150922
0.590500,0.149798
0.591000,0.150028
0.591500,0.151851
0.592000,0.154898
0.592500,0.158332
0.593000,0.161169
0.593500,0.162652
0.594000,0.162544
0.594500,0.161213
0.595000,0.159492
0.595500,0.158360
0.596000,0.158567
0.596500,0.160348
0.597000,0.163336
0.597500,0.166708
0.598000,0.169502
0.598500,0.170989
0.599000,0.170953
0.599500,0.169771
0.600000,0.168265
0.600500,0.167379
0.601000,0.167811
0.601500,0.169741
0.602000,0.172755
0.602500,0.176008
0.603000,0.178551
0.603500,0.179702
0.604000,0.179314
0.604500,0.177840
0.605000,0.176164
0.605500,0.175266
0.606000,0.175837
0.606500,0.178018
0.607000,0.181331
0.607500,0.184860
0.608000,0.187593
0.608500,0.188812
0.609000,0.188364
0.609500,0.186727
0.610000,0.184831
0.610500,0.183706
0.611000,0.184097
0.611500,0.186176
0.612000,0.189483
0.612500,0.193098
0.613000,0.195991
0.613500,0.197413
0.614000,0.197176
0.614500,0.195719
0.615000,0.193939
0.615500,0.192842
0.616000,0.193161
0.616500,0.195078
0.617000,0.198159
0.617500,0.201529
0.618000,0.204214
0.618500,0.205518
0.619000,0.205291
0.619500,0.203987
0.620000,0.202477
0.620500,0.201713
0.621000,0.202351
0.621500,0.204492
0.622000,0.207641
0.622500,0.210894
0.623000,0.213297
0.623500,0.214215
0.624000,0.213588
0.624500,0.211965
0.625000,0.210296
0.625500,0.209573
0.626000,0.210442
0.626500,0.212952
0.627000,0.216520
0.627500,0.220153
0.628000,0.222815
0.628500,0.223827
0.629000,0.223124
0.629500,0.221289
0.630000,0.219334
0.630500,0.218320
0.631000,0.218958
0.631500,0.221336
0.632000,0.224889
0.632500,0.228612
0.633000,0.231440
0.633500,0.232649
0.634000,0.232133
0.634500,0.230436
0.635000,0.228545
0.635500,0.227507
0.636000,0.228037
0.636500,0.230240
0.637000,0.233583
0.637500,0.237102
0.638000,0.239774
0.638500,0.240919
0.639000,0.240454
0.639500,0.238926
0.640000,0.237299
0.640500,0.236575
0.641000,0.237406
0.641500,0.239836
0.642000,0.243283
0.642500,0.246761
0.643000,0.249261
0.643500,0.250145
0.644000,0.249396
0.644500,0.247631
0.645000,0.245866
0.645500,0.245132
0.646000,0.246069
0.646500,0.248681
0.647000,0.252326
0.647500,0.255953
0.648000,0.258503
0.648500,0.259316
0.649000,0.258381
0.649500,0.256353
0.650000,0.254305
0.650500,0.253325
0.651000,0.254108
0.651500,0.256690
0.652000,0.260437
0.652500,0.264284
0.653000,0.267137
0.653500,0.268287
0.654000,0.267671
0.654500,0.265893
0.655000,0.263984
0.655500,0.263010
0.656000,0.263663
0.656500,0.266002
0.657000,0.269435
0.657500,0.272959
0.658000,0.275545
0.658500,0.276543
0.659000,0.275928
0.659500,0.274308
0.660000,0.272684
0.660500,0.272061
0.661000,0.273053
0.661500,0.275640
0.662000,0.279177
0.662500,0.282639
0.663000,0.285021
0.663500,0.285729
0.664000,0.284815
0.664500,0.282965
0.665000,0.281238
0.665500,0.280658
0.666000,0.281821
0.666500,0.284656
0.667000,0.288446
0.667500,0.292100
0.668000,0.294562
0.668500,0.295220
0.669000,0.294141
0.669500,0.292050
0.670000,0.290062
0.670500,0.289258
0.671000,0.290276
0.671500,0.293073
0.672000,0.296937
0.672500,0.300763
0.673000,0.303466
0.673500,0.304397
0.674000,0.303579
0.674500,0.301696
0.675000,0.299826
0.675500,0.299024
0.676000,0.299922
0.676500,0.302493
0.677000,0.306065
0.677500,0.309597
0.678000,0.312077
0.678500,0.312921
0.679000,0.312195
0.679500,0.310586
0.680000,0.309125
0.680500,0.308782
0.681000,0.310081
0.681500,0.312885
0.682000,0.316448
0.682500,0.319705
0.683000,0.321684
0.683500,0.321897
0.684000,0.320541
0.684500,0.318437
0.685000,0.316720
0.685500,0.316410
0.686000,0.318013
0.686500,0.321319
0.687000,0.325467
0.687500,0.329267
0.688000,0.331644
0.688500,0.332048
0.689000,0.330668
0.689500,0.328361
0.690000,0.326337
0.690500,0.325698
0.691000,0.327027
0.691500,0.330164
0.692000,0.334273
0.692500,0.338155
0.693000,0.340705
0.693500,0.341331
0.694000,0.340171
0.694500,0.338040
0.695000,0.336109
0.695500,0.335462
0.696000,0.336678
0.696500,0.339618
0.697000,0.343481
0.697500,0.347121
0.698000,0.349491
0.698500,0.350047
0.699000,0.348956
0.699500,0.347030
0.700000,0.345403
0.700500,0.345087
0.701000,0.346578
0.701500,0.349658
0.702000,0.353478
0.702500,0.356889
0.703000,0.358892
0.703500,0.359032
0.704000,0.357587
0.704500,0.355467
0.705000,0.353873
0.705500,0.353826
0.706000,0.355774
0.706500,0.359399
0.707000,0.363731
0.707500,0.367508
0.708000,0.369650
0.708500,0.369676
0.709000,0.367898
0.709500,0.365307
0.710000,0.363210
0.710500,0.362739
0.711000,0.364428
0.711500,0.368009
0.712000,0.372515
0.712500,0.376644
0.713000,0.379246
0.713500,0.379754
0.714000,0.378389
0.714500,0.376067
0.715000,0.374043
0.715500,0.373430
0.716000,0.374781
0.716500,0.377881
0.717000,0.381849
0.717500,0.385482
0.718000,0.387734
0.718500,0.388116
0.719000,0.386888
0.719500,0.384949
0.720000,0.383481
0.720500,0.383481
0.721000,0.385366
0.721500,0.388800
0.722000,0.392821
0.722500,0.396217
0.723000,0.397998
0.723500,0.397793
0.724000,0.396008
0.724500,0.393679
0.725000,0.392084
0.725500,0.392244
0.726000,0.394527
0.726500,0.398489
0.727000,0.403034
0.727500,0.406816
0.728000,0.408756
0.728500,0.408453
0.729000,0.406349
0.729500,0.403565
0.730000,0.401493
0.730500,0.401265
0.731000,0.403330
0.731500,0.407286
0.732000,0.412026
0.732500,0.416159
0.733000,0.418531
0.733500,0.418661
0.734000,0.416912
0.734500,0.414349
0.735000,0.412327
0.735500,0.411971
0.736000,0.413752
0.736500,0.417316
0.737000,0.421626
0.737500,0.425375
0.738000,0.427495
0.738500,0.427567
0.739000,0.425987
0.739500,0.423802
0.740000,0.422300
0.740500,0.422500
0.741000,0.424747
0.741500,0.428574
0.742000,0.432878
0.742500,0.436348
0.743000,0.437979
0.743500,0.437480
0.744000,0.435395
0.744500,0.432909
0.745000,0.431396
0.745500,0.431887
0.746000,0.434660
0.746500,0.439115
0.747000,0.443987
0.747500,0.447818
0.748000,0.449506
0.748500,0.448738
0.749000,0.446114
0.749500,0.442939
0.750000,0.440740
0.750500,0.440695
0.751000,0.443193
0.751500,0.447683
0.752000,0.452888
0.752500,0.457278
0.753000,0.459645
0.753500,0.459554
0.754000,0.457494
0.754500,0.454686
0.755000,0.452607
0.755500,0.452432
0.756000,0.454584
0.756500,0.458583
0.757000,0.463246
0.757500,0.467146
0.758000,0.469171
0.758500,0.468954
0.759000,0.467010
0.759500,0.464531
0.760000,0.462918
0.760500,0.463228
0.761000,0.465760
0.761500,0.469930
0.762000,0.474495
0.762500,0.478040
0.763000,0.479533
0.763500,0.478736
0.764000,0.476315
0.764500,0.473589
0.765000,0.472034
0.765500,0.472704
0.766000,0.475818
0.766500,0.480656
0.767000,0.485823
0.767500,0.489769
0.768000,0.491382
0.768500,0.490416
0.769000,0.487599
0.769500,0.484363
0.770000,0.482315
0.770500,0.482623
0.771000,0.485580
0.771500,0.490488
0.772000,0.495918
0.772500,0.500251
0.773000,0.502288
0.773500,0.501697
0.774000,0.499140
0.774500,0.496007
0.775000,0.493892
0.775500,0.493978
0.776000,0.496595
0.776500,0.501101
0.777000,0.506138
0.777500,0.510160
0.778000,0.512037
0.778500,0.511483
0.779000,0.509170
0.779500,0.506458
0.780000,0.504863
0.780500,0.505461
0.781000,0.508466
0.781500,0.513141
0.782000,0.518084
0.782500,0.521771
0.783000,0.523154
0.783500,0.522083
0.784000,0.519370
0.784500,0.516490
0.785000,0.515011
0.785500,0.515986
0.786000,0.519530
0.786500,0.524766
0.787000,0.530142
0.787500,0.534022
0.788000,0.535314
0.788500,0.533897
0.789000,0.530683
0.789500,0.527280
0.790000,0.525389
0.790500,0.526159
0.791000,0.529745
0.791500,0.535246
0.792000,0.541039
0.792500,0.545394
0.793000,0.547125
0.793500,0.546039
0.794000,0.543008
0.794500,0.539634
0.795000,0.537635
0.795500,0.538190
0.796000,0.541494
0.796500,0.546684
0.797000,0.552185
0.797500,0.556312
0.798000,0.557929
0.798500,0.556881
0.799000,0.554058
0.799500,0.551040
0.800000,0.549487
0.800500,0.550487
0.801000,0.554126
0.801500,0.559453
0.802000,0.564840
0.802500,0.568618
0.803000,0.569728
0.803500,0.568146
0.804000,0.564904
0.804500,0.561701
0.805000,0.560255
0.805500,0.561629
0.806000,0.565812
0.806500,0.571704
0.807000,0.577524
0.807500,0.581488
0.808000,0.582491
0.808500,0.580544
0.809000,0.576782
0.809500,0.573047
0.810000,0.571189
0.810500,0.572366
0.811000,0.576599
0.811500,0.582757
0.812000,0.588984
0.812500,0.593402
0.813000,0.594823
0.813500,0.593202
0.814000,0.589655
0.814500,0.586026
0.815000,0.584185
0.815500,0.585305
0.816000,0.589417
0.816500,0.595395
0.817000,0.601391
0.817500,0.605551
0.818000,0.606731
0.818500,0.604943
0.819000,0.601354
0.819500,0.597842
0.820000,0.596269
0.820500,0.597756
0.821000,0.602247
0.821500,0.608516
0.822000,0.614638
0.822500,0.618722
0.823000,0.619649
0.823500,0.617510
0.824000,0.613580
0.824500,0.609840
0.825000,0.608219
0.825500,0.609846
0.826000,0.614613
0.826500,0.621205
0.827000,0.627595
0.827500,0.631820
0.828000,0.632737
0.828500,0.630464
0.829000,0.626350
0.829500,0.622457
0.830000,0.620775
0.830500,0.622456
0.831000,0.627365
0.831500,0.634127
0.832000,0.640647
0.832500,0.644914
0.833000,0.645778
0.833500,0.643397
0.834000,0.639183
0.834500,0.635266
0.835000,0.633678
0.835500,0.635566
0.836000,0.640746
0.836500,0.647763
0.837000,0.654445
0.837500,0.658733
0.838000,0.659481
0.838500,0.656900
0.839000,0.652489
0.839500,0.648463
0.840000,0.646909
0.840500,0.648975
0.841000,0.654423
0.841500,0.661712
0.842000,0.668577
0.842500,0.672902
0.843000,0.673534
0.843500,0.670729
0.844000,0.666073
0.844500,0.661871
0.845000,0.660268
0.845500,0.662427
0.846000,0.668070
0.846500,0.675580
0.847000,0.682619
0.847500,0.687022
0.848000,0.687636
0.848500,0.684766
0.849000,0.680066
0.849500,0.675905
0.850000,0.674453
0.850500,0.676839
0.851000,0.682706
0.851500,0.690338
0.852000,0.697318
0.852500,0.701460
0.853000,0.701662
0.853500,0.698350
0.854000,0.693327
0.854500,0.689094
0.855000,0.687886
0.855500,0.690807
0.856000,0.697380
0.856500,0.705719
0.857000,0.713229
0.857500,0.717600
0.858000,0.717699
0.858500,0.714016
0.859000,0.708496
0.859500,0.703808
0.860000,0.702329
0.860500,0.705232
0.861000,0.712027
0.861500,0.720738
0.862000,0.728647
0.862500,0.733330
0.863000,0.733584
0.863500,0.729893
0.864000,0.724249
0.864500,0.719391
0.865000,0.717758
0.865500,0.720554
0.866000,0.727279
0.866500,0.735926
0.867000,0.743750
0.867500,0.748328
0.868000,0.748503
0.868500,0.744832
0.869000,0.739377
0.869500,0.734909
0.870000,0.733825
0.870500,0.737212
0.871000,0.744405
0.871500,0.753231
0.872000,0.760841
0.872500,0.764822
0.873000,0.764148
0.873500,0.759613
0.874000,0.753546
0.874500,0.748940
0.875000,0.748294
0.875500,0.752632
0.876000,0.761076
0.876500,0.771144
0.877000,0.779678
0.877500,0.784041
0.878000,0.783139
0.878500,0.777864
0.879000,0.770782
0.879500,0.765180
0.880000,0.763821
0.880500,0.767888
0.881000,0.776523
0.881500,0.787135
0.882000,0.796375
0.882500,0.801407
0.883000,0.800989
0.883500,0.795952
0.884000,0.788885
0.884500,0.783152
0.885000,0.781604
0.885500,0.785485
0.886000,0.793954
0.886500,0.804406
0.887000,0.813481
0.887500,0.818359
0.888000,0.817860
0.888500,0.812908
0.889000,0.806177
0.889500,0.801063
0.890000,0.800351
0.890500,0.805125
0.891000,0.814314
0.891500,0.825093
0.892000,0.833962
0.892500,0.838111
0.893000,0.836539
0.893500,0.830475
0.894000,0.822945
0.894500,0.817628
0.895000,0.817428
0.895500,0.823335
0.896000,0.833992
0.896500,0.846171
0.897000,0.855984
0.897500,0.860364
0.898000,0.858262
0.898500,0.851088
0.899000,0.842216
0.899500,0.835722
0.900000,0.834843
0.900500,0.840728
0.901000,0.851980
0.901500,0.865158
0.902000,0.876070
0.902500,0.881367
0.903000,0.879824
0.903500,0.872822
0.904000,0.863835
0.904500,0.857109
0.905000,0.856035
0.905500,0.861843
0.906000,0.873117
0.906500,0.886327
0.907000,0.897186
0.907500,0.902304
0.908000,0.900512
0.908500,0.893334
0.909000,0.884417
0.909500,0.878132
0.910000,0.877869
0.910500,0.884707
0.911000,0.896955
0.911500,0.910772
0.912000,0.921639
0.912500,0.926114
0.913000,0.923192
0.913500,0.914753
0.914000,0.904874
0.914500,0.898293
0.915000,0.898570
0.915500,0.906684
0.916000,0.920597
0.916500,0.935987
0.917000,0.947874
0.917500,0.952532
0.918000,0.948958
0.918500,0.939306
0.919000,0.928114
0.919500,0.920601
0.920000,0.920661
0.920500,0.929350
0.921000,0.944426
0.921500,0.961167
0.922000,0.974151
0.922500,0.979333
0.923000,0.975628
0.923500,0.965372
0.924000,0.953467
0.924500,0.945527
0.925000,0.945719
0.925500,0.955128
0.926000,0.971289
0.926500,0.989090
0.927000,1.002706
0.927500,1.007846
0.928000,1.003443
0.928500,0.992118
0.929000,0.979235
0.929500,0.970864
0.930000,0.971445
0.930500,0.982041
0.931000,0.999862
0.931500,1.019280
0.932000,1.033956
0.932500,1.039278
0.933000,1.034180
0.933500,1.021614
0.934000,1.007483
0.934500,0.998401
0.935000,0.999152
0.935500,1.010805
0.936000,1.030234
0.936500,1.051264
0.937000,1.067009
0.937500,1.072534
0.938000,1.066807
0.938500,1.053166
0.939000,1.038099
0.939500,1.028775
0.940000,1.030258
0.940500,1.043491
0.941000,1.064837
0.941500,1.087405
0.942000,1.103673
0.942500,1.108418
0.943000,1.100823
0.943500,1.084909
0.944000,1.068098
0.944500,1.058402
0.945000,1.061313
0.945500,1.077581
0.946000,1.102784
0.946500,1.128889
0.947000,1.147270
0.947500,1.152032
0.948000,1.142380
0.948500,1.123075
0.949000,1.102757
0.949500,1.090726
0.950000,1.093375
0.950500,1.111657
0.951000,1.140574
0.951500,1.170949
0.952000,1.192855
0.952500,1.199403
0.953000,1.189465
0.953500,1.168231
0.954000,1.145344
0.954500,1.131283
0.955000,1.133336
0.955500,1.152722
0.956000,1.183993
0.956500,1.217013
0.957000,1.240795
0.957500,1.247764
0.958000,1.236788
0.958500,1.213773
0.959000,1.189530
0.959500,1.175658
0.960000,1.180010
0.960500,1.203496
0.961000,1.239509
0.961500,1.276288
0.962000,1.301380
0.962500,1.306526
0.963000,1.291072
0.963500,1.262514
0.964000,1.233866
0.964500,1.218762
0.965000,1.226122
0.965500,1.256455
0.966000,1.301317
0.966500,1.346238
0.967000,1.376134
0.967500,1.381189
0.968000,1.360913
0.968500,1.324735
0.969000,1.288750
0.969500,1.269722
0.970000,1.278565
0.970500,1.315827
0.971000,1.371023
0.971500,1.426245
0.972000,1.462835
0.972500,1.468659
0.973000,1.443169
0.973500,1.398177
0.974000,1.353825
0.974500,1.331099
0.975000,1.343623
0.975500,1.391919
0.976000,1.462485
0.976500,1.532320
0.977000,1.577431
0.977500,1.582205
0.978000,1.546065
0.978500,1.484627
0.979000,1.424608
0.979500,1.394063
0.980000,1.411494
0.980500,1.477971
0.981000,1.575554
0.981500,1.673011
0.982000,1.737120
0.982500,1.745522
0.983000,1.696250
0.983500,1.609995
0.984000,1.523695
0.984500,1.477310
0.985000,1.498389
0.985500,1.590247
0.986000,1.728656
0.986500,1.869028
0.987000,1.962220
0.987500,1.973599
0.988000,1.898276
0.988500,1.766235
0.989000,1.634337
0.989500,1.566927
0.990000,1.611232
0.990500,1.776424
0.991000,2.024779
0.991500,2.279819
0.992000,2.450581
0.992500,2.465138
0.993000,2.302275
0.993500,2.009550
0.994000,1.699304
0.994500,1.520838
0.995000,1.614930
0.995500,2.063470
0.996000,2.850085
0.996500,3.845691
0.997000,4.826419
0.997500,5.521911
0.998000,5.682548
0.998500,5.147594
0.999000,3.894878
0.999500,2.056951
1.000000,-0.102270
};
\end{axis}
\end{tikzpicture}

%% file: Fig_modified_sum.tex
\centering
\begin{tikzpicture}[]
\begin{axis}[ymin=0, ymax=5.5, xmin=0.96,xmax=1,
  ytick={0,1,...,5}, ytick pos=left,
  xtick={0,0.01,...,1}, xtick pos=left,
  xlabel={$x/L$},
  ylabel={$\extPot_B(x)/D$},
  legend pos=north west,
  legend cell align={left},
  legend style={draw=none}]
\addplot+[
 black, smooth,mark=empty,line width=1pt,
] table [x=x, y=y, col sep=comma] {
    x,  y,      
0.960000,1.115972
0.960506,1.143387
0.961013,1.171925
0.961519,1.201521
0.962025,1.232107
0.962532,1.263609
0.963038,1.295948
0.963544,1.329040
0.964051,1.362798
0.964557,1.397128
0.965063,1.431936
0.965570,1.467121
0.966076,1.502580
0.966582,1.538207
0.967089,1.573891
0.967595,1.609520
0.968101,1.644981
0.968608,1.680157
0.969114,1.714929
0.969620,1.749179
0.970127,1.782786
0.970633,1.815629
0.971139,1.847586
0.971646,1.878537
0.972152,1.908362
0.972658,1.936939
0.973165,1.964151
0.973671,1.989881
0.974177,2.014013
0.974684,2.036435
0.975190,2.057037
0.975696,2.075712
0.976203,2.092357
0.976709,2.106870
0.977215,2.119157
0.977722,2.129127
0.978228,2.136691
0.978734,2.141770
0.979241,2.144285
0.979747,2.144167
0.980253,2.141349
0.980759,2.135774
0.981266,2.127388
0.981772,2.116145
0.982278,2.102006
0.982785,2.084937
0.983291,2.064912
0.983797,2.041914
0.984304,2.015931
0.984810,1.986958
0.985316,1.954998
0.985823,1.920064
0.986329,1.882172
0.986835,1.841348
0.987342,1.797626
0.987848,1.751046
0.988354,1.701655
0.988861,1.649508
0.989367,1.594668
0.989873,1.537204
0.990380,1.477192
0.990886,1.414714
0.991392,1.349859
0.991899,1.282722
0.992405,1.213406
0.992911,1.142016
0.993418,1.068666
0.993924,0.993474
0.994430,0.916561
0.994937,0.838056
0.995443,0.758090
0.995949,0.676799
0.996456,0.594320
0.996962,0.510798
0.997468,0.426376
0.997975,0.341202
0.998481,0.255425
0.998987,0.169197
0.999494,0.082670
1.000000,-0.004003
};
\addlegendentry{$B=20$}
\addplot+[
  black, dashed,mark=empty,line width=1pt,
] table [x=x, y=y, col sep=comma] {
    x,  y,      
0.960000,1.478856
0.960506,1.511985
0.961013,1.541586
0.961519,1.567174
0.962025,1.588317
0.962532,1.604647
0.963038,1.615864
0.963544,1.621744
0.964051,1.622146
0.964557,1.617014
0.965063,1.606381
0.965570,1.590371
0.966076,1.569200
0.966582,1.543176
0.967089,1.512693
0.967595,1.478234
0.968101,1.440359
0.968608,1.399703
0.969114,1.356968
0.969620,1.312915
0.970127,1.268349
0.970633,1.224118
0.971139,1.181090
0.971646,1.140152
0.972152,1.102188
0.972658,1.068071
0.973165,1.038650
0.973671,1.014732
0.974177,0.997075
0.974684,0.986373
0.975190,0.983242
0.975696,0.988211
0.976203,1.001711
0.976709,1.024065
0.977215,1.055481
0.977722,1.096046
0.978228,1.145720
0.978734,1.204330
0.979241,1.271572
0.979747,1.347008
0.980253,1.430072
0.980759,1.520066
0.981266,1.616172
0.981772,1.717454
0.982278,1.822870
0.982785,1.931281
0.983291,2.041461
0.983797,2.152111
0.984304,2.261875
0.984810,2.369349
0.985316,2.473107
0.985823,2.571706
0.986329,2.663712
0.986835,2.747713
0.987342,2.822338
0.987848,2.886271
0.988354,2.938274
0.988861,2.977195
0.989367,3.001988
0.989873,3.011728
0.990380,3.005619
0.990886,2.983012
0.991392,2.943407
0.991899,2.886469
0.992405,2.812027
0.992911,2.720085
0.993418,2.610819
0.993924,2.484581
0.994430,2.341895
0.994937,2.183452
0.995443,2.010110
0.995949,1.822878
0.996456,1.622914
0.996962,1.411507
0.997468,1.190070
0.997975,0.960120
0.998481,0.723266
0.998987,0.481190
0.999494,0.235631
1.000000,-0.011637

};
\addlegendentry{$B=40$}
\addplot+[
  black, loosely dotted,mark=empty,line width=1pt,
] table [x=x, y=y, col sep=comma] {
    x,  y,      
0.960000,1.481353
0.960506,1.458632
0.961013,1.427628
0.961519,1.389494
0.962025,1.345678
0.962532,1.297879
0.963038,1.247994
0.963544,1.198053
0.964051,1.150147
0.964557,1.106356
0.965063,1.068672
0.965570,1.038925
0.966076,1.018712
0.966582,1.009331
0.967089,1.011727
0.967595,1.026446
0.968101,1.053605
0.968608,1.092873
0.969114,1.143469
0.969620,1.204177
0.970127,1.273375
0.970633,1.349084
0.971139,1.429020
0.971646,1.510672
0.972152,1.591384
0.972658,1.668443
0.973165,1.739174
0.973671,1.801041
0.974177,1.851738
0.974684,1.889283
0.975190,1.912101
0.975696,1.919094
0.976203,1.909703
0.976709,1.883945
0.977215,1.842442
0.977722,1.786421
0.978228,1.717701
0.978734,1.638656
0.979241,1.552155
0.979747,1.461495
0.980253,1.370299
0.980759,1.282417
0.981266,1.201805
0.981772,1.132400
0.982278,1.077987
0.982785,1.042073
0.983291,1.027756
0.983797,1.037608
0.984304,1.073565
0.984810,1.136839
0.985316,1.227845
0.985823,1.346144
0.986329,1.490422
0.986835,1.658484
0.987342,1.847280
0.987848,2.052951
0.988354,2.270907
0.988861,2.495924
0.989367,2.722264
0.989873,2.943813
0.990380,3.154235
0.990886,3.347132
0.991392,3.516217
0.991899,3.655483
0.992405,3.759369
0.992911,3.822921
0.993418,3.841937
0.993924,3.813094
0.994430,3.734058
0.994937,3.603564
0.995443,3.421471
0.995949,3.188794
0.996456,2.907690
0.996962,2.581433
0.997468,2.214346
0.997975,1.811710
0.998481,1.379643
0.998987,0.924965
0.999494,0.455033
1.000000,-0.022430
};
\addlegendentry{$B=60$}
\addplot+[
  black, dash dot,mark=empty,line width=1pt,
] table [x=x, y=y, col sep=comma] {
    x,  y,      
0.960000,1.192679
0.960506,1.146309
0.961013,1.105911
0.961519,1.074705
0.962025,1.055397
0.962532,1.049992
0.963038,1.059639
0.963544,1.084539
0.964051,1.123897
0.964557,1.175953
0.965063,1.238066
0.965570,1.306858
0.966076,1.378419
0.966582,1.448543
0.967089,1.512993
0.967595,1.567780
0.968101,1.609430
0.968608,1.635229
0.969114,1.643425
0.969620,1.633374
0.970127,1.605623
0.970633,1.561913
0.971139,1.505112
0.971646,1.439065
0.972152,1.368385
0.972658,1.298179
0.973165,1.233734
0.973671,1.180186
0.974177,1.142175
0.974684,1.123530
0.975190,1.126984
0.975696,1.153953
0.976203,1.204389
0.976709,1.276719
0.977215,1.367877
0.977722,1.473437
0.978228,1.587835
0.978734,1.704669
0.979241,1.817075
0.979747,1.918143
0.980253,2.001354
0.980759,2.061027
0.981266,2.092720
0.981772,2.093587
0.982278,2.062653
0.982785,2.000989
0.983291,1.911772
0.983797,1.800223
0.984304,1.673419
0.984810,1.539988
0.985316,1.409691
0.985823,1.292916
0.986329,1.200112
0.986835,1.141181
0.987342,1.124875
0.987848,1.158213
0.988354,1.245971
0.988861,1.390264
0.989367,1.590246
0.989873,1.841955
0.990380,2.138317
0.990886,2.469301
0.991392,2.822249
0.991899,3.182343
0.992405,3.533202
0.992911,3.857590
0.993418,4.138184
0.993924,4.358380
0.994430,4.503093
0.994937,4.559504
0.995443,4.517734
0.995949,4.371381
0.996456,4.117924
0.996962,3.758940
0.997468,3.300150
0.997975,2.751262
0.998481,2.125635
0.998987,1.439771
0.999494,0.712658
1.000000,-0.035002
};
\addlegendentry{$B=80$}
\addplot+[
 black, densely dotted,mark=empty,line width=1pt,
] table [x=x, y=y, col sep=comma] {
    x,  y,      
0.960000,1.150135
0.960506,1.170868
0.961013,1.200994
0.961519,1.238317
0.962025,1.279834
0.962532,1.322013
0.963038,1.361139
0.963544,1.393688
0.964051,1.416696
0.964557,1.428084
0.965063,1.426904
0.965570,1.413480
0.966076,1.389422
0.966582,1.357513
0.967089,1.321473
0.967595,1.285610
0.968101,1.254405
0.968608,1.232064
0.969114,1.222070
0.969620,1.226802
0.970127,1.247237
0.970633,1.282794
0.971139,1.331320
0.971646,1.389241
0.972152,1.451862
0.972658,1.513792
0.973165,1.569469
0.973671,1.613712
0.974177,1.642281
0.974684,1.652353
0.975190,1.642889
0.975696,1.614838
0.976203,1.571151
0.976709,1.516593
0.977215,1.457364
0.977722,1.400555
0.978228,1.353479
0.978734,1.322949
0.979241,1.314550
0.979747,1.332000
0.980253,1.376651
0.980759,1.447194
0.981266,1.539604
0.981772,1.647360
0.982278,1.761914
0.982785,1.873402
0.983291,1.971528
0.983797,2.046567
0.984304,2.090381
0.984810,2.097367
0.985316,2.065246
0.985823,1.995596
0.986329,1.894080
0.986835,1.770310
0.987342,1.637345
0.987848,1.510841
0.988354,1.407888
0.988861,1.345638
0.989367,1.339802
0.989873,1.403152
0.990380,1.544138
0.990886,1.765756
0.991392,2.064766
0.991899,2.431338
0.992405,2.849193
0.992911,3.296249
0.993418,3.745740
0.993924,4.167759
0.994430,4.531117
0.994937,4.805401
0.995443,4.963076
0.995949,4.981479
0.996456,4.844568
0.996962,4.544264
0.997468,4.081307
0.997975,3.465530
0.998481,2.715529
0.998987,1.857739
0.999494,0.924971
1.000000,-0.045489
};
\addlegendentry{$B=100$}
\end{axis}
\end{tikzpicture}

%% file: Fig_ABplot.tex
\begin{tikzpicture}[scale=0.9]
\begin{axis}[ymin=0, ymax=1.2, xmin=-0.1, xmax=1.1,
  ytick={0,0.5,1}, ytick pos=left,
  yticklabels={$0$,$U(x^*)$,$U_0$},
  xtick={0,0.45,0.9,1}, xtick pos=left,
  xticklabels={$0$,$x^*$,$\alpha L$,$L$},
  xlabel={$x$},
  ylabel={$U(x)$},
  legend pos=north west,
  legend style={draw=none}]
\addplot [line width=2pt,
 domain=0:0.9, 
 samples=100, 
 color=black,shorten >=-0.9pt]
 {10/9*x};
 \addplot [line width=2pt,
 domain=0.9:1, 
 samples=10, 
 color=black,]
 {-10*x+10};
 \addplot [line width=2pt,
 domain=1:1.1, 
 samples=10, 
 color=black,]
 {10/9*(x-1)};
 \addplot [line width=2pt,
 domain=-0.1:0, 
 samples=10, 
 color=black,]
 {-10*(x+1)+10};
\addplot [dotted,mark=empty] coordinates {
(0.45,0) (0.45,1.2)
};
\addplot [dotted,mark=empty] coordinates {
(-0.1,0.5) (1.1,0.5)
};
\end{axis}
\end{tikzpicture}

%% file: Fig_comparisonAB_75.tex
\begin{tikzpicture}
\begin{axis}[
anchor=origin,  
ymin=0, ymax=0.02, xmin=0, xmax=6,
  ytick={0,0.005,0.010,0.015739,0.0200}, ytick pos=left,
  yticklabels={$0$,$0.005$,$0.010$,$\currentOptim$,$0.020$},
  xtick={0,1,2,3,3.9,4.27,5,6},
  xticklabels={$0$,$1$,$2$,$3$,$\!\!\!\nuOptim$,$\quad\nuRocSmall$,$5$,$6$},
  xtick pos=left,
  ylabel={$J$},
  xlabel={$\nu$},
  legend pos=north west,
  legend style={draw=none},
  scaled y ticks=false,
  yticklabel style={
            /pgf/number format/fixed,
            /pgf/number format/precision=3,
            /pgf/number format/fixed zerofill
        },]
\addplot+[
  black, mark options={black, scale=0.5},
  only marks, 
  error bars/.cd, 
    y fixed,
    y dir=both, 
    y explicit
] table [x=x, y=y, col sep=comma] {
  x,   y,
0.000000,0.000000
0.050000,0.000000
0.100000,0.000001
0.150000,0.000005
0.200000,0.000012
0.250000,0.000023
0.300000,0.000039
0.350000,0.000062
0.400000,0.000092
0.450000,0.000131
0.500000,0.000178
0.550000,0.000236
0.600000,0.000304
0.650000,0.000383
0.700000,0.000474
0.750000,0.000578
0.800000,0.000695
0.850000,0.000824
0.900000,0.000968
0.950000,0.001125
1.000000,0.001296
1.050000,0.001480
1.100000,0.001679
1.150000,0.001891
1.200000,0.002117
1.250000,0.002356
1.300000,0.002608
1.350000,0.002872
1.400000,0.003148
1.450000,0.003436
1.500000,0.003735
1.550000,0.004044
1.600000,0.004362
1.650000,0.004689
1.700000,0.005023
1.750000,0.005365
1.800000,0.005713
1.850000,0.006067
1.900000,0.006424
1.950000,0.006786
2.000000,0.007149
2.050000,0.007515
2.100000,0.007881
2.150000,0.008247
2.200000,0.008612
2.250000,0.008974
2.300000,0.009334
2.350000,0.009690
2.400000,0.010041
2.450000,0.010387
2.500000,0.010726
2.550000,0.011058
2.600000,0.011383
2.650000,0.011699
2.700000,0.012007
2.750000,0.012304
2.800000,0.012592
2.850000,0.012868
2.900000,0.013134
2.950000,0.013387
3.000000,0.013629
3.050000,0.013859
3.100000,0.014075
3.150000,0.014279
3.200000,0.014470
3.250000,0.014648
3.300000,0.014812
3.350000,0.014962
3.400000,0.015099
3.450000,0.015223
3.500000,0.015333
3.550000,0.015429
3.600000,0.015512
3.650000,0.015582
3.700000,0.015639
3.750000,0.015683
3.800000,0.015714
3.850000,0.015732
3.900000,0.015739
3.950000,0.015733
4.000000,0.015716
4.050000,0.015687
4.100000,0.015647
4.150000,0.015595
4.200000,0.015532
4.250000,0.015456
4.300000,0.015363
4.350000,0.015244
4.400000,0.015077
4.450000,0.014809
4.500000,0.014320
4.550000,0.013338
4.600000,0.011249
4.650000,0.006686
4.700000,-0.003371
4.750000,-0.025532
4.800000,-0.074148
4.850000,-0.180119
4.900000,-0.409439
4.950000,-0.901941
};
\addlegendentry{Field theory}
\addplot+[ 
  black, smooth,mark=empty,
] table [x=x, y=y, col sep=comma] {
x, y
0.0,0
0.05,0
0.100000,0.000001
0.150000,0.000005
0.200000,0.000012
0.250000,0.000023
0.300000,0.000039
0.350000,0.000062
0.400000,0.000092
0.450000,0.000131
0.500000,0.000178
0.550000,0.000236
0.600000,0.000304
0.650000,0.000383
0.700000,0.000474
0.750000,0.000578
0.800000,0.000695
0.850000,0.000824
0.900000,0.000968
0.950000,0.001125
1.000000,0.001295
1.050000,0.001480
1.100000,0.001679
1.150000,0.001891
1.200000,0.002117
1.250000,0.002356
1.300000,0.002608
1.350000,0.002872
1.400000,0.003148
1.450000,0.003436
1.500000,0.003735
1.550000,0.004044
1.600000,0.004362
1.650000,0.004689
1.700000,0.005023
1.750000,0.005365
1.800000,0.005713
1.850000,0.006067
1.900000,0.006424
1.950000,0.006786
2.000000,0.007149
2.050000,0.007515
2.100000,0.007881
2.150000,0.008247
2.200000,0.008612
2.250000,0.008974
2.300000,0.009334
2.350000,0.009690
2.400000,0.010041
2.450000,0.010387
2.500000,0.010726
2.550000,0.011059
2.600000,0.011383
2.650000,0.011699
2.700000,0.012007
2.750000,0.012304
2.800000,0.012592
2.850000,0.012868
2.900000,0.013134
2.950000,0.013387
3.000000,0.013629
3.050000,0.013859
3.100000,0.014075
3.150000,0.014279
3.200000,0.014470
3.250000,0.014648
3.300000,0.014812
3.350000,0.014962
3.400000,0.015099
3.450000,0.015223
3.500000,0.015333
3.550000,0.015429
3.600000,0.015512
3.650000,0.015582
3.700000,0.015639
3.750000,0.015683
3.800000,0.015714
3.850000,0.015732
3.900000,0.015739
3.950000,0.015733
4.000000,0.015716
4.050000,0.015687
4.100000,0.015647
4.150000,0.015596
4.200000,0.015535
4.250000,0.015464
4.300000,0.015384
4.350000,0.015294
4.400000,0.015195
4.450000,0.015087
4.500000,0.014971
4.550000,0.014848
4.600000,0.014717
4.650000,0.014579
4.700000,0.014435
4.750000,0.014284
4.800000,0.014128
4.850000,0.013966
4.900000,0.013799
4.950000,0.013627
5.000000,0.013451
5.050000,0.013271
5.100000,0.013087
5.150000,0.012900
5.200000,0.012709
5.250000,0.012516
5.300000,0.012321
5.350000,0.012124
5.400000,0.011925
5.450000,0.011724
5.500000,0.011523
5.550000,0.011320
5.600000,0.011117
5.650000,0.010913
5.700000,0.010709
5.750000,0.010505
5.800000,0.010301
5.850000,0.010098
5.900000,0.009895
5.950000,0.009693
6.000000,0.009492
6.050000,0.009293
6.100000,0.009094
6.150000,0.008897
6.200000,0.008702
6.250000,0.008508
6.300000,0.008316
6.350000,0.008127
6.400000,0.007939
6.450000,0.007753
6.500000,0.007570
6.550000,0.007389
6.600000,0.007210
6.650000,0.007034
6.700000,0.006861
6.750000,0.006690
6.800000,0.006521
6.850000,0.006356
6.900000,0.006193
6.950000,0.006032
7.000000,0.005875
7.050000,0.005720
7.100000,0.005568
7.150000,0.005419
7.200000,0.005273
7.250000,0.005130
7.300000,0.004989
7.350000,0.004852
7.400000,0.004717
7.450000,0.004585
7.500000,0.004456
7.550000,0.004329
7.600000,0.004206
7.650000,0.004085
7.700000,0.003967
7.750000,0.003851
7.800000,0.003739
7.850000,0.003629
7.900000,0.003521
7.950000,0.003416
8.000000,0.003314
8.050000,0.003214
8.100000,0.003117
8.150000,0.003022
8.200000,0.002930
8.250000,0.002840
8.300000,0.002752
8.350000,0.002667
8.400000,0.002583
8.450000,0.002502
8.500000,0.002424
8.550000,0.002347
8.600000,0.002273
8.650000,0.002200
8.700000,0.002130
8.750000,0.002061
8.800000,0.001995
8.850000,0.001930
8.900000,0.001867
8.950000,0.001806
9.000000,0.001747
9.050000,0.001690
9.100000,0.001634
9.150000,0.001580
9.200000,0.001527
9.250000,0.001476
9.300000,0.001427
9.350000,0.001379
9.400000,0.001332
9.450000,0.001287
9.500000,0.001244
9.550000,0.001201
9.600000,0.001160
9.650000,0.001121
9.700000,0.001082
9.750000,0.001045
9.800000,0.001009
9.850000,0.000974
9.900000,0.000940
9.950000,0.000907
};
\addlegendentry{Exact}
\draw [dotted] (axis cs:4.272,0) --  (axis cs:4.272,0.02);
\addplot [dashed,mark=empty] coordinates {
(3.9,0) (3.9,0.015739)
};
\addplot [dashed,mark=empty] coordinates {
(0,0.015739) (10,0.015739)
};
\end{axis}
\end{tikzpicture}

%% file: density_U.tex
\centering
\begin{tikzpicture}[]
\begin{axis}[ymin=0, ymax=3, xmin=-0.05, xmax=1.05,
  ytick={0,1,...,3}, ytick pos=left,
  xtick={0,0.2,...,1}, xtick pos=left,
  xlabel={$x$},
  ylabel={$\rho(x)$},
  legend pos=north west,
  legend style={draw=none}]
\addplot+[line width=2pt,
  black, smooth,mark=empty,
] table [x=x, y=y, col sep=comma] {
    x,  y,      
0.000000000000000000e+00,1.468439021674871059e-01
1.010101010101010187e-02,1.467748913479900441e-01
2.020202020202020374e-02,1.466965784566975073e-01
3.030303030303030387e-02,1.466087155598022607e-01
4.040404040404040747e-02,1.465114711743530918e-01
5.050505050505051108e-02,1.464048578242840293e-01
6.060606060606060774e-02,1.462885771690270453e-01
7.070707070707071829e-02,1.461627991189121678e-01
8.080808080808081495e-02,1.460274202402279398e-01
9.090909090909091161e-02,1.458823533096135772e-01
1.010101010101010222e-01,1.457274379453504798e-01
1.111111111111111188e-01,1.455626122530737643e-01
1.212121212121212155e-01,1.453877179800731057e-01
1.313131313131313260e-01,1.452024337518394714e-01
1.414141414141414366e-01,1.450064466867577451e-01
1.515151515151515194e-01,1.447992040848967710e-01
1.616161616161616299e-01,1.445799814911697123e-01
1.717171717171717404e-01,1.443474523246733032e-01
1.818181818181818232e-01,1.440992025500290319e-01
1.919191919191919338e-01,1.438308407426555080e-01
2.020202020202020443e-01,1.435329986021738335e-01
2.121212121212121271e-01,1.431824117014775799e-01
2.222222222222222376e-01,1.427080987289922887e-01
2.323232323232323482e-01,1.418012276757934509e-01
2.424242424242424310e-01,1.375604522232516114e-01
2.525252525252525415e-01,2.361034128062598825e+00
2.626262626262626521e-01,2.813180464129301672e+00
2.727272727272727626e-01,2.836452139311842924e+00
2.828282828282828731e-01,2.839498799442043531e+00
2.929292929292929282e-01,2.838666091212521447e+00
3.030303030303030387e-01,2.836743559414144578e+00
3.131313131313131493e-01,2.834514319811262339e+00
3.232323232323232598e-01,2.832266163681842475e+00
3.333333333333333703e-01,2.830124488151441575e+00
3.434343434343434809e-01,2.828150712217788243e+00
3.535353535353535914e-01,2.826377661356551663e+00
3.636363636363636465e-01,2.824824201884913233e+00
3.737373737373737570e-01,2.823501691580784456e+00
3.838383838383838675e-01,2.822417294075626248e+00
3.939393939393939781e-01,2.821575728172775133e+00
4.040404040404040886e-01,2.820980181229611805e+00
4.141414141414141992e-01,2.820633215863375565e+00
4.242424242424242542e-01,2.820536746649526805e+00
4.343434343434343647e-01,2.820692688728544617e+00
4.444444444444444753e-01,2.821103250118936590e+00
4.545454545454545858e-01,2.821772284650115203e+00
4.646464646464646964e-01,2.822707964368050604e+00
4.747474747474748069e-01,2.823934571611625088e+00
4.848484848484848619e-01,2.825566663208636697e+00
4.949494949494949725e-01,2.828917055035277794e+00
5.050505050505050830e-01,2.664384195316793491e+00
5.151515151515152491e-01,2.360664163408503935e+00
5.252525252525253041e-01,2.092825042234059740e+00
5.353535353535353591e-01,1.855626346117487557e+00
5.454545454545455252e-01,1.645464315212550988e+00
5.555555555555555802e-01,1.459223685882028443e+00
5.656565656565657463e-01,1.294160339203442867e+00
5.757575757575758013e-01,1.147849904818929900e+00
5.858585858585858563e-01,1.018147873463855158e+00
5.959595959595960224e-01,9.031569397938634669e-01
6.060606060606060774e-01,8.011980574411311506e-01
6.161616161616162435e-01,7.107848628735163876e-01
6.262626262626262985e-01,6.306018849491564771e-01
6.363636363636364646e-01,5.594842571767092432e-01
6.464646464646465196e-01,4.964007345546289640e-01
6.565656565656565746e-01,4.404380564953938038e-01
6.666666666666667407e-01,3.907873994920400551e-01
6.767676767676767957e-01,3.467324348738856643e-01
6.868686868686869618e-01,3.076383209178072398e-01
6.969696969696970168e-01,2.729428135385588150e-01
7.070707070707071829e-01,2.421471258991853837e-01
7.171717171717172379e-01,2.148082766067520577e-01
7.272727272727272929e-01,1.905308710674158135e-01
7.373737373737374590e-01,1.689517912163807090e-01
7.474747474747475140e-01,1.496630129832979550e-01
7.575757575757576801e-01,1.458031830000098283e-01
7.676767676767677351e-01,1.458839494733994713e-01
7.777777777777777901e-01,1.460082842357926791e-01
7.878787878787879562e-01,1.461364830526432623e-01
7.979797979797980112e-01,1.462600431688644287e-01
8.080808080808081773e-01,1.463764990467107951e-01
8.181818181818182323e-01,1.464844865072312263e-01
8.282828282828283983e-01,1.465837067690782258e-01
8.383838383838384534e-01,1.466738482160866719e-01
8.484848484848485084e-01,1.467546746177353367e-01
8.585858585858586745e-01,1.468262791436989179e-01
8.686868686868687295e-01,1.468884131139281957e-01
8.787878787878788955e-01,1.469412199161709842e-01
8.888888888888889506e-01,1.469846180782499179e-01
8.989898989898991166e-01,1.470186282346934437e-01
9.090909090909091717e-01,1.470432831441313981e-01
9.191919191919192267e-01,1.470584841912022933e-01
9.292929292929293927e-01,1.470644417820328176e-01
9.393939393939394478e-01,1.470609823210345113e-01
9.494949494949496138e-01,1.470482497461703952e-01
9.595959595959596689e-01,1.470261347598961765e-01
9.696969696969697239e-01,1.469945274522376499e-01
9.797979797979798899e-01,1.469537268153827958e-01
9.898989898989899450e-01,1.469035240391857455e-01
1.000000000000000000e+00,1.468439021674900480e-01

};
\end{axis}
\end{tikzpicture}

%% file: density_minus_U.tex
\centering
\begin{tikzpicture}[]
\begin{axis}[ymin=0.0, ymax=3, xmin=-0.05, xmax=1.05,
  ytick={0,1,...,3}, ytick pos=left,
  xtick={0,0.2,...,1}, xtick pos=left,
  xlabel={$x$},
  ylabel={$\rho(x)$},
  legend pos=north west,
  legend style={draw=none}]
\addplot+[line width=2pt,
  black, smooth,mark=empty,
] table [x=x, y=y, col sep=comma] {
    x,  y,      
0.000000000000000000e+00,1.663281959699258516e+00
1.010101010101010187e-02,1.663986135439233127e+00
2.020202020202020374e-02,1.664825493455017735e+00
3.030303030303030387e-02,1.665800468664327294e+00
4.040404040404040747e-02,1.666911580518608771e+00
5.050505050505051108e-02,1.668159311948003376e+00
6.060606060606060774e-02,1.669544328133254041e+00
7.070707070707071829e-02,1.671067454005546793e+00
8.080808080808081495e-02,1.672729645867867987e+00
9.090909090909091161e-02,1.674532073489193973e+00
1.010101010101010222e-01,1.676476142789197654e+00
1.111111111111111188e-01,1.678563614649882574e+00
1.212121212121212155e-01,1.680796833875876528e+00
1.313131313131313260e-01,1.683178936668616510e+00
1.414141414141414366e-01,1.685714323882111065e+00
1.515151515151515194e-01,1.688409352495956250e+00
1.616161616161616299e-01,1.691273841274681722e+00
1.717171717171717404e-01,1.694323852012018072e+00
1.818181818181818232e-01,1.697587509280521489e+00
1.919191919191919338e-01,1.701118405576197201e+00
2.020202020202020443e-01,1.705030614744386774e+00
2.121212121212121271e-01,1.709605284628740174e+00
2.222222222222222376e-01,1.715702887236629337e+00
2.323232323232323482e-01,1.727107114444694558e+00
2.424242424242424310e-01,1.780803276199268925e+00
2.525252525252525415e-01,1.039539500314781151e-01
2.626262626262626521e-01,8.743156623621899104e-02
2.727272727272727626e-01,8.685297171937490790e-02
2.828282828282828731e-01,8.688852749669442876e-02
2.929292929292929282e-01,8.703469708454071585e-02
3.030303030303030387e-01,8.720678060786916930e-02
3.131313131313131493e-01,8.738114956837453318e-02
3.232323232323232598e-01,8.754909732972160352e-02
3.333333333333333703e-01,8.770687103366628379e-02
3.434343434343434809e-01,8.785257011578923025e-02
3.535353535353535914e-01,8.798526472711980240e-02
3.636363636363636465e-01,8.810438569492375083e-02
3.737373737373737570e-01,8.820959705352865443e-02
3.838383838383838675e-01,8.830067441757634417e-02
3.939393939393939781e-01,8.837746764251327991e-02
4.040404040404040886e-01,8.843988309713224361e-02
4.141414141414141992e-01,8.848786412216688713e-02
4.242424242424242542e-01,8.852130666431431982e-02
4.343434343434343647e-01,8.854013787901467147e-02
4.444444444444444753e-01,8.854431344652477498e-02
4.545454545454545858e-01,8.853368173584461043e-02
4.646464646464646964e-01,8.850804644441313818e-02
4.747474747474748069e-01,8.846654689484430290e-02
4.848484848484848619e-01,8.840560678663762773e-02
4.949494949494949725e-01,8.828432016448335695e-02
5.050505050505050830e-01,9.369986266716502643e-02
5.151515151515152491e-01,1.056882145666451223e-01
5.252525252525253041e-01,1.191257694613233192e-01
5.353535353535353591e-01,1.342433595063813878e-01
5.454545454545455252e-01,1.512580905540885645e-01
5.555555555555555802e-01,1.704111551210637232e-01
5.656565656565657463e-01,1.919737212947300042e-01
5.757575757575758013e-01,2.162512472153007292e-01
5.858585858585858563e-01,2.435880676003294443e-01
5.959595959595960224e-01,2.743724563994427434e-01
6.060606060606060774e-01,3.090421516417679038e-01
6.161616161616162435e-01,3.480908818716091369e-01
6.262626262626262985e-01,3.920754805162938172e-01
6.363636363636364646e-01,4.416242504626740617e-01
6.464646464646465196e-01,4.974460425647312389e-01
6.565656565656565746e-01,5.603406773130260587e-01
6.666666666666667407e-01,6.312110499913414774e-01
6.767676767676767957e-01,7.110764931804756106e-01
6.868686868686869618e-01,8.010884431536048345e-01
6.969696969696970168e-01,9.025482051137376782e-01
7.070707070707071829e-01,1.016928672221132368e+00
7.171717171717172379e-01,1.145902827903636068e+00
7.272727272727272929e-01,1.291391023446174513e+00
7.373737373737374590e-01,1.455701649234010242e+00
7.474747474747475140e-01,1.642525766036575741e+00
7.575757575757576801e-01,1.685255379989781810e+00
7.676767676767677351e-01,1.683618864683168237e+00
7.777777777777777901e-01,1.681503373977317750e+00
7.878787878787879562e-01,1.679368306860015370e+00
7.979797979797980112e-01,1.677310156551637954e+00
8.080808080808081773e-01,1.675360372091971417e+00
8.181818181818182323e-01,1.673531771124646905e+00
8.282828282828283983e-01,1.671830418556766729e+00
8.383838383838384534e-01,1.670259406790410628e+00
8.484848484848485084e-01,1.668820479123177103e+00
8.585858585858586745e-01,1.667514545930272041e+00
8.686868686868687295e-01,1.666342011336522377e+00
8.787878787878788955e-01,1.665303164997819607e+00
8.888888888888889506e-01,1.664398156111801308e+00
8.989898989898991166e-01,1.663627050183570821e+00
9.090909090909091717e-01,1.662989780984702515e+00
9.191919191919192267e-01,1.662486376452393833e+00
9.292929292929293927e-01,1.662116803782710761e+00
9.393939393939394478e-01,1.661881104530422926e+00
9.494949494949496138e-01,1.661779292109798023e+00
9.595959595959596689e-01,1.661811412514059505e+00
9.696969696969697239e-01,1.661977527174854918e+00
9.797979797979798899e-01,1.662277860949515462e+00
9.898989898989899450e-01,1.662712605403870780e+00
1.000000000000000000e+00,1.663281959699258739e+00
};
\end{axis}
\end{tikzpicture}

%% file: Fig_reversalplot.tex
\begin{tikzpicture}[scale=0.9]
\begin{axis}[ymin=0, ymax=1.2, xmin=0, xmax=1,
  ytick={0,0.333333,0.6666666,1}, ytick pos=left,
  yticklabels={$0$,$1D$,$2D$,$3D$},
  xtick={0,1}, xtick pos=left,
  xticklabels={$0$,$L$},
  xlabel={$x$},
  ylabel={$U(x)$},
  legend pos=north west,
  legend style={draw=none}]
\addplot [line width=2pt,
 domain=0:0.25, 
 samples=100, 
 color=black,]
 {0};
 \addplot [line width=2pt,
 domain=0.25:0.5, 
 samples=10, 
 color=black,shorten >=-0.9pt,shorten <=-0.9pt]
 {1};
 \addplot[line width=2pt] coordinates{
                    (0.25,0)
                    (0.25,1)};
 \addplot[line width=2pt] coordinates{
                    (0.5,1)
                    (0.75,0)};
  \addplot[line width=2pt] coordinates{
                    (0.75,0)
                    (1,0)};
\end{axis}
\end{tikzpicture}

%% file: main.bbl
\begin{thebibliography}{49}%
\makeatletter
\providecommand \@ifxundefined [1]{%
 \@ifx{#1\undefined}
}%
\providecommand \@ifnum [1]{%
 \ifnum #1\expandafter \@firstoftwo
 \else \expandafter \@secondoftwo
 \fi
}%
\providecommand \@ifx [1]{%
 \ifx #1\expandafter \@firstoftwo
 \else \expandafter \@secondoftwo
 \fi
}%
\providecommand \natexlab [1]{#1}%
\providecommand \enquote  [1]{``#1''}%
\providecommand \bibnamefont  [1]{#1}%
\providecommand \bibfnamefont [1]{#1}%
\providecommand \citenamefont [1]{#1}%
\providecommand \href@noop [0]{\@secondoftwo}%
\providecommand \href [0]{\begingroup \@sanitize@url \@href}%
\providecommand \@href[1]{\@@startlink{#1}\@@href}%
\providecommand \@@href[1]{\endgroup#1\@@endlink}%
\providecommand \@sanitize@url [0]{\catcode `\\12\catcode `\$12\catcode
  `\&12\catcode `\#12\catcode `\^12\catcode `\_12\catcode `\%12\relax}%
\providecommand \@@startlink[1]{}%
\providecommand \@@endlink[0]{}%
\providecommand \url  [0]{\begingroup\@sanitize@url \@url }%
\providecommand \@url [1]{\endgroup\@href {#1}{\urlprefix }}%
\providecommand \urlprefix  [0]{URL }%
\providecommand \Eprint [0]{\href }%
\providecommand \doibase [0]{https://doi.org/}%
\providecommand \selectlanguage [0]{\@gobble}%
\providecommand \bibinfo  [0]{\@secondoftwo}%
\providecommand \bibfield  [0]{\@secondoftwo}%
\providecommand \translation [1]{[#1]}%
\providecommand \BibitemOpen [0]{}%
\providecommand \bibitemStop [0]{}%
\providecommand \bibitemNoStop [0]{.\EOS\space}%
\providecommand \EOS [0]{\spacefactor3000\relax}%
\providecommand \BibitemShut  [1]{\csname bibitem#1\endcsname}%
\let\auto@bib@innerbib\@empty
\bibitem [{\citenamefont {Marchetti}\ \emph {et~al.}(2013)\citenamefont
  {Marchetti}, \citenamefont {Joanny}, \citenamefont {Ramaswamy}, \citenamefont
  {Liverpool}, \citenamefont {Prost}, \citenamefont {Rao},\ and\ \citenamefont
  {Simha}}]{Marchetti2013Jul}%
  \BibitemOpen
  \bibfield  {author} {\bibinfo {author} {\bibfnamefont {M.~C.}\ \bibnamefont
  {Marchetti}}, \bibinfo {author} {\bibfnamefont {J.~F.}\ \bibnamefont
  {Joanny}}, \bibinfo {author} {\bibfnamefont {S.}~\bibnamefont {Ramaswamy}},
  \bibinfo {author} {\bibfnamefont {T.~B.}\ \bibnamefont {Liverpool}}, \bibinfo
  {author} {\bibfnamefont {J.}~\bibnamefont {Prost}}, \bibinfo {author}
  {\bibfnamefont {M.}~\bibnamefont {Rao}},\ and\ \bibinfo {author}
  {\bibfnamefont {R.~A.}\ \bibnamefont {Simha}},\ }\href
  {https://doi.org/10.1103/RevModPhys.85.1143} {\bibfield  {journal} {\bibinfo
  {journal} {Rev. Mod. Phys.}\ }\textbf {\bibinfo {volume} {85}},\ \bibinfo
  {pages} {1143} (\bibinfo {year} {2013})}\BibitemShut {NoStop}%
\bibitem [{\citenamefont {Doering}\ and\ \citenamefont
  {Gadoua}(1992)}]{Doering1992Oct}%
  \BibitemOpen
  \bibfield  {author} {\bibinfo {author} {\bibfnamefont {C.~R.}\ \bibnamefont
  {Doering}}\ and\ \bibinfo {author} {\bibfnamefont {J.~C.}\ \bibnamefont
  {Gadoua}},\ }\href {https://doi.org/10.1103/PhysRevLett.69.2318} {\bibfield
  {journal} {\bibinfo  {journal} {Phys. Rev. Lett.}\ }\textbf {\bibinfo
  {volume} {69}},\ \bibinfo {pages} {2318} (\bibinfo {year}
  {1992})}\BibitemShut {NoStop}%
\bibitem [{\citenamefont {Magnasco}(1993)}]{Magnasco1993Sep}%
  \BibitemOpen
  \bibfield  {author} {\bibinfo {author} {\bibfnamefont {M.~O.}\ \bibnamefont
  {Magnasco}},\ }\href {https://doi.org/10.1103/PhysRevLett.71.1477} {\bibfield
   {journal} {\bibinfo  {journal} {Phys. Rev. Lett.}\ }\textbf {\bibinfo
  {volume} {71}},\ \bibinfo {pages} {1477} (\bibinfo {year}
  {1993})}\BibitemShut {NoStop}%
\bibitem [{\citenamefont {Astumian}\ and\ \citenamefont
  {Bier}(1994)}]{Astumian1994Mar}%
  \BibitemOpen
  \bibfield  {author} {\bibinfo {author} {\bibfnamefont {R.~D.}\ \bibnamefont
  {Astumian}}\ and\ \bibinfo {author} {\bibfnamefont {M.}~\bibnamefont
  {Bier}},\ }\href {https://doi.org/10.1103/PhysRevLett.72.1766} {\bibfield
  {journal} {\bibinfo  {journal} {Phys. Rev. Lett.}\ }\textbf {\bibinfo
  {volume} {72}},\ \bibinfo {pages} {1766} (\bibinfo {year}
  {1994})}\BibitemShut {NoStop}%
\bibitem [{\citenamefont {Pavliotis}(2005)}]{Pavliotis2005Sep}%
  \BibitemOpen
  \bibfield  {author} {\bibinfo {author} {\bibfnamefont {G.~A.}\ \bibnamefont
  {Pavliotis}},\ }\href {https://doi.org/10.1016/j.physleta.2005.06.115}
  {\bibfield  {journal} {\bibinfo  {journal} {Phys. Lett. A}\ }\textbf
  {\bibinfo {volume} {344}},\ \bibinfo {pages} {331} (\bibinfo {year}
  {2005})}\BibitemShut {NoStop}%
\bibitem [{\citenamefont {Galajda}\ \emph {et~al.}(2007)\citenamefont
  {Galajda}, \citenamefont {Keymer}, \citenamefont {Chaikin},\ and\
  \citenamefont {Austin}}]{Galajda2007Dec}%
  \BibitemOpen
  \bibfield  {author} {\bibinfo {author} {\bibfnamefont {P.}~\bibnamefont
  {Galajda}}, \bibinfo {author} {\bibfnamefont {J.}~\bibnamefont {Keymer}},
  \bibinfo {author} {\bibfnamefont {P.}~\bibnamefont {Chaikin}},\ and\ \bibinfo
  {author} {\bibfnamefont {R.}~\bibnamefont {Austin}},\ }\href
  {https://doi.org/10.1128/JB.01033-07} {\bibfield  {journal} {\bibinfo
  {journal} {J. Bacteriol.}\ }\textbf {\bibinfo {volume} {189}},\ \bibinfo
  {pages} {8704} (\bibinfo {year} {2007})},\ \Eprint
  {https://arxiv.org/abs/17890308} {17890308} \BibitemShut {NoStop}%
\bibitem [{\citenamefont {Martin}\ \emph {et~al.}(2021)\citenamefont {Martin},
  \citenamefont {O'Byrne}, \citenamefont {Cates}, \citenamefont {Fodor},
  \citenamefont {Nardini}, \citenamefont {Tailleur},\ and\ \citenamefont {van
  Wijland}}]{Martin2021Mar}%
  \BibitemOpen
  \bibfield  {author} {\bibinfo {author} {\bibfnamefont {D.}~\bibnamefont
  {Martin}}, \bibinfo {author} {\bibfnamefont {J.}~\bibnamefont {O'Byrne}},
  \bibinfo {author} {\bibfnamefont {M.~E.}\ \bibnamefont {Cates}}, \bibinfo
  {author} {\bibfnamefont {{\ifmmode\acute{E}\else\'{E}\fi}.}~\bibnamefont
  {Fodor}}, \bibinfo {author} {\bibfnamefont {C.}~\bibnamefont {Nardini}},
  \bibinfo {author} {\bibfnamefont {J.}~\bibnamefont {Tailleur}},\ and\
  \bibinfo {author} {\bibfnamefont {F.}~\bibnamefont {van Wijland}},\ }\href
  {https://doi.org/10.1103/PhysRevE.103.032607} {\bibfield  {journal} {\bibinfo
   {journal} {Phys. Rev. E}\ }\textbf {\bibinfo {volume} {103}},\ \bibinfo
  {pages} {032607} (\bibinfo {year} {2021})}\BibitemShut {NoStop}%
\bibitem [{\citenamefont {Angelani}\ \emph {et~al.}(2011)\citenamefont
  {Angelani}, \citenamefont {Costanzo},\ and\ \citenamefont
  {Di~Leonardo}}]{Angelani2011Dec}%
  \BibitemOpen
  \bibfield  {author} {\bibinfo {author} {\bibfnamefont {L.}~\bibnamefont
  {Angelani}}, \bibinfo {author} {\bibfnamefont {A.}~\bibnamefont {Costanzo}},\
  and\ \bibinfo {author} {\bibfnamefont {R.}~\bibnamefont {Di~Leonardo}},\
  }\href {https://doi.org/10.1209/0295-5075/96/68002} {\bibfield  {journal}
  {\bibinfo  {journal} {EPL}\ }\textbf {\bibinfo {volume} {96}},\ \bibinfo
  {pages} {68002} (\bibinfo {year} {2011})}\BibitemShut {NoStop}%
\bibitem [{\citenamefont {Baek}(2019)}]{Baek2019}%
  \BibitemOpen
  \bibfield  {author} {\bibinfo {author} {\bibfnamefont {Y.}~\bibnamefont
  {Baek}},\ }\href@noop {} {\bibinfo {title} {{A multiscale approach to
  Brownian motors}}} (\bibinfo {year} {2019}),\ \bibinfo {note} {private
  communication}\BibitemShut {NoStop}%
\bibitem [{\citenamefont {Angelani}\ \emph {et~al.}(2009)\citenamefont
  {Angelani}, \citenamefont {Di~Leonardo},\ and\ \citenamefont
  {Ruocco}}]{Angelani2009Jan}%
  \BibitemOpen
  \bibfield  {author} {\bibinfo {author} {\bibfnamefont {L.}~\bibnamefont
  {Angelani}}, \bibinfo {author} {\bibfnamefont {R.}~\bibnamefont
  {Di~Leonardo}},\ and\ \bibinfo {author} {\bibfnamefont {G.}~\bibnamefont
  {Ruocco}},\ }\href {https://doi.org/10.1103/PhysRevLett.102.048104}
  {\bibfield  {journal} {\bibinfo  {journal} {Phys. Rev. Lett.}\ }\textbf
  {\bibinfo {volume} {102}},\ \bibinfo {pages} {048104} (\bibinfo {year}
  {2009})}\BibitemShut {NoStop}%
\bibitem [{\citenamefont {Koumakis}\ \emph {et~al.}(2014)\citenamefont
  {Koumakis}, \citenamefont {Maggi},\ and\ \citenamefont
  {Di~Leonardo}}]{Koumakis2014Jul}%
  \BibitemOpen
  \bibfield  {author} {\bibinfo {author} {\bibfnamefont {N.}~\bibnamefont
  {Koumakis}}, \bibinfo {author} {\bibfnamefont {C.}~\bibnamefont {Maggi}},\
  and\ \bibinfo {author} {\bibfnamefont {R.}~\bibnamefont {Di~Leonardo}},\
  }\href {https://doi.org/10.1039/C4SM00665H} {\bibfield  {journal} {\bibinfo
  {journal} {Soft Matter}\ }\textbf {\bibinfo {volume} {10}},\ \bibinfo {pages}
  {5695} (\bibinfo {year} {2014})}\BibitemShut {NoStop}%
\bibitem [{\citenamefont {Di~Leonardo}\ \emph {et~al.}(2010)\citenamefont
  {Di~Leonardo}, \citenamefont {Angelani}, \citenamefont {Dell{'}Arciprete},
  \citenamefont {Ruocco}, \citenamefont {Iebba}, \citenamefont {Schippa},
  \citenamefont {Conte}, \citenamefont {Mecarini}, \citenamefont {De~Angelis},\
  and\ \citenamefont {Di~Fabrizio}}]{DiLeonardo2010May}%
  \BibitemOpen
  \bibfield  {author} {\bibinfo {author} {\bibfnamefont {R.}~\bibnamefont
  {Di~Leonardo}}, \bibinfo {author} {\bibfnamefont {L.}~\bibnamefont
  {Angelani}}, \bibinfo {author} {\bibfnamefont {D.}~\bibnamefont
  {Dell{'}Arciprete}}, \bibinfo {author} {\bibfnamefont {G.}~\bibnamefont
  {Ruocco}}, \bibinfo {author} {\bibfnamefont {V.}~\bibnamefont {Iebba}},
  \bibinfo {author} {\bibfnamefont {S.}~\bibnamefont {Schippa}}, \bibinfo
  {author} {\bibfnamefont {M.~P.}\ \bibnamefont {Conte}}, \bibinfo {author}
  {\bibfnamefont {F.}~\bibnamefont {Mecarini}}, \bibinfo {author}
  {\bibfnamefont {F.}~\bibnamefont {De~Angelis}},\ and\ \bibinfo {author}
  {\bibfnamefont {E.}~\bibnamefont {Di~Fabrizio}},\ }\href
  {https://doi.org/10.1073/pnas.0910426107} {\bibfield  {journal} {\bibinfo
  {journal} {Proc. Natl. Acad. Sci. U.S.A.}\ }\textbf {\bibinfo {volume}
  {107}},\ \bibinfo {pages} {9541} (\bibinfo {year} {2010})}\BibitemShut
  {NoStop}%
\bibitem [{\citenamefont {Berthier}\ and\ \citenamefont
  {Kurchan}(2013)}]{Berthier2013May}%
  \BibitemOpen
  \bibfield  {author} {\bibinfo {author} {\bibfnamefont {L.}~\bibnamefont
  {Berthier}}\ and\ \bibinfo {author} {\bibfnamefont {J.}~\bibnamefont
  {Kurchan}},\ }\href {https://doi.org/10.1038/nphys2592} {\bibfield  {journal}
  {\bibinfo  {journal} {Nat. Phys.}\ }\textbf {\bibinfo {volume} {9}},\
  \bibinfo {pages} {310} (\bibinfo {year} {2013})}\BibitemShut {NoStop}%
\bibitem [{\citenamefont {Hagan}\ \emph {et~al.}(1989)\citenamefont {Hagan},
  \citenamefont {Doering},\ and\ \citenamefont {Levermore}}]{Hagan1989Oct}%
  \BibitemOpen
  \bibfield  {author} {\bibinfo {author} {\bibfnamefont {P.~S.}\ \bibnamefont
  {Hagan}}, \bibinfo {author} {\bibfnamefont {C.~R.}\ \bibnamefont {Doering}},\
  and\ \bibinfo {author} {\bibfnamefont {C.~D.}\ \bibnamefont {Levermore}},\
  }\href {http://www.jstor.org/stable/2102034} {\bibfield  {journal} {\bibinfo
  {journal} {SIAM J. Appl. Math.}\ }\textbf {\bibinfo {volume} {49}},\ \bibinfo
  {pages} {1480} (\bibinfo {year} {1989})}\BibitemShut {NoStop}%
\bibitem [{\citenamefont {Lindner}\ \emph {et~al.}(2001)\citenamefont
  {Lindner}, \citenamefont {Kostur},\ and\ \citenamefont
  {Schimansky-Geier}}]{Lindner2001Mar}%
  \BibitemOpen
  \bibfield  {author} {\bibinfo {author} {\bibfnamefont {B.}~\bibnamefont
  {Lindner}}, \bibinfo {author} {\bibfnamefont {M.}~\bibnamefont {Kostur}},\
  and\ \bibinfo {author} {\bibfnamefont {L.}~\bibnamefont {Schimansky-Geier}},\
  }\href {https://doi.org/10.1142/S0219477501000056} {\bibfield  {journal}
  {\bibinfo  {journal} {Fluct. Noise Lett.}\ }\textbf {\bibinfo {volume}
  {01}},\ \bibinfo {pages} {R25} (\bibinfo {year} {2001})}\BibitemShut
  {NoStop}%
\bibitem [{\citenamefont {Ghosh}\ \emph {et~al.}(2013)\citenamefont {Ghosh},
  \citenamefont {Misko}, \citenamefont {Marchesoni},\ and\ \citenamefont
  {Nori}}]{Ghosh2013Jun}%
  \BibitemOpen
  \bibfield  {author} {\bibinfo {author} {\bibfnamefont {P.~K.}\ \bibnamefont
  {Ghosh}}, \bibinfo {author} {\bibfnamefont {V.~R.}\ \bibnamefont {Misko}},
  \bibinfo {author} {\bibfnamefont {F.}~\bibnamefont {Marchesoni}},\ and\
  \bibinfo {author} {\bibfnamefont {F.}~\bibnamefont {Nori}},\ }\href
  {https://doi.org/10.1103/PhysRevLett.110.268301} {\bibfield  {journal}
  {\bibinfo  {journal} {Phys. Rev. Lett.}\ }\textbf {\bibinfo {volume} {110}},\
  \bibinfo {pages} {268301} (\bibinfo {year} {2013})}\BibitemShut {NoStop}%
\bibitem [{\citenamefont {Bressloff}\ and\ \citenamefont
  {Newby}(2013)}]{Bressloff2013Jan}%
  \BibitemOpen
  \bibfield  {author} {\bibinfo {author} {\bibfnamefont {P.~C.}\ \bibnamefont
  {Bressloff}}\ and\ \bibinfo {author} {\bibfnamefont {J.~M.}\ \bibnamefont
  {Newby}},\ }\href {https://doi.org/10.1103/RevModPhys.85.135} {\bibfield
  {journal} {\bibinfo  {journal} {Rev. Mod. Phys.}\ }\textbf {\bibinfo {volume}
  {85}},\ \bibinfo {pages} {135} (\bibinfo {year} {2013})}\BibitemShut
  {NoStop}%
\bibitem [{\citenamefont {Brenner}(1990)}]{Brenner1990Dec}%
  \BibitemOpen
  \bibfield  {author} {\bibinfo {author} {\bibfnamefont {H.}~\bibnamefont
  {Brenner}},\ }\href {https://doi.org/10.1021/la00102a001} {\bibfield
  {journal} {\bibinfo  {journal} {Langmuir}\ }\textbf {\bibinfo {volume} {6}},\
  \bibinfo {pages} {1715} (\bibinfo {year} {1990})}\BibitemShut {NoStop}%
\bibitem [{\citenamefont {Woillez}\ \emph {et~al.}(2019)\citenamefont
  {Woillez}, \citenamefont {Zhao}, \citenamefont {Kafri}, \citenamefont
  {Lecomte},\ and\ \citenamefont {Tailleur}}]{Woillez2019Jun}%
  \BibitemOpen
  \bibfield  {author} {\bibinfo {author} {\bibfnamefont {E.}~\bibnamefont
  {Woillez}}, \bibinfo {author} {\bibfnamefont {Y.}~\bibnamefont {Zhao}},
  \bibinfo {author} {\bibfnamefont {Y.}~\bibnamefont {Kafri}}, \bibinfo
  {author} {\bibfnamefont {V.}~\bibnamefont {Lecomte}},\ and\ \bibinfo {author}
  {\bibfnamefont {J.}~\bibnamefont {Tailleur}},\ }\href
  {https://doi.org/10.1103/PhysRevLett.122.258001} {\bibfield  {journal}
  {\bibinfo  {journal} {Phys. Rev. Lett.}\ }\textbf {\bibinfo {volume} {122}},\
  \bibinfo {pages} {258001} (\bibinfo {year} {2019})}\BibitemShut {NoStop}%
\bibitem [{\citenamefont {Bijnens}\ and\ \citenamefont
  {Maes}(2021)}]{Bijnens2021Mar}%
  \BibitemOpen
  \bibfield  {author} {\bibinfo {author} {\bibfnamefont {B.}~\bibnamefont
  {Bijnens}}\ and\ \bibinfo {author} {\bibfnamefont {C.}~\bibnamefont {Maes}},\
  }\href {https://doi.org/10.1088/1742-5468/abe29e} {\bibfield  {journal}
  {\bibinfo  {journal} {J. Stat. Mech.: Theory Exp.}\ }\textbf {\bibinfo
  {volume} {2021}}\bibinfo  {number} { (3)},\ \bibinfo {pages}
  {033206}}\BibitemShut {NoStop}%
\bibitem [{\citenamefont {Walter}\ \emph {et~al.}(2021)\citenamefont {Walter},
  \citenamefont {Pruessner},\ and\ \citenamefont {Salbreux}}]{Walter2021Jan}%
  \BibitemOpen
\bibfield  {number} {  }\bibfield  {author} {\bibinfo {author} {\bibfnamefont
  {B.}~\bibnamefont {Walter}}, \bibinfo {author} {\bibfnamefont
  {G.}~\bibnamefont {Pruessner}},\ and\ \bibinfo {author} {\bibfnamefont
  {G.}~\bibnamefont {Salbreux}},\ }\href
  {https://doi.org/10.1103/PhysRevResearch.3.013075} {\bibfield  {journal}
  {\bibinfo  {journal} {Phys. Rev. Res.}\ }\textbf {\bibinfo {volume} {3}},\
  \bibinfo {pages} {013075} (\bibinfo {year} {2021})}\BibitemShut {NoStop}%
\bibitem [{\citenamefont {Tjhung}\ \emph {et~al.}(2018)\citenamefont {Tjhung},
  \citenamefont {Nardini},\ and\ \citenamefont {Cates}}]{Tjhung2018Sep}%
  \BibitemOpen
  \bibfield  {author} {\bibinfo {author} {\bibfnamefont {E.}~\bibnamefont
  {Tjhung}}, \bibinfo {author} {\bibfnamefont {C.}~\bibnamefont {Nardini}},\
  and\ \bibinfo {author} {\bibfnamefont {M.~E.}\ \bibnamefont {Cates}},\ }\href
  {https://doi.org/10.1103/PhysRevX.8.031080} {\bibfield  {journal} {\bibinfo
  {journal} {Phys. Rev. X}\ }\textbf {\bibinfo {volume} {8}},\ \bibinfo {pages}
  {031080} (\bibinfo {year} {2018})}\BibitemShut {NoStop}%
\bibitem [{\citenamefont {Berger}\ \emph {et~al.}(2009)\citenamefont {Berger},
  \citenamefont {Schmiedl},\ and\ \citenamefont {Seifert}}]{Berger2009Mar}%
  \BibitemOpen
  \bibfield  {author} {\bibinfo {author} {\bibfnamefont {F.}~\bibnamefont
  {Berger}}, \bibinfo {author} {\bibfnamefont {T.}~\bibnamefont {Schmiedl}},\
  and\ \bibinfo {author} {\bibfnamefont {U.}~\bibnamefont {Seifert}},\ }\href
  {https://doi.org/10.1103/PhysRevE.79.031118} {\bibfield  {journal} {\bibinfo
  {journal} {Phys. Rev. E}\ }\textbf {\bibinfo {volume} {79}},\ \bibinfo
  {pages} {031118} (\bibinfo {year} {2009})}\BibitemShut {NoStop}%
\bibitem [{\citenamefont {Tailleur}\ and\ \citenamefont
  {Cates}(2008)}]{Tailleur2008May}%
  \BibitemOpen
  \bibfield  {author} {\bibinfo {author} {\bibfnamefont {J.}~\bibnamefont
  {Tailleur}}\ and\ \bibinfo {author} {\bibfnamefont {M.~E.}\ \bibnamefont
  {Cates}},\ }\href {https://doi.org/10.1103/PhysRevLett.100.218103} {\bibfield
   {journal} {\bibinfo  {journal} {Phys. Rev. Lett.}\ }\textbf {\bibinfo
  {volume} {100}},\ \bibinfo {pages} {218103} (\bibinfo {year}
  {2008})}\BibitemShut {NoStop}%
\bibitem [{\citenamefont {Zhang}\ and\ \citenamefont
  {Pruessner}(2021)}]{Zhang2021Jun}%
  \BibitemOpen
  \bibfield  {author} {\bibinfo {author} {\bibfnamefont {Z.}~\bibnamefont
  {Zhang}}\ and\ \bibinfo {author} {\bibfnamefont {G.}~\bibnamefont
  {Pruessner}},\ }\href {https://arxiv.org/abs/2106.07383v1} {\bibfield
  {journal} {\bibinfo  {journal} {arXiv}\ } (\bibinfo {year} {2021})},\ \Eprint
  {https://arxiv.org/abs/2106.07383} {2106.07383} \BibitemShut {NoStop}%
\bibitem [{\citenamefont {Feynman}\ \emph {et~al.}(1963)\citenamefont
  {Feynman}, \citenamefont {Leighton},\ and\ \citenamefont
  {Sands}}]{Feynman1963}%
  \BibitemOpen
  \bibfield  {author} {\bibinfo {author} {\bibfnamefont {R.}~\bibnamefont
  {Feynman}}, \bibinfo {author} {\bibfnamefont {R.}~\bibnamefont {Leighton}},\
  and\ \bibinfo {author} {\bibfnamefont {M.}~\bibnamefont {Sands}},\
  }\href@noop {} {\emph {\bibinfo {title} {{The Feynman Lecturers in
  Physics}}}}\ (\bibinfo  {publisher} {Addison-Wesley, Reading},\ \bibinfo
  {year} {1963})\BibitemShut {NoStop}%
\bibitem [{\citenamefont {Cates}(2012)}]{Cates2012Mar}%
  \BibitemOpen
  \bibfield  {author} {\bibinfo {author} {\bibfnamefont {M.~E.}\ \bibnamefont
  {Cates}},\ }\href {https://doi.org/10.1088/0034-4885/75/4/042601} {\bibfield
  {journal} {\bibinfo  {journal} {Rep. Prog. Phys.}\ }\textbf {\bibinfo
  {volume} {75}},\ \bibinfo {pages} {042601} (\bibinfo {year}
  {2012})}\BibitemShut {NoStop}%
\bibitem [{\citenamefont {Coppola}\ and\ \citenamefont
  {Kantsler}(2021)}]{Coppola2021Jul}%
  \BibitemOpen
  \bibfield  {author} {\bibinfo {author} {\bibfnamefont {S.}~\bibnamefont
  {Coppola}}\ and\ \bibinfo {author} {\bibfnamefont {V.}~\bibnamefont
  {Kantsler}},\ }\href {https://doi.org/10.1103/PhysRevE.104.014602} {\bibfield
   {journal} {\bibinfo  {journal} {Phys. Rev. E}\ }\textbf {\bibinfo {volume}
  {104}},\ \bibinfo {pages} {014602} (\bibinfo {year} {2021})}\BibitemShut
  {NoStop}%
\bibitem [{\citenamefont {Doi}(1976)}]{Doi1976Sep}%
  \BibitemOpen
  \bibfield  {author} {\bibinfo {author} {\bibfnamefont {M.}~\bibnamefont
  {Doi}},\ }\href {https://doi.org/10.1088/0305-4470/9/9/008} {\bibfield
  {journal} {\bibinfo  {journal} {J. Phys. A: Math. Gen.}\ }\textbf {\bibinfo
  {volume} {9}},\ \bibinfo {pages} {1465} (\bibinfo {year} {1976})}\BibitemShut
  {NoStop}%
\bibitem [{\citenamefont {Peliti}(1985)}]{Peliti1985Sep}%
  \BibitemOpen
  \bibfield  {author} {\bibinfo {author} {\bibfnamefont {L.}~\bibnamefont
  {Peliti}},\ }\href {https://doi.org/10.1051/jphys:019850046090146900}
  {\bibfield  {journal} {\bibinfo  {journal} {J. Phys.}\ }\textbf {\bibinfo
  {volume} {46}},\ \bibinfo {pages} {1469} (\bibinfo {year}
  {1985})}\BibitemShut {NoStop}%
\bibitem [{\citenamefont {Cardy}\ \emph {et~al.}(2008)\citenamefont {Cardy},
  \citenamefont {Falkovich},\ and\ \citenamefont {Gawedzki}}]{Cardy2008Dec}%
  \BibitemOpen
  \bibfield  {author} {\bibinfo {author} {\bibfnamefont {J.}~\bibnamefont
  {Cardy}}, \bibinfo {author} {\bibfnamefont {G.}~\bibnamefont {Falkovich}},\
  and\ \bibinfo {author} {\bibfnamefont {K.}~\bibnamefont {Gawedzki}},\
  }\href@noop {} {\emph {\bibinfo {title} {{Non-equilibrium Statistical
  Mechanics and Turbulence}}}}\ (\bibinfo  {publisher} {Cambridge University
  Press},\ \bibinfo {address} {Cambridge, England, UK},\ \bibinfo {year}
  {2008})\BibitemShut {NoStop}%
\bibitem [{\citenamefont
  {T{\ifmmode\ddot{a}\else\"{a}\fi}uber}(2014)}]{Tauber2014Mar}%
  \BibitemOpen
  \bibfield  {author} {\bibinfo {author} {\bibfnamefont {U.~C.}\ \bibnamefont
  {T{\ifmmode\ddot{a}\else\"{a}\fi}uber}},\ }\href@noop {} {\emph {\bibinfo
  {title} {{Critical Dynamics: A Field Theory Approach to Equilibrium and
  Non-Equilibrium Scaling Behavior}}}}\ (\bibinfo  {publisher} {Cambridge
  University Press},\ \bibinfo {address} {Cambridge, England, UK},\ \bibinfo
  {year} {2014})\BibitemShut {NoStop}%
\bibitem [{\citenamefont {Garcia-Millian}\ and\ \citenamefont
  {Pruessner}(2022)}]{Garcia-MillanPruessner:2022}%
  \BibitemOpen
  \bibfield  {author} {\bibinfo {author} {\bibfnamefont {R.}~\bibnamefont
  {Garcia-Millian}}\ and\ \bibinfo {author} {\bibfnamefont {G.}~\bibnamefont
  {Pruessner}},\ }\href@noop {} {\bibfield  {journal} {\bibinfo  {journal} {To
  be published}\ } (\bibinfo {year} {2022})}\BibitemShut {NoStop}%
\bibitem [{\citenamefont {Reimann}(2001)}]{Reimann2001May}%
  \BibitemOpen
  \bibfield  {author} {\bibinfo {author} {\bibfnamefont {P.}~\bibnamefont
  {Reimann}},\ }\href {https://doi.org/10.1103/PhysRevLett.86.4992} {\bibfield
  {journal} {\bibinfo  {journal} {Phys. Rev. Lett.}\ }\textbf {\bibinfo
  {volume} {86}},\ \bibinfo {pages} {4992} (\bibinfo {year}
  {2001})}\BibitemShut {NoStop}%
\bibitem [{\citenamefont {Razin}(2020)}]{Razin2020Sep}%
  \BibitemOpen
  \bibfield  {author} {\bibinfo {author} {\bibfnamefont {N.}~\bibnamefont
  {Razin}},\ }\href {https://doi.org/10.1103/PhysRevE.102.030103} {\bibfield
  {journal} {\bibinfo  {journal} {Phys. Rev. E}\ }\textbf {\bibinfo {volume}
  {102}},\ \bibinfo {pages} {030103} (\bibinfo {year} {2020})}\BibitemShut
  {NoStop}%
\bibitem [{\citenamefont {Press}\ \emph {et~al.}(2007)\citenamefont {Press},
  \citenamefont {Teukolsky}, \citenamefont {Vetterling},\ and\ \citenamefont
  {Flannery}}]{Press2007Sep}%
  \BibitemOpen
  \bibfield  {author} {\bibinfo {author} {\bibfnamefont {W.~H.}\ \bibnamefont
  {Press}}, \bibinfo {author} {\bibfnamefont {S.~A.}\ \bibnamefont
  {Teukolsky}}, \bibinfo {author} {\bibfnamefont {W.~T.}\ \bibnamefont
  {Vetterling}},\ and\ \bibinfo {author} {\bibfnamefont {B.~P.}\ \bibnamefont
  {Flannery}},\ }\href
  {https://www.cambridge.org/gb/academic/subjects/mathematics/numerical-recipes/numerical-recipes-art-scientific-computing-3rd-edition?format=HB&utm_source=shortlink&utm_medium=shortlink&utm_campaign=numericalrecipes}
  {\emph {\bibinfo {title} {{Numerical Recipes}}}}\ (\bibinfo  {publisher}
  {Cambridge University Press},\ \bibinfo {address} {Cambridge, England, UK},\
  \bibinfo {year} {2007})\BibitemShut {NoStop}%
\bibitem [{\citenamefont {Xin}\ \emph {et~al.}(2014)\citenamefont {Xin},
  \citenamefont {Liu},\ and\ \citenamefont {Li}}]{Xin2014Oct}%
  \BibitemOpen
  \bibfield  {author} {\bibinfo {author} {\bibfnamefont {H.}~\bibnamefont
  {Xin}}, \bibinfo {author} {\bibfnamefont {Q.}~\bibnamefont {Liu}},\ and\
  \bibinfo {author} {\bibfnamefont {B.}~\bibnamefont {Li}},\ }\href
  {https://doi.org/10.1038/srep06576} {\bibfield  {journal} {\bibinfo
  {journal} {Sci. Rep.}\ }\textbf {\bibinfo {volume} {4}},\ \bibinfo {pages}
  {1} (\bibinfo {year} {2014})}\BibitemShut {NoStop}%
\bibitem [{\citenamefont {Fallesen}\ \emph {et~al.}(2017)\citenamefont
  {Fallesen}, \citenamefont {Roostalu}, \citenamefont {Duellberg},
  \citenamefont {Pruessner},\ and\ \citenamefont {Surrey}}]{Fallesen2017Nov}%
  \BibitemOpen
  \bibfield  {author} {\bibinfo {author} {\bibfnamefont {T.}~\bibnamefont
  {Fallesen}}, \bibinfo {author} {\bibfnamefont {J.}~\bibnamefont {Roostalu}},
  \bibinfo {author} {\bibfnamefont {C.}~\bibnamefont {Duellberg}}, \bibinfo
  {author} {\bibfnamefont {G.}~\bibnamefont {Pruessner}},\ and\ \bibinfo
  {author} {\bibfnamefont {T.}~\bibnamefont {Surrey}},\ }\href
  {https://doi.org/10.1016/j.bpj.2017.09.006} {\bibfield  {journal} {\bibinfo
  {journal} {Biophys. J.}\ }\textbf {\bibinfo {volume} {113}},\ \bibinfo
  {pages} {2055} (\bibinfo {year} {2017})},\ \Eprint
  {https://arxiv.org/abs/29117528} {29117528} \BibitemShut {NoStop}%
\bibitem [{\citenamefont {Vizsnyiczai}\ \emph {et~al.}(2020)\citenamefont
  {Vizsnyiczai}, \citenamefont {Frangipane}, \citenamefont {Bianchi},
  \citenamefont {Saglimbeni}, \citenamefont {Dell{'}Arciprete},\ and\
  \citenamefont {Di~Leonardo}}]{Vizsnyiczai2020May}%
  \BibitemOpen
  \bibfield  {author} {\bibinfo {author} {\bibfnamefont {G.}~\bibnamefont
  {Vizsnyiczai}}, \bibinfo {author} {\bibfnamefont {G.}~\bibnamefont
  {Frangipane}}, \bibinfo {author} {\bibfnamefont {S.}~\bibnamefont {Bianchi}},
  \bibinfo {author} {\bibfnamefont {F.}~\bibnamefont {Saglimbeni}}, \bibinfo
  {author} {\bibfnamefont {D.}~\bibnamefont {Dell{'}Arciprete}},\ and\ \bibinfo
  {author} {\bibfnamefont {R.}~\bibnamefont {Di~Leonardo}},\ }\href
  {https://doi.org/10.1038/s41467-020-15711-0} {\bibfield  {journal} {\bibinfo
  {journal} {Nat. Commun.}\ }\textbf {\bibinfo {volume} {11}},\ \bibinfo
  {pages} {1} (\bibinfo {year} {2020})}\BibitemShut {NoStop}%
\bibitem [{\citenamefont {Sipos}\ \emph {et~al.}(2015)\citenamefont {Sipos},
  \citenamefont {Nagy}, \citenamefont {Di~Leonardo},\ and\ \citenamefont
  {Galajda}}]{Sipos2015Jun}%
  \BibitemOpen
  \bibfield  {author} {\bibinfo {author} {\bibfnamefont {O.}~\bibnamefont
  {Sipos}}, \bibinfo {author} {\bibfnamefont {K.}~\bibnamefont {Nagy}},
  \bibinfo {author} {\bibfnamefont {R.}~\bibnamefont {Di~Leonardo}},\ and\
  \bibinfo {author} {\bibfnamefont {P.}~\bibnamefont {Galajda}},\ }\href
  {https://doi.org/10.1103/PhysRevLett.114.258104} {\bibfield  {journal}
  {\bibinfo  {journal} {Phys. Rev. Lett.}\ }\textbf {\bibinfo {volume} {114}},\
  \bibinfo {pages} {258104} (\bibinfo {year} {2015})}\BibitemShut {NoStop}%
\bibitem [{\citenamefont {Singh}\ \emph {et~al.}(2017)\citenamefont {Singh},
  \citenamefont {Patteson}, \citenamefont {Purohit},\ and\ \citenamefont
  {Arratia}}]{Singh2017Oct}%
  \BibitemOpen
  \bibfield  {author} {\bibinfo {author} {\bibfnamefont {J.}~\bibnamefont
  {Singh}}, \bibinfo {author} {\bibfnamefont {A.~E.}\ \bibnamefont {Patteson}},
  \bibinfo {author} {\bibfnamefont {P.~K.}\ \bibnamefont {Purohit}},\ and\
  \bibinfo {author} {\bibfnamefont {P.~E.}\ \bibnamefont {Arratia}},\ }\href
  {https://arxiv.org/abs/1710.04068v1} {\bibfield  {journal} {\bibinfo
  {journal} {arXiv}\ } (\bibinfo {year} {2017})},\ \Eprint
  {https://arxiv.org/abs/1710.04068} {1710.04068} \BibitemShut {NoStop}%
\bibitem [{\citenamefont {Dyer}\ and\ \citenamefont
  {Ball}(2021)}]{Dyer2021May}%
  \BibitemOpen
  \bibfield  {author} {\bibinfo {author} {\bibfnamefont {O.~T.}\ \bibnamefont
  {Dyer}}\ and\ \bibinfo {author} {\bibfnamefont {R.~C.}\ \bibnamefont
  {Ball}},\ }\href {https://doi.org/10.1063/5.0049386} {\bibfield  {journal}
  {\bibinfo  {journal} {Phys. Fluids}\ }\textbf {\bibinfo {volume} {33}},\
  \bibinfo {pages} {051904} (\bibinfo {year} {2021})}\BibitemShut {NoStop}%
\bibitem [{\citenamefont {Gaspard}(2004)}]{Gaspard2004Nov}%
  \BibitemOpen
  \bibfield  {author} {\bibinfo {author} {\bibfnamefont {P.}~\bibnamefont
  {Gaspard}},\ }\href {https://doi.org/10.1007/s10955-004-3455-1} {\bibfield
  {journal} {\bibinfo  {journal} {J. Stat. Phys.}\ }\textbf {\bibinfo {volume}
  {117}},\ \bibinfo {pages} {599} (\bibinfo {year} {2004})}\BibitemShut
  {NoStop}%
\bibitem [{\citenamefont {Cocconi}\ \emph {et~al.}(2020)\citenamefont
  {Cocconi}, \citenamefont {Garcia-Millan}, \citenamefont {Zhen}, \citenamefont
  {Buturca},\ and\ \citenamefont {Pruessner}}]{Cocconi2020Nov}%
  \BibitemOpen
  \bibfield  {author} {\bibinfo {author} {\bibfnamefont {L.}~\bibnamefont
  {Cocconi}}, \bibinfo {author} {\bibfnamefont {R.}~\bibnamefont
  {Garcia-Millan}}, \bibinfo {author} {\bibfnamefont {Z.}~\bibnamefont {Zhen}},
  \bibinfo {author} {\bibfnamefont {B.}~\bibnamefont {Buturca}},\ and\ \bibinfo
  {author} {\bibfnamefont {G.}~\bibnamefont {Pruessner}},\ }\href
  {https://doi.org/10.3390/e22111252} {\bibfield  {journal} {\bibinfo
  {journal} {Entropy}\ }\textbf {\bibinfo {volume} {22}},\ \bibinfo {pages}
  {1252} (\bibinfo {year} {2020})}\BibitemShut {NoStop}%
\bibitem [{\citenamefont {Garcia-Millan}\ and\ \citenamefont
  {Pruessner}(2021)}]{Garcia-Millan2021Jun}%
  \BibitemOpen
  \bibfield  {author} {\bibinfo {author} {\bibfnamefont {R.}~\bibnamefont
  {Garcia-Millan}}\ and\ \bibinfo {author} {\bibfnamefont {G.}~\bibnamefont
  {Pruessner}},\ }\href {https://doi.org/10.1088/1742-5468/ac014d} {\bibfield
  {journal} {\bibinfo  {journal} {J. Stat. Mech.: Theory Exp.}\ }\textbf
  {\bibinfo {volume} {2021}}\bibinfo  {number} { (6)},\ \bibinfo {pages}
  {063203}}\BibitemShut {NoStop}%
\bibitem [{\citenamefont {Pietzonka}\ \emph {et~al.}(2019)\citenamefont
  {Pietzonka}, \citenamefont {Fodor}, \citenamefont {Lohrmann}, \citenamefont
  {Cates},\ and\ \citenamefont {Seifert}}]{Pietzonka2019Nov}%
  \BibitemOpen
\bibfield  {number} {  }\bibfield  {author} {\bibinfo {author} {\bibfnamefont
  {P.}~\bibnamefont {Pietzonka}}, \bibinfo {author} {\bibfnamefont
  {{\ifmmode\acute{E}\else\'{E}\fi}.}~\bibnamefont {Fodor}}, \bibinfo {author}
  {\bibfnamefont {C.}~\bibnamefont {Lohrmann}}, \bibinfo {author}
  {\bibfnamefont {M.~E.}\ \bibnamefont {Cates}},\ and\ \bibinfo {author}
  {\bibfnamefont {U.}~\bibnamefont {Seifert}},\ }\href
  {https://doi.org/10.1103/PhysRevX.9.041032} {\bibfield  {journal} {\bibinfo
  {journal} {Phys. Rev. X}\ }\textbf {\bibinfo {volume} {9}},\ \bibinfo {pages}
  {041032} (\bibinfo {year} {2019})}\BibitemShut {NoStop}%
\bibitem [{\citenamefont {Gibbs}(1898)}]{Gibbs1898Dec}%
  \BibitemOpen
  \bibfield  {author} {\bibinfo {author} {\bibfnamefont {J.~W.}\ \bibnamefont
  {Gibbs}},\ }\href {https://doi.org/10.1038/059200b0} {\bibfield  {journal}
  {\bibinfo  {journal} {Nature}\ }\textbf {\bibinfo {volume} {59}},\ \bibinfo
  {pages} {200} (\bibinfo {year} {1898})}\BibitemShut {NoStop}%
\bibitem [{\citenamefont {T{\ifmmode\ddot{a}\else\"{a}\fi}uber}\ \emph
  {et~al.}(2005)\citenamefont {T{\ifmmode\ddot{a}\else\"{a}\fi}uber},
  \citenamefont {Howard},\ and\ \citenamefont {Vollmayr-Lee}}]{Tauber2005Apr}%
  \BibitemOpen
  \bibfield  {author} {\bibinfo {author} {\bibfnamefont {U.~C.}\ \bibnamefont
  {T{\ifmmode\ddot{a}\else\"{a}\fi}uber}}, \bibinfo {author} {\bibfnamefont
  {M.}~\bibnamefont {Howard}},\ and\ \bibinfo {author} {\bibfnamefont {B.~P.}\
  \bibnamefont {Vollmayr-Lee}},\ }\href
  {https://doi.org/10.1088/0305-4470/38/17/r01} {\bibfield  {journal} {\bibinfo
   {journal} {J. Phys. A: Math. Gen.}\ }\textbf {\bibinfo {volume} {38}},\
  \bibinfo {pages} {R79} (\bibinfo {year} {2005})}\BibitemShut {NoStop}%
\bibitem [{\citenamefont {Inc.}(2021)}]{Mathematica}%
  \BibitemOpen
  \bibfield  {author} {\bibinfo {author} {\bibfnamefont {W.~R.}\ \bibnamefont
  {Inc.}},\ }\href {https://www.wolfram.com/mathematica} {\bibinfo {title}
  {Mathematica, {V}ersion 13.0.0}} (\bibinfo {year} {2021}),\ \bibinfo {note}
  {champaign, IL, 2021}\BibitemShut {NoStop}%
\end{thebibliography}%
